\documentclass[aps,prb,twocolumn,amsmath,amssymb,superscriptaddress,floatfix,10pt]{revtex4-1}

\usepackage{booktabs}
\usepackage{graphicx}
\usepackage{amsmath}
\usepackage{amstext}
\usepackage{amssymb}
\usepackage{romannum}
\usepackage[colorlinks,citecolor=blue]{hyperref}
\usepackage{amsfonts}
\usepackage{longtable,booktabs}
\usepackage{url}
\usepackage{subfigure}%
\usepackage{dsfont}
\usepackage{datetime}

\usepackage{amsbsy}
\usepackage{dcolumn}
\usepackage{amsthm}
\usepackage{bm}
\usepackage{esint}
\usepackage{multirow}
\usepackage{hyperref}
\hypersetup{
    colorlinks=true,
    linkcolor=magenta,
    filecolor=cyan,
    urlcolor=blue,
}
\usepackage{cleveref}
\usepackage{comment}
\usepackage{mathrsfs}
\usepackage{color}
\newcounter{defcounter}
\usepackage{amsthm}

  \usepackage{extarrows}
 \usepackage{graphicx}
    \usepackage{amssymb}
    \usepackage{amsthm}
    \usepackage{mathtools}
    \usepackage{adjustbox}

\usepackage{hyperref}
\hypersetup{
    colorlinks=true,
    linkcolor=blue,
    filecolor=blue,
    urlcolor=blue,
}

\def\l{\lambda}

\def\a{\alpha}

\def\p{\partial}
\def\m{\mathcal}
\def\D{\Delta}
\def\d{\delta}
\def\T{\Theta}
\def\s{\sigma}

\def\t{\theta}

\def\R{\Romannum}

\def\b{\boldsymbol}

\begin{document}
\pagenumbering{arabic}
\title{A coupled wire description of surface \texorpdfstring{$ADE$}{ADE} topological orders}

\author{Bo Han}
\affiliation{Department of Physics, University of Illinois at Urbana-Champaign, Urbana, IL 61801, USA}
\author{Jeffrey C. Y. Teo}
\affiliation{Department of Physics, University of Virginia, Charlottesville, VA 22904, USA}
\date{\today}

\begin{abstract}
Symmetry-protected and symmetry-enriched topological (SPT/SET) phases in three dimensions are quantum systems that support non-trivial two-dimensional (2D) surface states. These surface states develop finite excitation energy gaps when the relevant symmetries are broken. On the other hand, one-dimensional (1D) gapless modes can populate along interfaces that separate adjacent gapped surface domains with distinct symmetry-breaking orders. A surface strip pattern in general reduces the low-energy SPT/SET surface degrees of freedom onto a 2D array of gapless 1D channels. These channels can be coupled to one another by quasiparticle tunneling, and these interwire interactions collectively provide an effective description of the surface state. In this paper, we study a general class of symmetry-preserving or breaking SPT/SET surface states that admit finite excitation energy gaps and Abelian topological orders via the coupled-wire construction. In particular, we focus on the prototype Abelian surface topological orders that fall under the $ADE$ classification of simply-laced Lie algebras. We also elaborate on the emergent symmetry and duality properties of the coupled-wire models.
\end{abstract}


\maketitle


\section{Introduction}
Topological phases of matter have been drawing attention in the past decade of both condensed matter and high energy physicists since the discovery of topological insulators (TIs) and superconductors (TSCs)~\cite{QiZhangreview11,HasanKane10}. They are attractive not only because they present exotic properties in theory, but also because some of these phenomena can be verified in materials. They introduce new frontiers in previously well-studied physical concepts, such as quantum phase transitions in condensed matter physics and quantum anomalies in high energy physics.

Topological phases are quantum phases that do not adiabatically connect to trivial ones. The ground states of these phases are quantum mechanically entangled to an extent that any deformation path connecting a topological state and a trivial state must go through a quantum phase transition where the bulk excitation energy gap closes. For example, a topological insulating phase must be separated from a normal insulating phase by a gapless Dirac/Weyl (semi)metallic phase or critical point~\cite{Murakami2007}. This is intimately related to the fact that, generically in three-dimensional (3D) real space, a topological material and a normal one are distinctly separated by an anomalous two-dimensional (2D) surface. For example, the gapless Dirac surface state provides a definitive measurable signature of a topological insulator~\cite{Xia:2009uq}. Some topological phases require the presence of symmetries. For example, topological insulators rely on time reversal symmetry, which protects the Kramers degeneracy of the surface Dirac point, and charge conservation, which disallows pairing. In general, symmetries provide a finer classification of topological phases by forbidding deformation paths that violate them. These phases are referred to as symmetry-protected or symmetry-enriched topological (SPT/SET) phases depending on whether the 3D bulk support integral or fractional quasiparticle excitations. However, this paper will not focus on the distinction between SPT and SET, and its general results will be applicable to both situations.

The surface of an SPT/SET state can obtain a finite excitation energy gap by (a) breaking the relevant symmetry, or (b) developing a surface topological order that supports fractional surface quasiparticle excitations that are absent in the bulk. For example, the Dirac surface state of a topological insulator can acquire a finite Dirac mass by breaking time reversal or a superconducting pairing gap by breaking charge conservation. On the other hand, it can gain a many-body energy gap while preserving all symmetries. However, the symmetric surface must carry topological order, such as the $T$-Pfaffian, that supports quasiparticle and charge fractionalization~\cite{WangPotterSenthilgapTI13,MetlitskiKaneFisher13b,ChenFidkowskiVishwanath14,BondersonNayakQi13}. The main focus of this paper is to develop an exactly solvable model technique in describing a collection of prototype classes of Abelian SPT/SET surface states.

We will focus on three classes of surface states that correspond to the $ADE$ classification of simply-laced Lie algebra~\cite{bigyellowbook}. These simple affine Lie algebras at level 1 were explored as conformal field theories that effectively describe the (1+1)-dimensional [(1+1)D] boundary edge states of $2+1$D Abelian topological phases~\cite{khan2014,KhanTeoHughesVishveshwara16}. In this paper, we discover a relationship between the $ADE$ classification and SPT/SET surface states. The $A$ class corresponds to a series of charge U(1) conserving gapped surface states that live on the symmetry breaking boundary surfaces of topological (crystalline) insulators~\cite{Ando_Fu_TCI_review} or fractional topological insulators~\cite{MaciejkoQiKarchZhang10}. The $D$ class corresponds to a series of superconducting gapped surface states of topological superconductors~\cite{SchnyderRyuFurusakiLudwig08,Kitaevtable08}.  The $E$ class corresponds to three exceptional surface states of a topological paramagnet~\cite{VishwanathSenthil12,WangPotterSenthil13}. For simplicity, we only consider Abelian surface topological orders, whose quasiparticle excitations can be fractional but cannot support non-local quantum information storage. The non-simply-laced simple Lie algebras in the $B,C,F,G$ series correspond to non-Abelian surface topological orders, and will not be addressed in this paper.

We will explore these correspondences using the exactly solvable coupled-wire model technique on the surface SPTs/SETs. In general, coupled-wire models may have several advantages compared with the more conventional pure field-theoretic approaches. One can write microscopic many-body interacting Hamiltonians explicitly in terms of local electronic degrees of freedom. In many situations, these Hamiltonians can be theoretically designed in a way so that they are exactly solvable and do not require a mean-field approximation. In addition, one can also perform explicit symmetry and duality transformations on the local fields and study the topological properties of the ground states, quasiparticle excitations as well as their braiding statistics. 

Generalizing sliding Luttinger liquid theories~\cite{OHernLubenskyToner99,EmeryFradkinKivelsonLubensky00,VishwanathCarpentier01,SondhiYang01,MukhopadhyayKaneLubensky01}, the coupled wire construction was first developed in Ref.~\onlinecite{Kane2002CoupledWire} to study the Laughlin~\cite{Laughlin83} and Haldane-Halperin hierarchy~\cite{Haldane83,Halperin84} fractional quantum Hall (FQH) states. Later this construction was applied to non-Abelian FQH states~\cite{Teo2014CoupledWire,SagiOregSternHalperin15,KaneSternHalperin17,Kane2018Z4orbifold,Klinovaja2014qshe,Klinovaja2015qahe}, anyon systems~\cite{khan2014,OregSelaStern14,Hong2017TCIsurface}, spin liquids~\cite{MengNeupertGreiterThomale15}, studies on duality~\cite{Mross2016DiracQED3} and many other areas in two spatial dimensions. Recently, the coupled wire construction has also been applied to study three spatial dimensional Abelian and non-Abelian topological systems~\cite{Meng15,IadecolaNeupertChamonMudry16,IadecolaNeupertChamonMudry17}, Dirac (semi)metals~\cite{Raza2017DSMcoupledwire}, Weyl (semi)metals~\cite{Vazifeh13}, Dirac superconductors~\cite{ParkRazaGilbertTeo2018} and other strongly correlated fractional topological systems~\cite{MengGrushinShtengelBardarson16}. 

\begin{figure}[htbp]
\includegraphics[width=0.4\textwidth]{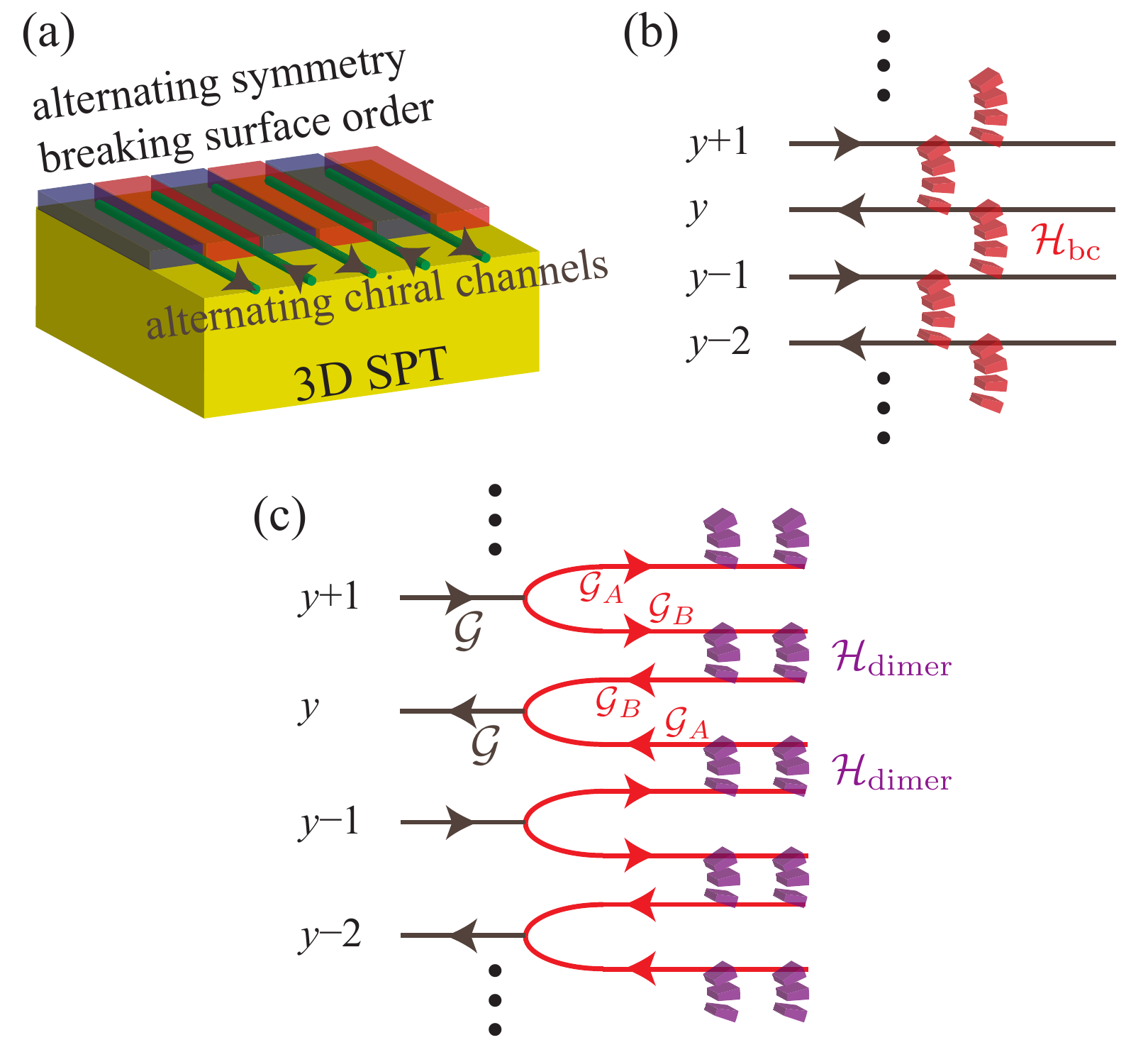}
\caption{Coupled wire description of a topological surface state. (a) Emergence of surface channels through alternating symmetry breaking. (b) Gapless surface state resulting from uniform competing inter-channel backscattering $\mathcal{H}_{\mathrm{bc}}$. (c) Surface gapping through channel bipartition and non-competing inter-channel dimerization $\mathcal{H}_{\mathrm{dimer}}$.}\label{fig:surfaceCW}
\end{figure}

The application of the coupled-wire technique on the surface of an SPT/SET relies on an anisotropic reduction of low-energy surface degrees of freedom onto a 2D array of parallel 1D wires. The simplest example was demonstrated on the surface a topological insulator~\cite{MrossEssinAlicea15} with a magnetic surface stripe order with alternating magnetic orientations (see Fig.~\ref{fig:surfaceCW}). The Dirac surface state becomes massive in the interior of each magnetic strip. This leaves behind chiral Dirac channels with alternating propagating directions that live along the interfaces between strips where the magnetic order flips. A similar construction was also applied to the surface of topological superconductors~\cite{Sahoo2016coupledwireTSC}. In this paper, instead of deriving from the 3D bulk of an SPT/SET, we begin with the assumption that an array of chiral channels -- each described by certain conformal field theory (CFT) related to one of the $ADE$ affine Lie algebras at level 1 -- can be generated by similar alternating symmetry-breaking stripe order on the surface of an SPT/SET. This assumption can be verified in the three prototype examples of topological (crystalline) insulators, superconductors and paramagnets mentioned above. On the other hand, it may also be applicable to other more exotic types of SPT/SET such as fractional topological insulators and superconductors.

\subsection{Summary of results}\label{sec:introsummary}
Figure~\ref{fig:surfaceCW} summarizes the coupled wire models that describes the surface $ADE$ topological orders of SPTs/SETs. The surface state of a generic SPT/SET gains a finite excitation energy gap in the interior of each symmetry-breaking strip. The remaining gapless degrees of freedom are localized along 1D interfaces between adjacent strips with distinct symmetry-breaking orders. The low-energy degrees of freedom along each interface are effectively described by a conformal field theory (CFT), or more precisely, an affine Kac-Moody current algebra (also known as an affine Lie algebra~\cite{bigyellowbook} or Wess-Zumino-Witten (WZW) theory~\cite{WessZumino71,WittenWZW}). In single-body mean-field topological band insulators and superconductors, the gapless modes along these line interfaces, or line defects in general, were completely classified~\cite{TeoKane}. Such an interface hosts a number of copies of chiral Dirac (or Majorana) fermions that propagate in a single-direction and is described by a $U(N)_1$ (respectively~$SO(N)_1$) current algebra. However, our surface wire construction does not only restrict to the non-interacting case. It also applies to general SPTs/SETs such as fractional topological insulators, which lead to fractional surface parton Dirac $U(N)_1/\mathbb{Z}_N$ orbifold channels~\cite{Sahoo2017FTI,SirotaSahooChoTeo18}, and topological paramagnets, which lead to surface $E_8$ channels~\cite{BurnellChenFidkowskiVishwanath13}. 

In this paper, we explore the possible surface interactions that lead to non-trivial Abelian surface topological orders regardless of whether the interactions preserve or break the relevant symmetries of the underlying SPT/SET. In other words, the surface topological orders are not necessarily anomalous and for some cases, are realizable in non-holographic pure 2D systems. Instead, we are interested in surface states that facilitate non-trivial quasiparticle fractionalization through surface many-body interactions. The coupled wire construction provides an exact solvable description of such interactions. The oscillating symmetry-breaking surface stripe order requires the propagating directions of the gapless interface channels to alternate. The gapping interactions are theoretically constructed (see Fig.~\ref{fig:surfaceCW}(c)) by first decomposing the current algebra $\mathcal{G}$ along each interface channel into two decoupled fractional components \begin{align}\mathcal{G}\sim\mathcal{G}_A\times\mathcal{G}_B,\label{Gdecomposition}\end{align} and subsequently backscattering the two current components to adjacent interfaces in opposite directions \begin{align}\mathcal{H}=u\sum_{\mbox{\footnotesize $y$ even}}{\bf J}^y_{\mathcal{G}_A}\cdot{\bf J}^{y+1}_{\mathcal{G}_A}+u\sum_{\mbox{\footnotesize $y$ odd}}{\bf J}^y_{\mathcal{G}_B}\cdot{\bf J}^{y+1}_{\mathcal{G}_B}.\end{align} The collection of backscattering interaction between fractional Kac-Moody currents is a $2+1$D generalization of the $1+1$D AKLT spin chain~\cite{AKLT1,AKLT2}, and leads to fractional gapped quasiparticle excitations. We apply the models to the $A$, $D$ and $E$ series, where the decomposition \eqref{Gdecomposition} is given by \eqref{UNpartition} for the $A$ classes, \eqref{eqn:SOdecom1} for the $D$ classes, and \eqref{E8decomposition} for exceptional $E$ classes.

In addition to the exactly solvable model, the coupled wire construction also provides an explicit description of symmetries and dualities. Although time reversal symmetry is necessarily broken by each chiral channel, the array of channels with alternating propagating directions collectively recovers an emergent antiferromagnetic time reversal (AFTR) symmetry, which accompanies local time reversal with a half-translation $y\to y+1$. We will elaborate on how the AFTR symmetry is preserved in the $D$ class and how it is broken in the $A$ and $E$ classes. On the other hand, duality is also a central theme in theoretical physics. It is a powerful technique that relates distinct theories with no $a priori$ common origins. For example, the order and disorder (i.e.~low- and high- temperature) phases of the 2D classical Ising model are related by the Kramers-Wannier duality.~\cite{Baxterbook} Duality provides a field theoretical mapping between weakly and strongly interacting phases. Recently, there has been some work on non-supersymmetric dualities at the field theoretical level~\cite{Hsin2016SUduality,Aharony2017SOSPduality,Seiberg2016Dualityweb,Karch2016PVduality}and the concept of duality has also been established in a coupled wire description of composite Dirac fermions~\cite{Mross2016DiracQED3}. In this paper, we perform similar constructions to the gapped surface $ADE$ topological orders. Although it is mentioned in the Introduction, Table~\ref{tab:3dbulk} summarizes the 3D bulk SPT/SET phases corresponding to the ADE classifications of surface topological orders discussed in this paper. For a coupled wire construction of these 3D bulk systems, we will discuss it in a separate paper. 

\begin{table}[h]
\label{tab:3dbulk}
\centering
\begin{tabular}{|c|c|c|}
\hline \hline
Classification & 3d bulk SPT/SET & Section in this paper  \\
\hline
Class A & TCI, FTI & Sec.~\ref{sec:A-series}  \\
Class D & TSC & Sec.~\ref{sec:Dseries} \\
Class E & TP, $E_8$ QH~\cite{LopesQuitoHanTeo2019E8G2F4} & Sec.\ref{sec:E-series} \\
\hline \hline
\end{tabular}
\caption{3D bulk SPT/SET topological phases corresponding to the surface topological orders of ADE classifications discussed in this paper. TCI=topological crystalline insulator, FTI=fractional topological insulator, TSC= topological superconductor, TP=topological paramagnet~\cite{VishwanathSenthil12,WangPotterSenthil13}, QH = quantum Hall.}
\end{table}

The outline of this paper is as follows. In Sec.~\ref{sec:Generalcoupledwire}, we explicitly demonstrate the coupled wire construction in two simple and specific examples, and elaborate on the central themes that can be systematically carried over to the general scenarios. Sec.~\ref{sec:Dirac-QED3} briefly reviews the coupled-wire derivation of the duality between free Dirac fermion and QED$_3$ proposed in Ref.~\onlinecite{Mross2016DiracQED3}. Sec.~\ref{sec:Dseries} reviews the coupled wire models for surface Majorana fermions discussed in Ref.~\onlinecite{Sahoo2016coupledwireTSC} and discuss their duality properties. Next, we introduce the topological orders and duality properties of the $A$ and $E$ classes systematically in Sec.~\ref{sec:A-series} and \ref{sec:E-series} respectively. Sec.~\ref{sec:DisCon} concludes this paper and provides further discussions as well as future directions. Appendix~\ref{append:HaldaneCondition} is a brief review of the Haldane's nullity gapping condition~\cite{Haldane1995Haldanecondition} for bosonized sine-Gordon models. Appendix~\ref{append:Liealgebra} contains the relevant background information of the {\it ADE} classifications and their representations. 

\section{General coupled wire construction of surface gapping interactions}\label{sec:Generalcoupledwire}
The coupled wire construction provided exactly solvable many-body interacting models of surface states of symmetry protected topological (SPT) phases. Examples include the $T$-Pfaffian surface state of a topological insulator~\cite{MrossEssinAlicea15}, and the $SO(3)_3$-like surface state of a topological superconductor~\cite{Sahoo2016coupledwireTSC}. These surface states preserve the relevant symmetries of the SPT phase. The $T$-Pfaffian surface state~\cite{MetlitskiKaneFisher13b,WangPotterSenthilgapTI13,ChenFidkowskiVishwanath14,BondersonNayakQi13} preserves time-reversal and charge conservation, while the $SO(3)_3$-like superconducting surface preserve time-reversal~\cite{LukaszChenVishwanath}. They arise as a consequence of strong many-body interaction beyond the single-body mean field description. The massless Dirac (Majorana) fermion on the surface of a topological insulator (respectively superconductor) cannot acquire a single-body mass term without breaking the relevant symmetries. In general, the surface state of a SPT phase can only develop a finite excitation energy gap while preserving symmetries by many-body interactions that introduce additional surface topological order. This allows fractional quasiparticle surface excitations to emerge that carry fractional properties, such as electric charge and exchange statistics. For example, the $T$-Pfaffian surface state supports excitations with fractionally quantized electric charge in units of $e/4$.

The coupled wire description of topological surface states is based on an anisotropic surface arrangement where the relevant symmetries emerge in the long wavelength low energy limit. The surface of a topological insulator (superconductor) can be mimicked by an array of 1D chiral Dirac (respectively Majorana) channels with alternating propagating directions (see Fig.~\ref{fig:surfaceCW}). Electronic quasiparticles propagate continuously along each channel and tunnel discretely from one wire to the next. The inter-channel tunneling amplitude is suppressed by an energy barrier, which comes from symmetry breaking interactions that remove or integrate out low-energy electronic degrees of freedom in the surface strips between channels. For example, the symmetry breaking interactions are given by the Dirac (Majorana) mass on the surface of a topological insulator (respectively superconductor). The symmetry breaking interactions correspond to order parameters, such as magnetization or pairing phase. These symmetry breaking order parameters alternate from strips to strips (see Fig.~\ref{fig:surfaceCW}(a)). For example, the surface magnetization flips between adjacent strips. Consequently, the 1D interface, where the Dirac mass changes sign, bounds the chiral Dirac mode in low-energy. Similarly, the pairing phase conjugates from one strip to the next, and therefore the interface between adjacent surface strips hosts the chiral Majorana mode. Symmetry is restored in an ``antiferromagnetic" manner because the order parameters are conjugated by the symmetry between neighboring strips and the propagating directions are reversed by the symmetry between neighboring channels.

The coupled-wire Hamiltonian consists of the kinetic energy of each chiral channel $\mathcal{H}_{\mathrm{KE}}^y$ and backscattering coupling potentials $\mathcal{H}_{\mathrm{bc}}^{y+1/2}$ between neighboring channels, where each channel is labeled by an integer $y$ that represents its vertical position in the array (see Fig.~\ref{fig:surfaceCW}(b) and (c)). The antiferromagnetic symmetry requires 
the inter-channel backscatterings to have uniform strength. In other words, symmetry forbids inter-channel dimerization, where counter-propagating channels are pairwise coupled. Under a dimerization where the strength of $\mathcal{H}_{\mathrm{bc}}^{y+1/2}$ alternates between even and odd $y$, the surface state acquires a symmetry breaking energy gap. Similar to the Su-Schrieffer-Heeger model~\cite{SSH}, there are two topologically distinct gapped phases -- one where $\mathcal{H}_{\mathrm{bc}}^{y+1/2}$ is stronger for even $y$ and channels are paired between $y=2n$ and $2n+1$, and the other where $\mathcal{H}_{\mathrm{bc}}^{y+1/2}$ is stronger for odd $y$ and channels are paired between $y=2n-1$ and $2n$. The critical point that separates these two phases has uniform $\mathcal{H}_{\mathrm{bc}}^{y+1/2}$ (see Fig.~\ref{fig:surfaceCW}(b)). It preserves the relevant symmetry and has vanishing energy gap. For example, the array of chiral Dirac (Majorana) channels under uniform inter-channel coupling recovers the massless Dirac (respectively Majorana) fermions on the surface of a topological insulator (respectively superconductor).

The uniform backscattering model that preserves the antiferromagnetic symmetry is gapless because adjacent backscattering terms compete. Moreover, the antiferromagnetic symmetry forbids any channel dimerization. On the other hand, if each channel can be fractionalized and bipartitioned into two decoupled components, then they can be backscattered and dimerized in opposite directions (see Fig.~\ref{fig:surfaceCW}(c)). This is a higher dimensional analog of the Haldane integral spin chain~\cite{Haldanespinchain1,Haldanespinchain2} and the AKLT spin chain~\cite{AKLT1,AKLT2}, where the integral spin on each site is fractionalized into a pair of half-integral spins and they are independently dimerized with neighboring ones. The backscattering of these fractional degrees of freedom is now non-competing because they act on orthogonal Hilbert spaces. Moreover, the antiferromagnetic symmetry is preserved if the dimerization strength $\mathcal{H}_{\mathrm{dimer}}^{y+1/2}$ is uniform. The channel fractionalization is stabilized by the many-body inter-channel backscattering $\mathcal{H}_{\mathrm{dimer}}^{y+1/2}$, which is a combination of products of local electronic operators.

In this paper, we consider a variety of SPT phases, whose surface state can be mapped into an array of integral electronic channels. The SPT phase could be protected by certain combinations of global symmetries such as time-reversal and local symmetries represented by a continuous group. Instead of elaborating on the 3D SPT phases, we target surface topological order and begin with the general assumption that the surface array of chiral channels is supported by some unknown 3D SPT bulk. In particular, we focus on situations where these channels can be bosonized. Before inter-channel coupling, each channel can be described in low-energy by a conformal field theory (\hypertarget{CFT}{CFT}), which falls under the {\it ADE} classification of affine Lie algebra~\cite{bigyellowbook} at level one. The $A$-series consists of the Lie algebras $A_r=SU(r+1)$, where $r$ is the rank of the algebra. The $D$-series consists of $D_r=SO(2r)$, and the $E$-series consists of the exceptional $E_6$, $E_7$ and $E_8$. These algebras form the fractional degrees of freedom under the bipartition of channels. Their general construction will be discussed in upcoming sections. In this section, we present the main ideas in the coupled wire construction by demonstrating the $A_3=SU(4)$ and $D_3=SO(6)$ case.

\subsection{\texorpdfstring{$SO(6)$}{SO(6)} and \texorpdfstring{$U(4)$}{U(4)} as illustrative examples}
In this subsection, we take the $SO(6)$ and $U(4)$ surface models as examples to illustrate the coupled wire construction. In particular, we demonstrate the inter-channel backscattering sine-Gordon interactions. The ground state of each of these interactions exhibits an angle order parameter, which is the ground state expectation value of the angle variable in the sine-Gordon potential. These angle order parameters can take discrete values in a lattice, which will be referred to as the ``Haldane's dual lattice". We also present the fractional gapped excitations that corresponds to deconfined kinks of the sine-Gordon interactions. These excitations can be created or destroyed by bosonized vertex operators, whose exponents lie also in the dual lattice.

We begin with the $SO(6)_1$ model. This model can be supported by the surface of a class DIII topological superconductor~\cite{SchnyderRyuFurusakiLudwig08,Kitaevtable08} with topological index $N=12$. The surface carries 12 massless Majorana fermions, which cannot be turned massive without breaking time reversal symmetry. The surface state can be mimicked by a coupled-wire model previously provided in Ref.~\onlinecite{Sahoo2016coupledwireTSC}. An antiferromagnetic surface pair density wave, where the surface is decorated by an array of parallel strips with alternating time-reversal breaking pairing phases $\varphi=\pm\pi/2$, supports an array of chiral Majorana interfaces. Each is sandwiched between adjacent strips with time-reversal conjugate Majorana mass, and carries 12 chiral Majorana $\psi^1_y,\ldots,\psi^{12}_y$, where $y$ labels the interface.

We group the Majorana fermions in two collections $\psi^{A,i}_y=\psi^i_y$ and $\psi^{B,i}_y=\psi^{6+i}_y$, for $i=1,\ldots,6$. Each collection generates a $SO(6)$ Wess-Zumino-Witten (WZW) algebra (also known as Kac-Moody or affine Lie algebra) at level one. The algebra consists of current operators \begin{align}J^{C,jk}_y=i\psi^{C,j}_y\psi^{C,k}_y\label{JMajcurrent}\end{align} for $1\leq j<k\leq6$ and $C=A,B$. 
We first pair Majorana fermions into Dirac fermions $c^{C,j}_y=(\psi^{C,2j-1}_y+i\psi^{C,2j}_y)/\sqrt{2}$, for $j=1,2,3$, and bosonize each Dirac fermion $c^{C,j}_y\sim e^{i\phi^{C,j}_y}$. The bosonized variables follow the action with Lagrangian density \begin{align}\mathcal{L}_0&=\sum_y\sum_{C=A,B}\left[\frac{(-1)^y}{2\pi}\sum_{j=1}^3\partial_t\phi^{C,j}_y\partial_x\phi^{C,j}_y\right.\nonumber\\&\;\;\;\left.+\sum_{j,j'=1}^3V_{jj'}\partial_x\phi^{C,j}_y\partial_x\phi^{C,j'}_y\right],\end{align} where $V_{jj'}$ is a non-universal velocity matrix. The alternating sign $(-1)^y$ signifies the alternating propagating directions of the channels. The action dictates the equal-time commutation relation \begin{align}\left[\phi^{C,j}_y(x),\partial_{x'}\phi^{C',j'}_{y'}(x')\right]=2\pi i\delta^{CC'}\delta^{jj'}\delta_{yy'}\delta(x-x')\label{ETCR1}\end{align} or equivalently the time-ordered correlation function \begin{align}\phi^{C,j}_y(z)\phi^{C',j'}_{y'}(z')=-\delta^{CC'}\delta^{jj'}\delta_{yy'}\log(z-z')+\ldots\end{align} up to non-singular terms and Klein factors, where $z\sim\tau+i(-1)^yx$ is the (anti)holomorphic complex space-time parameter.

The current operators \eqref{JMajcurrent} can be expressed in terms of the bosonized variables. There are 3 Cartan generators \begin{align}H^{C,j}_y=i\partial\phi^{C,j}_y\sim{c^{C,j}_y}^\dagger c^{C,j}_y=i\psi^{C,2j-1}_y\psi^{C,2j}_y\label{SO6Cartan}\end{align} that form a maximal set of mutually commuting Hermitian operators. In addition, there are 12 roots \begin{align}E^{C,\boldsymbol\alpha}_y=\exp\left(i\alpha_j\phi^{C,j}_y\right),\label{SO6roots}\end{align} which act as ladder operators on the root lattice. The root vectors $\boldsymbol\alpha=(\alpha_1,\alpha_2,\alpha_3)$ all have integral entries $\alpha_j=0,\pm1$ and length square $|\boldsymbol\alpha|^2=2$ so that there are two and only two non-zero entries. Each vertex operator $E^{C,\boldsymbol\alpha}_y$ can be expressed as a complex quadratic combination of Majorana fermions \eqref{JMajcurrent}. The Cartan generators and roots therefore generate the complexified $SO(6)$ WZW algebra for each channel $y$ and sector $C=A,B$. One can pick a set of three linearly independent simple roots \begin{align}R_{SO(6)}=\begin{pmatrix}--&\boldsymbol\alpha^1&--\\--&\boldsymbol\alpha^2&--\\--&\boldsymbol\alpha^3&--\end{pmatrix}=\begin{pmatrix}0&1&1\\1&-1&0\\0&1&-1\end{pmatrix}. \label{simplerootsSO6} \end{align} All 12 roots can be expressed as integral linear combination of the simple ones. The choice of simple roots recovers the Cartan matrix of $SO(6)$ by the inner product \begin{align}K_{SO(6)}=R_{SO(6)}R_{SO(6)}^T=\begin{pmatrix}2&-1&0\\-1&2&-1\\0&-1&2\end{pmatrix}.\label{KSO6}\end{align} The roots also generate and lie inside a face-centered cubic lattice $\mathrm{FCC}=\mathrm{span}_{\mathbb{Z}}\{\boldsymbol\alpha^1,\boldsymbol\alpha^2,\boldsymbol\alpha^3\}$ in three dimensions. We refer to this as the root lattice.

Now, we introduce the inter-channel backscattering sine-Gordon potential \begin{align}\mathcal{H}_{\mathrm{dimer}}&=-\frac{u}{2}\sum_y\sum_{\boldsymbol\alpha}E^{A,\boldsymbol\alpha}_yE^{B,-\boldsymbol\alpha}_{y+1}\nonumber\\&=-u\sum_y\sum_{\boldsymbol\alpha}\cos\left(\boldsymbol\alpha\cdot2\boldsymbol\Theta_{y+1/2}\right),\label{Hdimer}\end{align} where $2\Theta_{y+1/2}=(2\Theta^1_{y+1/2},2\Theta^2_{y+1/2},2\Theta^3_{y+1/2})$ and $2\Theta^j_{y+1/2}=\phi^{A,j}_y-\phi^{B,j}_{y+1}$. In a periodic cylinder geometry with $L=2l$ channels, there are $3L$ counter-propagating pairs of bosons and there are also $3L$ linearly independent sine-Gordon angle variables $\boldsymbol\alpha\cdot2\boldsymbol\Theta_{y+1/2}$. The angle variables satisfy the ``Haldane nullity" gapping condition~\cite{Haldane1995Haldanecondition} \begin{align}\left[\boldsymbol\alpha\cdot2\boldsymbol\Theta_{y+1/2}(x),\boldsymbol\alpha'\cdot2\boldsymbol\Theta_{y'+1/2}(x')\right]=0.\label{Haldanenullitygapping}\end{align} There are actually $12L$ sine-Gordon terms because there are 12 roots in $SO(6)$. However, only $3L$ of them are linearly independent, but the redundant sine-Gordon terms do not compete. Collectively, they pin the angle variables \begin{align}\boldsymbol\alpha\cdot\left\langle2\boldsymbol\Theta_{y+1/2}\right\rangle\in2\pi\mathbb{Z}\label{SGpinning1}\end{align} in the ground state, for all root vectors $\boldsymbol\alpha$. Since the roots generate a FCC lattice, eq.\eqref{SGpinning1} requires the ground state expectation values of the angle variables $\left\langle2\boldsymbol\Theta_{y+1/2}\right\rangle$ to lie in the body-centered cubic (BCC) reciprocal lattice \begin{gather}\begin{split}\mathcal{L}_{\boldsymbol\Theta}&\equiv\left\{2\boldsymbol\Theta:\boldsymbol\alpha\cdot2\boldsymbol\Theta\in2\pi\mathbb{Z}\right\}\\&=2\pi\mathrm{BCC}=\mathrm{span}_{\mathbb{Z}}\left\{2\pi\boldsymbol\beta_1,2\pi\boldsymbol\beta_2,2\pi\boldsymbol\beta_3\right\},\end{split}\label{BCC}\\R^\vee_{SO(6)}=\begin{pmatrix}--&\boldsymbol\beta_1&--\\--&\boldsymbol\beta_2&--\\--&\boldsymbol\beta_3&--\end{pmatrix}=\begin{pmatrix}1/2&1/2&1/2\\1&0&0\\1/2&1/2&-1/2\end{pmatrix}.\end{gather} Here, $\boldsymbol\beta_I=\frac{1}{2}\varepsilon_{IJK}\boldsymbol\alpha^J\times\boldsymbol\alpha^K/\left[\boldsymbol\alpha^1\cdot(\boldsymbol\alpha^2\times\boldsymbol\alpha^3)\right]$ are the simple dual roots so that $\boldsymbol\alpha^I\cdot\boldsymbol\beta_J=\delta^I_J$. In Lie algebra language, $\b{\beta}_I$ are called {\it fundamental weights}. In the following discussion, we use the terms ``simple dual roots'', ``primitive reciprocal vectors'' and ``fundamental weights'' interchangeably. We refer to the lattice $\mathcal{L}_{\boldsymbol\Theta}$ of simultaneous minima of the sine-Gordon potentials as the ``Haldane's dual lattice". In Lie algebra language, $\text{span}_{\mathbb{Z}} \{ \b{\beta}_1,\b{\beta}_2,\b{\beta}_3 \}$ are called weight lattice. To comply with physics community, we use ``Haldane's dual lattice'' in the following discussions.

The inter-channel backscattering interactions \eqref{Hdimer} therefore freeze the angle-variables and introduce an finite excitation energy gap. Deconfined excitations are of the form of kinks where the expectation value $\left\langle2\boldsymbol\Theta_{y+1/2}(x)\right\rangle$ jumps discontinuously along $x$ from one lattice value to another. They can be represented using fractional vertex operators \begin{align}V^{C,\boldsymbol\gamma}_y(x_0)=\exp\left[i\gamma_j\phi^{C,j}_y(x_0)\right]\label{vertexprimary1}\end{align} that corresponds to a primary field of $SO(6)_1$, where $\boldsymbol\gamma=(\gamma_1,\gamma_2,\gamma_3)$ can take non-integral entries. For example, the vertex operator $V^{A,\boldsymbol\gamma}_y(x_0)$ creates a kink for $\left\langle2\boldsymbol\Theta_{y+1/2}(x)\right\rangle$ at $x_0$ because \begin{align}&V^{A,\boldsymbol\gamma}_y(x_0)^\dagger2\partial_x\Theta^j_{y+1/2}(x)V^{A,\boldsymbol\gamma}_y(x_0)\nonumber\\&=2\partial_x\Theta^j_{y+1/2}(x)+i\left[\gamma_k\phi^{A,k}_y(x_0),2\partial_x\Theta^j_{y+1/2}(x)\right]\nonumber\\&=2\partial_x\Theta^j_{y+1/2}(x)-2\pi(-1)^y\gamma_j\delta(x_0-x)\label{kinkcommrel}\end{align} from the equal-time commutation relation \eqref{ETCR1}. Integrating the above equation $x$ near $x_0$, we see the vertex operator creates a discontinuity for $\left\langle2\boldsymbol\Theta_{y+1/2}(x)\right\rangle$, where it jumps by $-2\pi\boldsymbol\gamma$ from $x<x_0$ to $x>x_0$. The excitation is deconfined if the angle-variable on both sides of $x_0$ minimizes all the sine-Gordon potentials in \eqref{Hdimer}. Otherwise, it will cost a linearly diverging energy to pull apart from its anti-partner. This restricts the jump of height of the kink $2\pi\boldsymbol\gamma$ to also live in the Haldane's dual lattice $\mathcal{L}_{\boldsymbol\Theta}$. In other words, deconfined excitations are represented by vertex operators $V^{C,\boldsymbol\gamma}_y$ \eqref{vertexprimary1} where $\boldsymbol\gamma$ lives in the BCC lattice \eqref{BCC}. Similarly, we have
\begin{align}
&V^{B,\boldsymbol\gamma}_y(x_0)^\dagger2\partial_x\Theta^j_{y-1/2}(x)V^{B,\boldsymbol\gamma}_y(x_0) \nonumber \\
&= 2 \p_x \T^j_{y-1/2}(x) + 2\pi (-1)^y \gamma_j \d(x_0 - x). \label{kinkcommrel2}
\end{align}
It shows that if $\b{\gamma}$ is one of the reciprocal vectors in the BCC lattice \eqref{BCC}, then $V^{C,\b{\gamma}}_y$ creates a deconfined quasiparticle excitation in the form of a kink of the sine-Gordon angle order parameter $\left\langle2\boldsymbol\Theta_{y-1/2}(x)\right\rangle$. 

\begin{figure}[htbp]
\centering\includegraphics[width=0.35\textwidth]{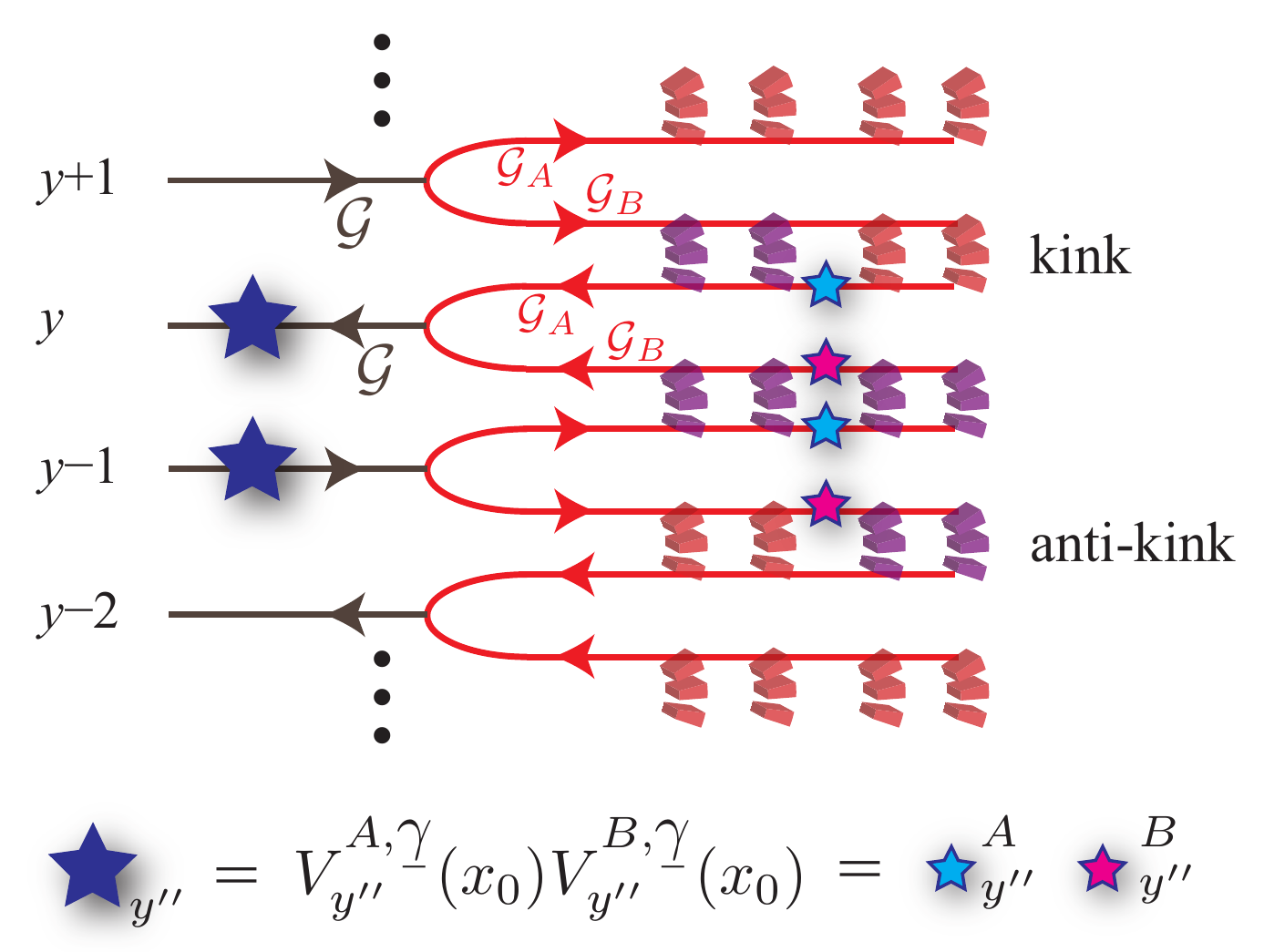}
\caption{A string of ``quasi-local" operators \eqref{eqn:ystring} creates a pair of fractional surface excitations in the form of a kink and anti-kink pair of the sine-Gordon order parameter $\left\langle2\boldsymbol\Theta_{y-1/2}(x)\right\rangle$.}\label{fig:kinks}
\end{figure}

It is crucial to recognize that in general the kink excitations may be {\em fractional}, in which case they must come in kink and anti-kink pairs. The notion of ``quasi-locality" is set by the 3D SPT/SET bulk, which may already support long-range entangled topological order and carry non-trivial quasiparticle and quasi-string excitations. We will address this issue soon after the description of $SO(6)$ primary fields and Wilson strings below. At the moment, we consider ``quasi-local" surface vertex operators that consist of a product of both the $A$ and $B$ sectors. We see that the combination $V^{A,\boldsymbol\gamma}_y(x_0)V^{B,\boldsymbol\gamma}_y(x_0)$ creates a kink-antikink pair in $\left\langle2\boldsymbol\Theta_{y+1/2}(x)\right\rangle$ and $\left\langle2\boldsymbol\Theta_{y-1/2}(x)\right\rangle$. The kink and anti-kink can be separated vertically by applying the string of vertex operators
\begin{align}
\chi^{A,\b{\gamma}}_{y,y'}(x_0) &= \prod_{y''=y}^{y'}V^{A,\b{\gamma}}_{y''} (x_0) V^{B,\b{\gamma}}_{y''}(x_0), \label{eqn:ystring}
\end{align}
on the ground state, where $y'>y$. This create a kink and anti-kink pair in $\left\langle2\boldsymbol\Theta_{y'+1/2}(x)\right\rangle$ and $\left\langle2\boldsymbol\Theta_{y-1/2}(x)\right\rangle$ without creating extra kinks in between (see Fig.~\ref{fig:kinks}). This is because the effect of $V^{A,\b{\gamma}}_{y''} (x_0)$ and $V^{B,\b{\gamma}}_{y''+1}(x_0)$ cancels. Physically what happens is that a pair of kink-antikink excitations are created in each wire in between and consequently the quasiparticle is transported, which is explicitly shown in Eq.~(\ref{kinkcommrel}) and (\ref{kinkcommrel2}) In this sense, these excitations are deconfined along the $y$ direction. It should be noticed that the kink-anti-kink pair can only be created by the operator string \eqref{eqn:ystring}, which is constructed by the series of ``quasi-local" operators $V^{A,\b{\gamma}}_{y''} (x_0) V^{B,\b{\gamma}}_{y''}(x_0)$. They cannot be created by $V^{A,\b{\gamma}}_{y'} (x_0) V^{B,\b{\gamma}}_{y}(x_0)$ alone without a string in between because of surface locality. We will address the surface ``quasi-locality" later. In addition to Eq.~(\ref{Hdimer}), the full Hamiltonian also involves velocity terms from its kinetic part and possibly backscattering of the Cartan generators. Although they do not in general commute with the quasiparticle string (\ref{eqn:ystring}), this does not affect the deconfinement of a quasiparticle pair. This is because when acting on the ground state, the open string (\ref{eqn:ystring}) only creates a kink-anti-kink pair while leaving the angle-variable order parameter $\langle2\boldsymbol\Theta_{y+1/2}(x)\rangle$ locally constant except at the kinks. Since $\langle\partial_x\b{\Theta}_{y+1/2}(x)\rangle=0$ except at the end of the strings, velocity terms do not contribute a linear-diverging confining energy.

The quasiparticle kinks can be moved in the $x$-direction by applying
\begin{align}
\rho_y(x,x_0) &= e^{i \int_{x_0}^x \gamma_j \p_{x'} \phi^{C,j}_y(x')},
\end{align}
which moves a quasiparticle excitation from $x_0$ to $x$ on the same wire, without creating extra kinks in between. Together with \eqref{eqn:ystring}, they describe the two-dimensional local motion of the quasiparticle kinks.

These deconfined excitation operators form representations of the $SO(6)_1$ affine Lie algebra. They obey the operator product expansion with the current generators \eqref{SO6Cartan} and \eqref{SO6roots} \begin{align}H^{C,j}_y(z)V^{C,\boldsymbol\gamma}_y(z')&=\frac{\gamma_j}{z-z'}V^{C,\boldsymbol\gamma}_y(z')+\ldots,\nonumber\\E^{C,\boldsymbol\alpha}_y(z)V^{C,\boldsymbol\gamma}_y(z')&=(z-z')^{\boldsymbol\alpha\cdot\boldsymbol\gamma}V^{C,\boldsymbol\alpha+\boldsymbol\gamma}_y(z')+\ldots.\end{align} In particular, primary fields are vertex operators with bounded singularities $\boldsymbol\alpha\cdot\boldsymbol\gamma\geq-1$. More precisely, each primary field is represented by a super-selection sector of vertex operators $\{V^{C,\boldsymbol\gamma^1}_y,\ldots,V^{C,\boldsymbol\gamma^r}_y\}$ that transform under \begin{align}E^{C,\boldsymbol\alpha}_y(z)V^{C,\boldsymbol\gamma^a}_y(z')&=\frac{(E^{\boldsymbol\alpha}_\rho)^a_b}{z-z'}V^{C,\boldsymbol\gamma^b}_y(z')+\ldots\label{SO6EVOPE}\end{align} where $E^{\boldsymbol\alpha}_\rho$ is the $r$-dimensional irreducible matrix representation of the root $E^{\boldsymbol\alpha}$ of $SO(6)$. The current operators $E^{\boldsymbol\alpha}$ are therefore raising and lowering operators that rotate $V^{\boldsymbol\gamma}\to V^{\boldsymbol\alpha+\boldsymbol\gamma}$ if $\boldsymbol\alpha\cdot\boldsymbol\gamma=-1$. The singular factor $1/(z-z')$ reflects the unit scaling dimension of the current operators, and higher order non-singular terms are non-universal.

The $SO(6)$ affine Lie algebra at level 1 has four primary fields labeled by $1,\psi,s_+,s_-$. They corresponds to the trivial, vector, even and odd spinor representations of $SO(6)$ respectively. We now show their corresponding super-sectors of vertex operators. The primary field $\psi$ at wire $y$ and sector $C=A,B$ is generated by $\{e^{\pm i\phi^{C,1}_y},e^{\pm i\phi^{C,2}_y},e^{\pm i\phi^{C,3}_y}\}$, which form the 6-dimensional vector representation of $SO(6)$. These vertex operators can also be decomposed into real and imaginary components $e^{i\phi^{C,j}_y}=\psi^{C,2j-1}_y+i\psi^{C,2j}_y$, where $\psi^{C,1}_y,\ldots,\psi^{C,6}_y$ are Majorana fermions with spin (i.e.~conformal scaling dimension) $h_\psi=1/2$. The even/odd twist primary fields $s_\pm$ are generated by $e^{i\boldsymbol\varepsilon\cdot\boldsymbol\phi/2}$, where $\boldsymbol\varepsilon=(\varepsilon_1,\varepsilon_2,\varepsilon_3)$ and $\varepsilon_j=\pm1$. $\boldsymbol\varepsilon$ is even (odd) if $\varepsilon_1\varepsilon_2\varepsilon_3=+1$ (resp.~$-1$). The collection of even (odd) vertices form the even (resp.~odd) spinor representation of $SO(6)$. These vertices operators have spin $h_{s_\pm}=3/8$. 

Using eq.\eqref{kinkcommrel}, the vector primary field $\psi^A_y$ at $x_0$ creates an $2\pi$ kink of the sine-Gordon angle variable so that \begin{align}\langle2\boldsymbol\Theta_{y+1/2}(x_0+\delta)\rangle-\langle2\boldsymbol\Theta_{y+1/2}(x_0-\delta)\rangle=-2\pi(-1)^y{\bf e}_j,\label{2pikinkSO6} \end{align} where the expectation values are taken with respect to the excited state $e^{i\phi^{A,j}_y(x_0)}|GS\rangle$. On the other hand, the spinor primary fields $(s_\pm)^A_y$ at $x_0$ creates a $\pi$ kink where \begin{align}\langle2\boldsymbol\Theta_{y+1/2}(x_0+\delta)\rangle-\langle2\boldsymbol\Theta_{y+1/2}(x_0-\delta)\rangle=-\pi(-1)^y\boldsymbol\varepsilon. \label{pikinkSO6} \end{align} Since the ``heights" of the kinks, which are given by the right hand side of the two equations above, belong to the Haldane's dual lattice $\mathcal{L}_{\boldsymbol\Theta}$ (see eq.\eqref{BCC}), the primary fields correspond to deconfined excitations that only cost a finite amount of energy to create and do not cost energy to move.

At this point, it is essential to address the surface ``quasi-locality"' and take into account the 3D bulk SPT/SET state that supports the surface state. The 12 Majorana fermions $\psi^{A,1}_y,\ldots,\psi^{A,6}_y$ and $\psi^{B,1}_y,\ldots,\psi^{B,6}_y$ associates a $SO(12)_1$ WZW algebra along each wire $y$. The primary fields in the $SO(12)_1$ CFT are quasiparticle excitations that are supported by the 3D bulk, and should {\em not} be treated as fractional excitations allowed by the surface gapping interactions. For the purpose of describing the surface topological order, primary fields in $SO(12)_1$ should be regarded as ``quasi-local" in the sense that such an excitation can be present without having a partner on the surface. This is because its partner can exist in the 3D bulk. On the other hand, the surface backscattering potential \eqref{Hdimer} allows additional fractional excitations that must come in pairs on the boundary surface. These are quasiparticles that do not connect to any bulk excitations.

The $SO(12)_1$ WZW algebra that associates to the ``quasi-local" primary field excitations is generated by the Cartan operators $H_y^{A,j}$, $H_y^{B,j}$ defined in \eqref{SO6Cartan} as well as the the 60 roots \begin{align}E^{\boldsymbol\lambda}_y=\exp\left[i\left(\lambda^A_j\phi^{A,j}_y+\lambda^B_j\phi^{B,j}_y\right)\right]\end{align} where the root vectors $\boldsymbol\lambda=(\lambda^A_1,\lambda^A_2,\lambda^A_3,\lambda^B_1,\lambda^B_2,\lambda^B_3)$ have integral entries $\lambda^C_j=0,\pm1$ and length square $|\boldsymbol\lambda|^2=2$ so that there are two and only two non-zero entries. The simple roots can be chosen to be \begin{align}R_{SO(12)}=\begin{pmatrix}--&\boldsymbol\lambda^1&--\\--&\boldsymbol\lambda^2&--\\--&\boldsymbol\lambda^3&--\\--&\boldsymbol\lambda^4&--\\--&\boldsymbol\lambda^5&--\\--&\boldsymbol\lambda^6&--\end{pmatrix}=\begin{pmatrix}1&-1&0&0&0&0\\0&1&-1&0&0&0\\0&0&1&-1&0&0\\0&0&0&1&-1&0\\0&0&0&0&1&-1\\0&0&0&0&1&1\end{pmatrix}. \label{simplerootsSO12} \end{align} The ``quasi-local" surface excitations that connect to the 3D bulk are represented by the vertex operator \begin{align}V^{\bf l}_y(x_0)=\exp\left[i\left(l^A_j\phi^{A,j}_y(x_0)+l^B_j\phi^{B,j}_y(x_0)\right)\right]\label{SO12primary}\end{align} where the weight vectors ${\bf l}=(l^A_1,l^A_2,l^A_3,l^B_1,l^B_2,l^B_3)$ satisfy \begin{align}\boldsymbol\lambda \cdot {\bf l}\in\mathbb{Z}\end{align} for all $SO(12)$ roots $\boldsymbol\lambda$. The weight vectors are integral combinations of the simple dual roots or fundamental weights \begin{gather}R^\vee_{SO(6)}=\begin{pmatrix}--&{\bf l}_1&--\\\vdots&\vdots&\vdots\\--&{\bf l}_6&--\end{pmatrix}=\left(\begin{smallmatrix}1 & 0 & 0 & 0 & 0 & 0 \\
 1 & 1 & 0 & 0 & 0 & 0 \\
 1 & 1 & 1 & 0 & 0 & 0 \\
 1 & 1 & 1 & 1 & 0 & 0 \\
 \frac{1}{2} & \frac{1}{2} & \frac{1}{2} & \frac{1}{2} & \frac{1}{2} & -\frac{1}{2} \\
 \frac{1}{2} & \frac{1}{2} & \frac{1}{2} & \frac{1}{2} & \frac{1}{2} & \frac{1}{2}\end{smallmatrix}\right),\end{gather} which obey $\b{\lambda}^I\cdot{\bf l}_J=\delta^I_J$. The entries of a general weight vector ${\bf l}$ are either all integers or all half-integers.

\begin{figure}[htbp]
\centering\includegraphics[width=0.3\textwidth]{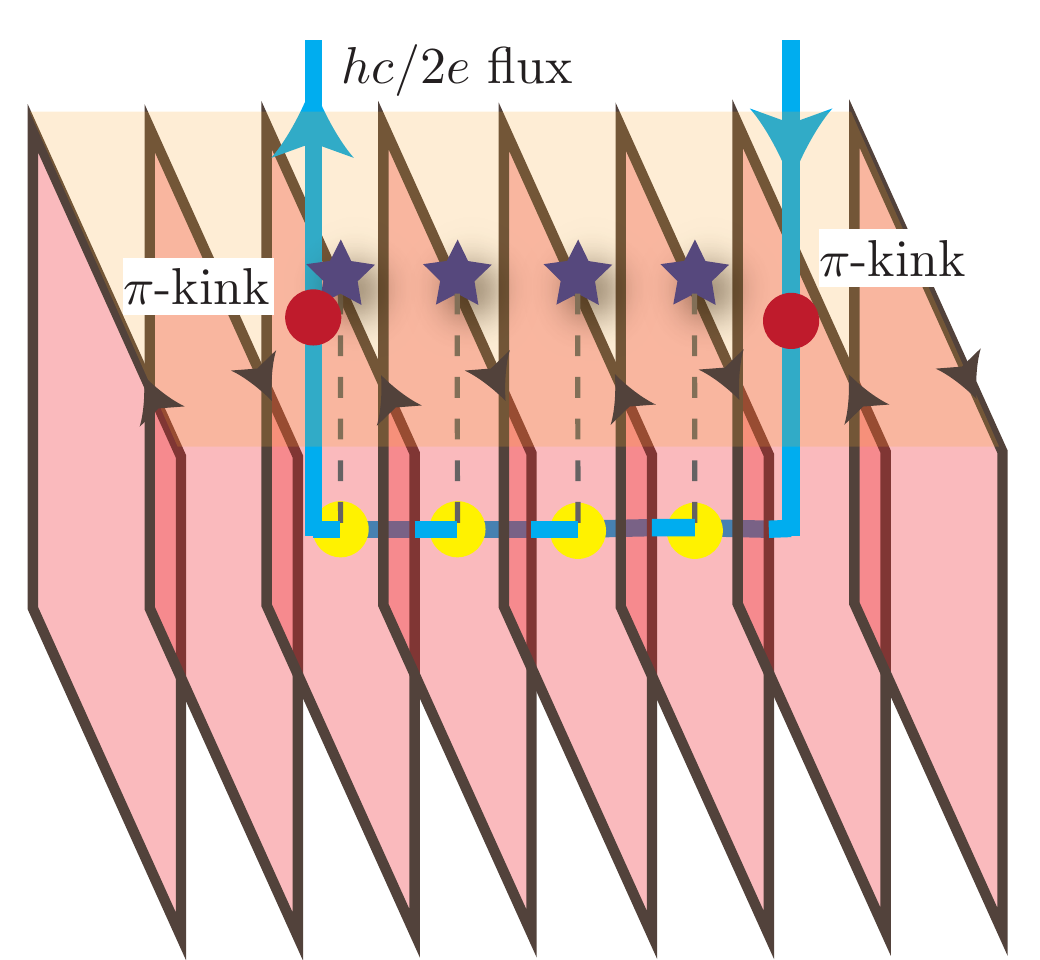}
\caption{A $hc/2e$ flux vortex in the topological superconducting bulk associates to a string of vertex operators on the surface (represented by the blue stars) and create a pair of $\pi$-kink excitations (red dots).}\label{fig:piflux}
\end{figure}

It is useful to notice that there is a tensor product structure (referred to as conformal embedding or level rank duality in the CFT context~\cite{bigyellowbook}) \begin{align}SO(12)_1\supseteq SO(6)_1\times SO(6)_1\end{align} that splits the ``quasi-local" $SO(12)_1$ primary fields \eqref{SO12primary} into the fractional $SO(6)_1^A$ and $SO(6)_1^B$ components \begin{align}V^{\bf l}_y(x_0)&\sim\exp\left(il^A_j\phi^{A,j}_y(x_0)\right)\exp\left(il^B_j\phi^{B,j}_y(x_0)\right)\nonumber\\&=V^{A,{\bf l}_A}_yV^{B,{\bf l}_B}_y.\end{align} In particular, if $\boldsymbol\gamma=(\gamma_1,\gamma_2,\gamma_3)$ lies inside the BCC Haldane dual lattice \eqref{BCC}, then the combination $V^{A,\boldsymbol\gamma}_yV^{B,\boldsymbol\gamma}_y$ is a $SO(12)_1$ primary field and therefore represents a ``quasi-local" excitation that connects to the 3D bulk. This shows that the vertex operator string \eqref{eqn:ystring} composes of ``quasi-local" excitations. For example, in the class DIII topological superconductor case, a $hc/2e$ flux vortex inside the bulk corresponds to the vertex $V^{\boldsymbol\varepsilon}_y$ for each layer $y$ that interests flux vortex, where $\varepsilon=(1/2,\ldots,1/2)$. It associates to the vertex operator string $\prod_{y=y_0}^{y_1}V^{\boldsymbol\varepsilon}_y$ on the surface, and create a pair of $\pi$-kink quasiparticle excitations (see figure~\ref{fig:piflux}). Each vertex operator $V^{\boldsymbol\varepsilon}_y$ is ``quasi-local" as it connects to the bulk, but the $\pi$-kink excitations are fractional. They are supported by the surface backscattering interactions and can only exist on the boundary surface.

Next, we illustrate the $U(4)_1$ model. The array of wire is now supported on the surface of some three dimensional symmetry protected topological state (see Fig.~\ref{fig:surfaceCW}(a)), and each wire hosts eight Dirac fermions. The 3D SPT state can be a topological crystalline insulator~\cite{Ando_Fu_TCI_review} with mirror Chern number 8 that supports 8 massless surface Dirac cones. It can be a topological paramagnet~\cite{VishwanathSenthil12,WangPotterSenthil13} that supports 8 neutral Dirac fermions along a time reversal breaking domain wall. Alternatively, it can also be a fractional bosonic topological insulator where a local boson is fractionalized into 8 parton Dirac fermions and the surface hosts 8 parton Dirac cones. In this paper, we do not focus on the origin of the wire array, but instead we concentrate on its symmetric gapping interactions.

Here, the 8 Dirac fermions of each wire are decomposed into two groups $c^A_j=c_j\sim e^{i\phi^A_j}$ and $c^B_j=c_{4+j}\sim e^{i\phi^B_j}$, for $j=1,2,3,4$. Each sector is described by a $U(4)$ Kac-Moody conformal field theory at level 1. The bosonized variables follow the action with Lagrangian density \begin{align}\mathcal{L}_0&=\sum_y\sum_{C=A,B}\left[\frac{(-1)^y}{2\pi}\sum_{j=1}^4\partial_t\phi^{C,j}_y\partial_x\phi^{C,j}_y\right.\nonumber\\&\;\;\;\left.+\sum_{j,j'=1}^4V_{jj'}\partial_x\phi^{C,j}_y\partial_x\phi^{C,j'}_y\right],\end{align} where $V_{jj'}$ is a non-universal velocity matrix. We further decompose each sector $C=A,B$ into \begin{align}U(4)_1\sim U(1)_4\times SU(4)_1.\end{align} $U(1)_4$ represent the diagonal component and is generated by the bosonized variable \begin{align}4\phi^C_{\rho,y}=\boldsymbol\alpha^0\cdot\boldsymbol\phi^C_y=\phi^{C,1}_y+\ldots\phi^{C,4}_y,\end{align} where $\boldsymbol\alpha^0=(1,1,1,1)$. Although in this paper we do not focus on charge conservation, for the charge preserving SPT states, the $U(1)_4$ sector is solely responsible for electric charge transport. The $SU(4)$ Kac-Moody current algebra at level 1 is generated by the 3 Cartan generators \begin{align}H^{C,j}_y=i\partial\phi^{C,j}_y-i\partial\phi^{C,j+1}_y\end{align} for $j=1,2,3$, and the 12 roots \begin{align}E^{C,\boldsymbol\alpha}_y=\exp\left(i\alpha_j\phi^{C,j}_y\right)\end{align} where the root vectors $\boldsymbol\alpha=(\alpha_1,\alpha_2,\alpha_3,\alpha_4)\in\Delta_{SU(4)}$ has entries $\alpha_j=0,\pm1$, length square $|\boldsymbol\alpha|^2=2$ and is traceless $\alpha_1+\alpha_2+\alpha_3+\alpha_4=0$. The $SU(4)_1$ represents electrically neutral degrees of freedom if the SPT state preserves charge symmetry. It also completely decoupled from $U(1)_4$ as all the roots $\boldsymbol\alpha$ are orthogonal to $\boldsymbol\alpha^0$.

One can pick the simple roots of $SU(4)$ to be \begin{align}R_{SU(4)}=\begin{pmatrix}--&\boldsymbol\alpha^1&--\\--&\boldsymbol\alpha^2&--\\--&\boldsymbol\alpha^3&--\end{pmatrix}=\begin{pmatrix}1&-1&0&0\\0&1&-1&0\\0&0&1&-1\end{pmatrix}.\end{align} This recovers the Cartan matrix of $SU(4)$ \begin{align}K_{SU(4)}=R_{SU(4)}R_{SU(4)}^T=\begin{pmatrix}2&-1&0\\-1&2&-1\\0&-1&2\end{pmatrix},\end{align} which is identical to that of $SO(6)$ (see eq.\eqref{KSO6}). Consequently, as an affine Lie algebra or a Kac-Moody algebra, $SU(4)$ and $SO(6)$ are equivalent. For instance, they have the identical dimension $d=15$ and rank $r=3$. The root structures of the two are also isomorphic except the $SO(6)$ roots are presented in three dimensions whereas the $SU(4)$ ones are presented in a 3D orthogonal complement of $(1,1,1,1)$ in four dimensions. The equivalence implies the $SU(4)$ roots span a face-centered cubic root lattice $\mathrm{FCC}=\mathrm{span}_{\mathbb{Z}}\{\boldsymbol\alpha^1,\boldsymbol\alpha^2,\boldsymbol\alpha^3\}$.

The inter-channel backscattering sine-Gordon potential (see also Fig.~\ref{fig:surfaceCW}(a)) is \begin{align}\mathcal{H}_{\mathrm{dimer}}&=\mathcal{H}^{U(1)_4}+\mathcal{H}^{SU(4)_1},\label{HdimerU4}\\\mathcal{H}^{U(1)_4}&=-u\sum_y\cos\left(4\phi^A_{\rho,y}-4\phi^B_{\rho,y+1}\right)\nonumber\\&=-u\sum_y\cos\left(2\Theta^1_{y+1/2}+\ldots+2\Theta^4_{y+1/2}\right),\nonumber\\\mathcal{H}^{SU(4)_1}&=-\frac{u}{2}\sum_y\sum_{\boldsymbol\alpha}E^{A,\boldsymbol\alpha}_yE^{B,-\boldsymbol\alpha}_{y+1}\nonumber\\&=-u\sum_y\sum_{\boldsymbol\alpha}\cos\left(\boldsymbol\alpha\cdot2\boldsymbol\Theta_{y+1/2}\right),\nonumber\end{align} where $2\Theta_{y+1/2}=(2\Theta^1_{y+1/2},2\Theta^2_{y+1/2},2\Theta^3_{y+1/2},2\Theta^4_{y+1/2})$ and $2\Theta^j_{y+1/2}=\phi^{A,j}_y-\phi^{B,j}_{y+1}$. Similar to the $SO(6)_1$ Hamiltonian \eqref{Hdimer}, the backscattering term here also introduces a finite excitation energy gap. The angle variables of the sine-Gordon Hamiltonian obey the Haldane nullity gapping condition (c.f.~\eqref{Haldanenullitygapping}). The $SU(4)$ current-current backscattering provides more than enough gapping terms, and linearly dependent redundant terms are non-competing if $u>0$. The ground state expectation values of the angle variables $\langle2\boldsymbol\Theta_{y+1/2}\rangle$ belongs in the ``Haldane's dual lattice" \begin{align}\mathcal{L}_{\boldsymbol\Theta}&\equiv\left\{2\boldsymbol\Theta:\boldsymbol\alpha\cdot2\boldsymbol\Theta,\boldsymbol\alpha^0\cdot2\boldsymbol\Theta\in2\pi\mathbb{Z}\right\}\end{align} so that the sine-Gordon energy \eqref{HdimerU4} is minimized. The dual lattice can be decomposed into two orthogonal components \begin{align}\mathcal{L}_{\boldsymbol\Theta}&=\mathcal{L}_{\boldsymbol\Theta}^{U(1)}+\mathcal{L}_{\boldsymbol\Theta}^{SU(4)} \label{eqn:BCCU(4)},\\\mathcal{L}_{\boldsymbol\Theta}^{U(1)}&=\mathrm{span}_{\mathbb{Z}}\{2\pi\boldsymbol\beta_0\},\nonumber\\\mathcal{L}_{\boldsymbol\Theta}^{SU(4)}&=2\pi\mathrm{BCC}=\mathrm{span}_{\mathbb{Z}}\{2\pi\boldsymbol\beta_1,2\pi\boldsymbol\beta_2,2\pi\boldsymbol\beta_3\},\nonumber\end{align} where the primitive reciprocal vectors of $\mathcal{L}_{\boldsymbol\Theta}^{U(1)}$ and $\mathcal{L}_{\boldsymbol\Theta}^{SU(4)}$ are \begin{align}\boldsymbol\beta_\mu&=\frac{1}{3!}\varepsilon_{\mu\nu\lambda\sigma}\frac{\boldsymbol\alpha^\nu\wedge\boldsymbol\alpha^\lambda\wedge\boldsymbol\alpha^\sigma}{\boldsymbol\alpha^0\cdot(\boldsymbol\alpha^1\wedge\boldsymbol\alpha^2\wedge\boldsymbol\alpha^3)},\\\boldsymbol\beta_0&=\frac{1}{4}(1,1,1,1),\nonumber\\R^\vee_{SU(4)}&=\begin{pmatrix}--&\boldsymbol\beta_1&--\\--&\boldsymbol\beta_2&--\\--&\boldsymbol\beta_3&--\end{pmatrix}=\frac{1}{4}\begin{pmatrix}3&-1&-1&-1\\2&2&-2&-2\\1&1&1&-3\end{pmatrix}.\nonumber\end{align}

Similar to the $SO(6)_1$ case, the deconfined excitations of the sine-Gordon model \eqref{HdimerU4} are kinks of the angle variables where $\langle2\boldsymbol\Theta_{y+1/2}\rangle$ jumps discontinuously from one value to another in $\mathcal{L}_{\boldsymbol\Theta}$. The kinks can be created by fractional vertex operators $V^{C,\boldsymbol\gamma}_y=\exp\left[i\gamma_j\phi^{C,j}_y\right]$ (c.f.~\eqref{vertexprimary1}), where in this case the fractional lattice vectors are four dimensional $\boldsymbol\gamma=(\gamma_1,\gamma_2,\gamma_3,\gamma_4)$. Excitations can be decomposed into $U(1)_4$ and $SU(4)_1$ components that associates to kinks of $\mathcal{H}^{U(1)}$ and $\mathcal{H}^{SU(4)}$ in \eqref{HdimerU4} respectively. For $U(1)_4$, the primary fields $[n]_\rho$ are vertex operators $e^{in\phi^C_{\rho,y}}=e^{in(\phi^{C,1}_y+\ldots+\phi^{C,4}_y)/4}$, where $n$ is an integer. They carry spins (or conformal scaling dimensions) $h_{[n]_\rho}=n^2/8$.

For $SU(4)_1$, certain vertex operators can be grouped together into super-selection sectors $\{V^{C,\boldsymbol\gamma^1}_y,\ldots,V^{C,\boldsymbol\gamma^r}_y\}$ and corresponds to a primary field of $SU(4)_1$. Vertices of each super-sector transform among each other under the $SU(4)_1$ affine Lie algebra (c.f.~\eqref{SO6EVOPE}). As $SU(4)_1$ and $SO(6)_1$ are equivalent, there is a one-to-one correspondence between the primary fields. Using the same notation in $SO(6)_1$, the primary fields $1,\psi,s_+,s_-$ of $SU(4)_1$ corresponds to the trivial, vector, fundamental and anti-fundamental representations of $SU(4)$. The primary field $\psi$ corresponds to the super-sector of 6 vertex operators $e^{i\boldsymbol\gamma^\psi\cdot\boldsymbol\phi^C_y}$, where $\boldsymbol\gamma^\psi=(1,1,-1,-1)/2$ or any permutation of the entries. The super-sector of the primary field $s_\pm$ consists of the 4 vertex operators $e^{i\boldsymbol\gamma^{s_\pm}\cdot\boldsymbol\phi^C_y}$, where $\boldsymbol\gamma^{s_\pm}=\pm(3,-1,-1,-1)/4$ or any permutation of the entries. The spins (i.e.~conformal scaling dimensions) of the primary fields are $h_\psi=1/2$ and $h_{s_\pm}=3/8$, which unsurprisingly match that of the primary fields of $SO(6)_1$.

\begin{figure}[htbp]
\centering
\includegraphics[width=0.35\textwidth]{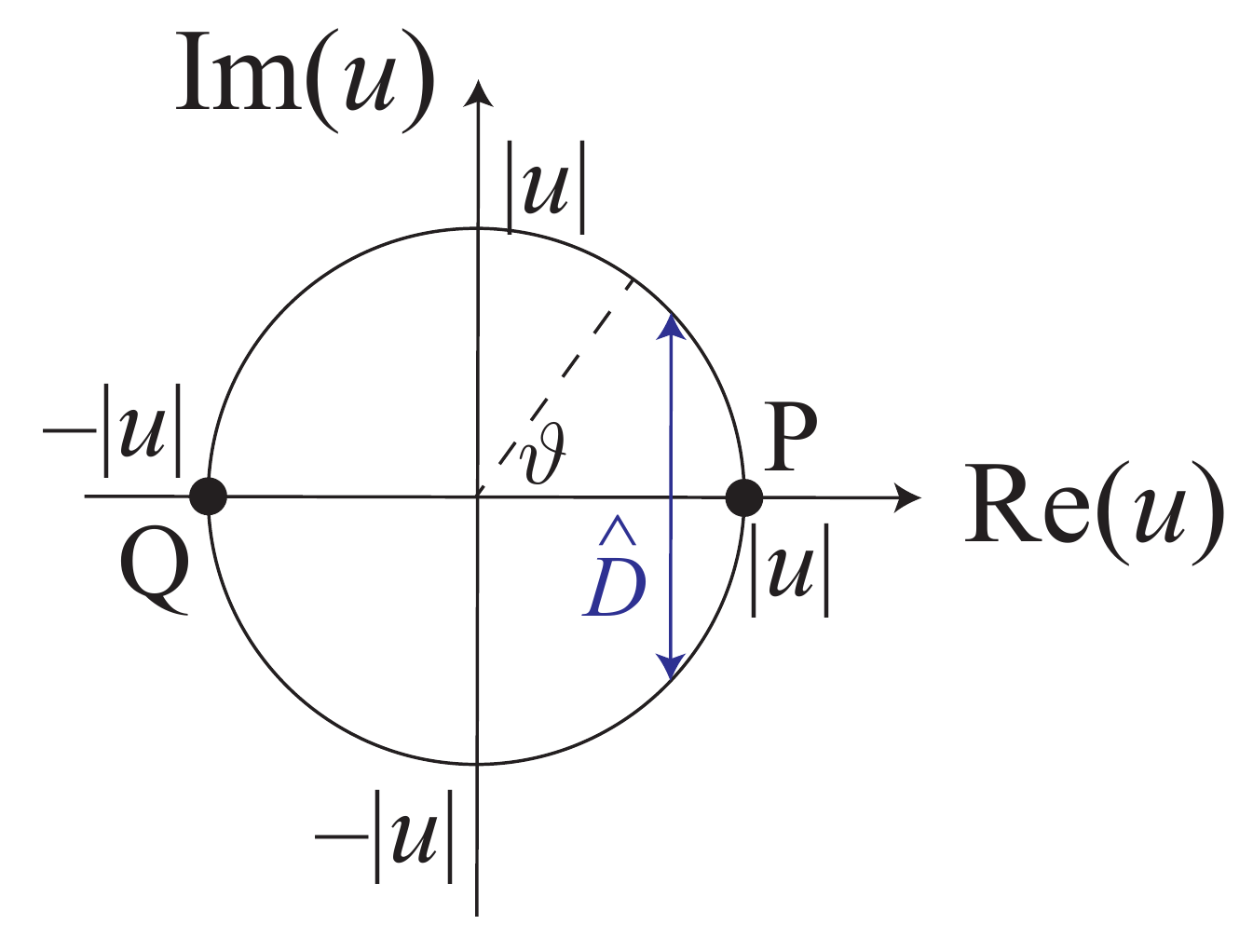}
\caption{Duality transformation of the sine-Gordon term on the $u$-plane. $\hat{D}$ is the duality operator. Under $\hat{D}$, points on the circle with radius $|u|$ is reflected with respect to the real axis. $P,Q$ are self-dual points. $P$ describes a gapless point, which can be seen in Fig.~\ref{fig:SGdual}(c). Other points on the circle describe gapped phases.}
\label{fig:uplane}
\end{figure}

\begin{figure}[htbp]
\centering\includegraphics[width=0.45\textwidth]{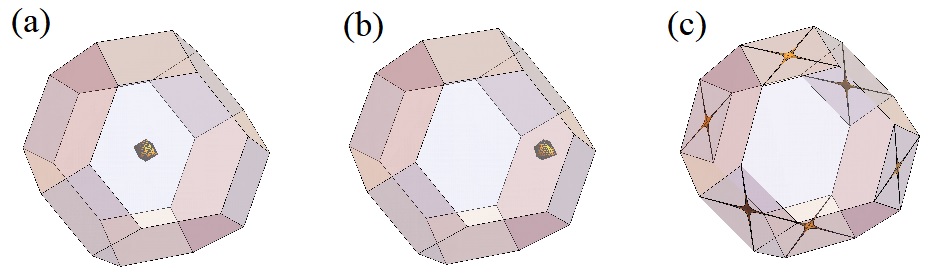}
\caption{The ground state expectation values of $\langle2\boldsymbol\Theta_{y+1/2}\rangle$ that minimize the sine-Gordon Hamiltonian \eqref{SU4SG} for (a) $\vartheta=0$, (b) $\vartheta=-3\pi/5$ and (c) $\vartheta=\pi$. The plots are taken over the fundamental region in $\mathbb{R}^3$ modulo the Haldane dual lattice $\mathcal{L}_{\boldsymbol\Theta}^{SU(4)}$ in \eqref{eqn:BCCU(4)}. The sine-Gordon Hamiltonian generically has a finite energy gap and a single minimum for $-\pi<\vartheta<\pi$. At $\vartheta=\pi$, there are gapless Goldstone modes on the boundary of the fundamental region.}
\label{fig:SGdual}
\end{figure}

Before we end this section, let us take a closer look at the sine-Gordon terms for $SU(4)_1$ sector. Usually we take $u>0$ such that the sine-Gordon terms are pinned at their respective minima to gap out the system from the renormalization group (RG) analysis. What if $u<0$ or even $u$ is a complex parameter? This is related to the duality properties of $ADE$ surface topological orders discussed later. So let us study the general structure of sine-Gordon terms when $u=|u|e^{i \vartheta}$ is complex valued. The general sine-Gordon is
\begin{align}
\m{H}^{SU(4)_1} &= -\frac{|u|}{2} \sum_y \sum_{\b{\a} \in \D_+} \left(E^{A,\b{\a}}_y E^{B,-\b{\a}}_{y+1} e^{i \vartheta} \right. \nonumber \\
& \quad \left.+ E^{A,-\b{\a}}_y E^{B,\b{\a}}_{y+1} e^{-i \vartheta} \right) \nonumber \\
&= -|u| \sum_y \sum_{\b{\a} \in \D_+} \cos{\left( \b{\a} \cdot 2\b{\T}_{y+1/2} + \vartheta \right)},
\label{SU4SG}\end{align}
where $\D_+$ is the set of positive roots. In this case, we find that as long as $\vartheta \neq \pi$, the system is gapped; when $\vartheta=\pi$ the system becomes gapless. Reversing the sign of $\b{\T}_{y+1/2}$ is equivalent to taking the complex conjugate of $u$, namely,
\begin{align}
2\b{\T}_{y+1/2} \to -2\b{\T}_{y+1/2} &\Leftrightarrow u \to u^* \Leftrightarrow \vartheta \to -\vartheta,
\end{align}
which is also equivalent to a reflection with respect to the real axis in the $u$ complex plane. The duality transformation on the $u$-plane is shown in Fig.~\ref{fig:uplane}. Since $SO(6)_1$ has the same root structure as $SU(4)_1$, the above analysis also works for $SO(6)_1$ theory. The ground state structure is shown in Fig.~\ref{fig:SGdual}.

\section{Review of free Dirac fermion/\texorpdfstring{QED$_3$}{QED3} duality}
\label{sec:Dirac-QED3}
In this section, we review the coupled wire derivation of the free Dirac fermion/QED$_3$ duality following Ref.~\onlinecite{Mross2016DiracQED3,Mross2017BosonFermionDuality}. Written explicitly, the duality says
\begin{align}
S_{\text{Dirac}} &= \int dx^3 \ i \bar{\Psi} \gamma^\mu (\p_\mu-iA_\mu) \Psi \nonumber \\
& \quad \updownarrow \nonumber \\
S_{\text{QED}_3} &= \int dx^3 \ i \bar{\tilde{\Psi}} \gamma^\mu (\p_\mu - i a_\mu) \tilde{\Psi} + \frac{1}{4\pi} \epsilon_{\mu \nu \rho} A_\mu \p_\nu a_\rho, \label{eqn:FFDual1}
\end{align}
where $a_\mu$ is a dynamical $U(1)$ gauge field and $A_\mu$ is a background $U(1)$ field. Since in (2+1)D, a single copy of Dirac fermion with unit charge suffers from the traditional ``parity'' anomaly, the duality is better understood to hold at the surface of a (3+1)D topological insulator. We add quotation marks for ``parity'' because strictly speaking, parity is in the connected component of the rotation group in (2+1)D. Therefore, the anomaly is better called as an anomaly of time-reversal symmetry $T$ or reflection symmetry $R$. Detailed  clarifications can be found in Ref.~\onlinecite{Witten2016ParityAnomaly}. Several derivations have been given from the field-theoretic perspectives. Specifically, what they have done is to start from the conjectured fermion/boson duality, which is the duality between a single free Dirac fermion and a complex boson  coupled to a dynamical $U(1)$ gauge field at the $O(2)$ Wilson-Fishier fixed point with quartic interactions.~\cite{Seiberg2016Dualityweb, Karch2016PVduality} Then they perform flux attachment to the original duality to obtain the fermion/fermion duality. The same can be performed at the coupled wire level, which may be clearer in the sense that one can see the explicit interactions at the microscopic level. We now review it below. 

Let us start from the array of 1D chiral electron wires, each aligned along the $x$-direction. The Hamiltonian can be written as
\begin{align}
H &= \sum_y \int dx \ v_x (-1)^y \psi^\dagger_y (-i \p_x) \psi_y \nonumber \\
& \qquad - v_y (-1)^y (\psi^\dagger_y \psi_{y+1} + \text{h.c.}), \label{eqn:Dirac1}
\end{align}
where in Eq.~(\ref{eqn:Dirac1}) $y$ is the wire label along the $y$-direction. Wires labeled by even $y$  carry right-moving electrons and odd $y$ carry left-moving electrons. The first term in Eq.~(\ref{eqn:Dirac1}) describes the kinetic energy of electrons and the second term describes uniform inter-wire hopping between neighboring wires (see Fig.~\ref{fig:surfaceCW}(b)). Using a two-component spinor $\Psi(x,y) =(\psi_{2y}(x), \psi_{2y+1}(x))^T$, Eq.~(\ref{eqn:Dirac1}) can be rewritten in the continuum limit as
\begin{align}
H &= \int dxdy \Psi^\dagger [ v_x \s^z (-i\p_x) + v_y \s^y (-i\p_y)] \Psi, \label{eqn:Dirac2}
\end{align}
where the sum $\sum_y$ is replaced by $\int dy$. Eq.~(\ref{eqn:Dirac2}) therefore recovers the effective Hamiltonian for a single copy of Dirac fermion in (2+1)D. Now, let us bosonize the Dirac fermion on each wire by $\psi_y = e^{i \phi_y}$, where $\phi_y$ is a chiral boson field satisfying the commutation relation
\begin{align}
[\phi_y(x), \phi_{y'}(x')] &= \d_{yy'} (-1)^y i \pi \mathrm{sgn}(x-x') \nonumber \\
& \quad + i \pi \mathrm{sgn}(y'-y), \label{eqn:DiracComm1}
\end{align}
where $\mathrm{sgn}(s) = s/|s|$ and $\mathrm{sgn}(0)=0$. The first and second lines of Eq.~(\ref{eqn:DiracComm1}) give the correct anticommutation relations of fermions in the same wire and between different wires, respectively. 
Written in terms of boson fields, the original Dirac action in Eq.~(\ref{eqn:FFDual1}) becomes
\begin{align}
S_{\text{Dirac}} &= \sum_y \int dx dt \ \big[ \frac{i (-1)^y}{4\pi} \p_x \phi_y \p_t \phi_y + \frac{v_x}{4\pi} (\p_x \phi_y)^2  \nonumber \\
& \quad + v_y (-1)^y \cos{(\phi_y - \phi_{y+1})} \big]. \label{eqn:DiracAction1}
\end{align}
Under renormalization group (RG) flow, this theory remains gapless due to the competition between neighboring sine-Gordon terms. 

\begin{figure}[htbp]
\centering\includegraphics[width=0.35\textwidth]{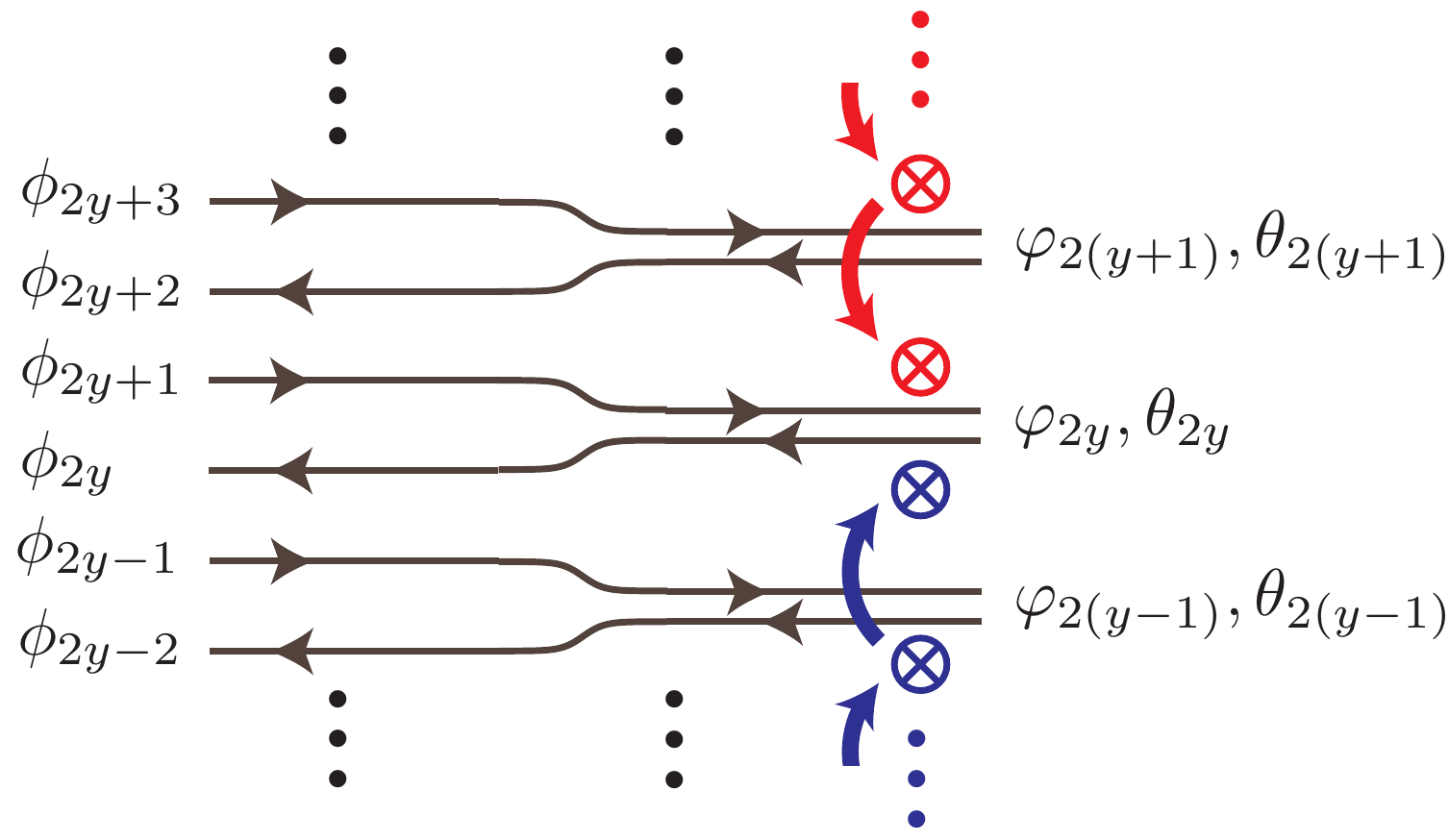}
\caption{Pictorial illustration of the duality transformation in Eq.~(\ref{eqn:DiracDual1}) or (\ref{eqn:DiracDual2}). Two flux quanta from $+\infty$ and $-\infty$ attached to each pair of wires.}\label{fig:duality}
\end{figure}

Now, let us perform the duality transformation
\begin{align}
\tilde{\phi}_y(x) &= \sum_{y'} \text{sgn}(y-y') (-1)^{y'} \phi_{y'}(x) \equiv \sum_{y'} D_{yy'} \phi_{y'} (x). \label{eqn:DiracDual1}
\end{align}
This duality transformation \eqref{eqn:DiracDual1} is a flux attachment (see Fig.~\ref{fig:duality}). Using the non-chiral basis between wire $2y$ and $2y+1$, $\varphi_{2y}, \t_{2y} = (\phi_{2y} \pm \phi_{2y+1})/2$, Eq.~(\ref{eqn:DiracDual1}) is equivalent to 
\begin{align}
\tilde{\psi}^\dagger_{2y/2y+1} &\sim \psi^\dagger_{2y+1/2y} \prod_{y'>y} e^{2i \t_{2y'}} \prod_{y'<y} e^{-2i\t_{2y'}}, \label{eqn:DiracDual2}
\end{align}
where $e^{2i\t_{2y}}$ brings a $2\pi$ phase slip in $\varphi_{2y}$. Eq.~(\ref{eqn:DiracDual2}) can be understood as bringing two fluxes from positive and negative infinities to the fermion at wire $2y/2y+1$. One can check that under duality (\ref{eqn:DiracDual1}), the equal-time commutation relation only changes by a sign
\begin{align}
[\tilde{\phi}_y(x), \tilde{\phi}_{y'}(x')] &= - [\phi_y(x), \phi_{y'}(x')].
\end{align}
Physically it means that the dual fermions $e^{i\tilde\phi_y}$ have opposite chiralities with the original ones. After duality transformation, the original action (\ref{eqn:DiracAction1}) for the Dirac fermion becomes
\begin{align}
\tilde{S}_{\text{Dirac}} &= \sum_y \int dx dt\ \Big\{ \frac{-i(-1)^y}{4\pi} \p_x \tilde{\phi}_y \p_t \tilde{\phi}_y + \frac{v_x}{4\pi} (\p_x D^{-1}_{yy'} \tilde{\phi}_{y'})^2 \nonumber \\
& \quad + v_y (-1)^y \cos{(\tilde{\phi}_y} - \tilde{\phi}_{y+1}) \Big\}. \label{eqn:DiracAction2}
\end{align}
One can see that in the dual action (\ref{eqn:DiracAction2}), the first and last terms have the same form as the original action (\ref{eqn:DiracAction1}). However, the second term is highly non-local. To resolve this, one introduces two Lagrangian multipliers $\tilde{a}_{0,y}, \tilde{a}_{1,y}$ on each wire and rewrite Eq.~(\ref{eqn:DiracAction2}) as
\begin{align}
\tilde{\m{L}}_{\text{Dirac}} &= \sum_y\frac{-i(-1)^y}{4\pi} \p_x \tilde{\phi}_y \p_t \tilde{\phi}_y + \m{L}_{\text{QED}_3},
\end{align}
where 
\begin{align}
\m{L}_{\text{QED}_3} &= \m{L}_0 + \m{L}_{\text{staggered-CS}} + \m{L}_{\text{Maxwell}} + \m{L}_{\text{tunnel}}, \nonumber \\
\m{L}_0 &= \sum_y\frac{i (-1)^y}{2\pi} \p_x \tilde{\phi}_y \tilde{a}_{0,y} + \sum_y\frac{\tilde{v}_x}{4\pi} (\p_x \tilde{\phi}_y - \tilde{a}_{1,y})^2, \nonumber \\
\m{L}_{\text{staggered-CS}} &= \sum_y\frac{i(-1)^y}{8\pi} (\D \tilde{a}_{0,y}) (\tilde{a}_{1,y+1} + \tilde{a}_{1,y}), \nonumber \\
\m{L}_{\text{Maxwell}} &= \sum_y\frac{1}{16\pi} \left[ \frac{1}{v_x} (\D \tilde{a}_{0,y})^2 + v_x (\D \tilde{a}_{1,y})^2 \right],
\end{align}
and $\D \tilde{a}_{i,y} \equiv \tilde{a}_{i,y+1} - \tilde{a}_{i,y}$. 
Now one can see that the dual Dirac theory is nothing but QED$_3$, where $\tilde{a}_{0,y}, \tilde{a}_{1,y}$ are now the emergent $U(1)$ gauge field under the gauge fixing $\tilde{a}_{2,y} =0$. The theory is invariant under the gauge transformation
\begin{align}
\tilde{\phi}_y &\to \tilde{\phi}_y + f_y, \nonumber \\
\tilde{a}_{0,y} &\to \tilde{a}_{0,y} + \p_t f_y, \nonumber \\
\tilde{a}_{1,y} &\to \tilde{a}_{1,y} + \p_x f_y, \nonumber \\
\tilde{a}_{2,y+1/2} &\to \tilde{a}_{2,y+1/2} + (f_{y+1}-f_y),
\end{align}
if we restore the $\tilde{a}_{2,y+1/2}$ component. Introducing these emergent gauge fields in the path integral only contributes an irrelevant overall multiplicative factor, which is unimportant. Thus the duality between a single Dirac fermion and QED$_3$ is established at the path integral level. 

Let us now take a look at how symmetries transform under duality. If we define time reversal (TR) symmetry and particle-hole (PH) symmetry on the basis $\Psi$ as
\begin{align}
\m{T} &: \Psi \to i \s^y \Psi, \quad \m{C}: \Psi \to i \sigma^y \Psi^\dagger,
\end{align}
then under the duality transformation (\ref{eqn:DiracDual1}) with some modifications to the transformation of $\phi$ variables,~\cite{Mross2017BosonFermionDuality} we have
\begin{align}
\tilde{\m{T}} &: \tilde{\Psi} \to i \s^y \tilde{\Psi}^\dagger, \quad \tilde{\m{C}}: \tilde{\Psi} \to i \s^y \tilde{\Psi}.
\end{align}
We see that TR and PH symmetries are exchanged under duality. In the following discussion of the surface topological orders of $ADE$ classifications, the generalization of the duality transformation for the single Dirac fermion will be utilized. 

\section{D-series: \texorpdfstring{$SO(N)_1$}{SO(N)} surface theory}\label{sec:Dseries}
\subsection{Surface massless Majorana fermions in a coupled wire model}
The coupled-wire model for D-series has been discussed in Ref.~\onlinecite{Sahoo2016coupledwireTSC} for the Majorana surfaces of topological superconductors. A particular case for $SO(6)$ was demonstrated in section~\ref{sec:Generalcoupledwire}. We here describe the general construction. The generic coupled wire Hamiltonian for $N$ copies of surface massless Majorana fermions is the sum
\begin{align}
\m{H}_0 +\m{H}_{\mathrm{bc}} &= \sum^\infty_{y=-\infty} i v_x (-1)^j \boldsymbol\psi^T_y \p_x \boldsymbol\psi_y\nonumber\\&\;\;\;+\sum^\infty_{y=-\infty}iv_y\boldsymbol\psi^T_y\boldsymbol\psi_{y+1}, \label{eqn:BareH0}
\end{align}
where the integer $y$ labels the wire in the vertical direction (see Fig.~\ref{fig:surfaceCW}), and $\boldsymbol\psi=(\psi^1,\ldots,\psi^N)$ is an $N$-component Majorana fermion. Majorana fermions on adjacent wires have opposite chiralities. The uniform non-dimerizing backscattering terms in $\mathcal{H}_{\mathrm{bc}}$ on the second line compete with neighboring ones, and the Hamiltonian describes $N$ massless Majorana fermions with linear dispersion in both the $x$ and $y$ directions. In this paper, we are interested in Abelian surface topological phases, and for this reason, we restrict $N=2r>4$. On each wire, Majorana fermion pairs form Dirac fermions, which can then be bosonized
\begin{align}
c_y^a &= \frac{1}{\sqrt{2}} \left( \psi_y^{2a-1}+i \psi^{2a}_y \right) \sim e^{i \phi^a_y}, \quad a=1,\ldots,r. \label{eqn:Bosonization1}
\end{align}
The bosons satisfy the equal-time commutation relation 
\begin{align}
\left[ \phi^a_y(x), \phi^{a'}_{y'}(x') \right] &= i\pi(-1)^y \big[ \d_{yy'} \d^{aa'} \mathrm{sgn}(x-x')  \nonumber \\
&  + \d_{yy'} \mathrm{sgn}(a-a') \big] + i\pi \mathrm{sgn}(y'-y), \label{eqn:Comm1}
\end{align}
where terms on the second line enforces mutual anticommutation product relations between Dirac fermions, and sgn$(x)=x/|x| = \pm 1$ for $x \neq 0$ and sgn(0)=0. The first line of Eq.~(\ref{eqn:Comm1}) is equivalent to the commutator between conjugate fields
\begin{align}
\left[ \p_x \phi^a_y(x), \phi^{a'}_{y'}(x') \right] &= 2i\pi (-1)^y \d_{yy'} \d^{aa'} \delta(x-x'),
\end{align}
which is dictated by the ``$p \dot{q}$'' term of the Lagrangian density
\begin{align}
\m{L}_0 &= \frac{1}{2\pi} \sum_y \sum^r_{a=1} (-1)^y \p_x \phi^a_y \p_t \phi^a_y.
\end{align}
The total Lagrangian density can be written in terms of boson fields as $\m{L} =\m{L}_0 - \m{H}_0 $, where 
\begin{align}
\m{H}_0 &= V_x \sum_y \sum^r_{a=1} \p_x \phi^a_y \p_x \phi^a_y \label{eqn:Hparallel}
\end{align}
is the non-universal sliding Luttinger liquid (SLL) component along each wire. The non-dimerizing backscattering terms in \eqref{eqn:BareH0} can also be bosonized, and take the form of $\mathcal{H}_{\mathrm{bc}}=V_y\sum_y\sum^r_{a=1}\cos\left(\phi^a_y-\phi^a_{y+1}\right)$. However, we suppress these single-body terms throughout this section for the following reason. The bosonized Hamiltonian density \eqref{eqn:Hparallel} has an additional local gauge symmetry \begin{align}\phi^a_y\to\phi^a_y+2\pi m^a_y\label{Dbosongauge}\end{align} where $m^1_y,\ldots,m^r_y$ are either all integers or all half-integers. This represents a local $\mathbb{Z}_2$ gauge symmetry that transforms the Majorana fermions according to \begin{align}\psi^j_y\to(-1)^{M_y}\psi^j_y,\label{Z2gaugesymm}\end{align} where $M_y\equiv2m^a_y$ modulo 2. Eq.~\eqref{Z2gaugesymm} is violated by the fermionic Hamiltonian density \eqref{eqn:BareH0}. Instead, the fermionic $\mathcal{H}_0$ and $\mathcal{H}_{\mathrm{bc}}$ in \eqref{eqn:BareH0} are only symmetric under a global $\mathbb{Z}_2$ symmetry where $m=m_y$ is uniform. Throughout this section, we focus on a {\em bosonic} coupled wire surface constructions that preserve the local $\mathbb{Z}_2$ symmetry \eqref{Z2gaugesymm}. For example, the model mimics the surface of a bosonic topological superconductor that supports emergent Majorana fermion coupled with a $\mathbb{Z}_2$ gauge theory. The vectors ${\bf m}_y=(m^1_y,\ldots,m^r_y)^T$ that correspond to the gauge transformation \eqref{Dbosongauge} live in a lattice \begin{align}\mathcal{L}^r_{\mathrm{gauge}}&=\left\{{\bf m}:2m^a\in\mathbb{Z},m^1\equiv\ldots\equiv m^r\mbox{ mod 1}\right\}\nonumber\\&=\mathrm{span}_{\mathbb{Z}}\left\{\frac{1}{2}\boldsymbol\varepsilon=\frac{1}{2}(\varepsilon^1,\ldots,\varepsilon^r)^T:\varepsilon^a=\pm1\right\}.\end{align} In this section, we focus on scenarios where $r=2n$ is even. In this case, we further restricts the gauge vectors ${\bf m}_y$ in \eqref{Dbosongauge} to live in the {\em even} lattice \begin{align}\mathcal{L}^{r,+}_{\mathrm{gauge}}=\mathrm{span}_{\mathbb{Z}}\left\{\frac{1}{2}\boldsymbol\varepsilon_+:\varepsilon_+^a=\pm1,\prod_{a=1}^r\varepsilon_+^a=+1\right\}\end{align} for $r=2n\geq4$. The $r=2$ case is special and corresponds to the decomposable algebra $SO(4)=SU(2)\times SU(2)$, where the even gauge lattice is $\mathcal{L}^{2,+}_{\mathrm{gauge}}=\mathrm{span}_{\mathbb{Z}}\left\{(1,-1)^T,(1/2,1/2)^T\right\}$.

Before moving on, we briefly comment on the symmetries of the model. If the surface state is supported by a bulk time-reversal symmetry-protected topology, then the coupled wire model exhibits an antiferromagnetic time-reversal (AFTR) symmetry,~\cite{Sahoo2016coupledwireTSC} which accompanies the time-reversal that flips the fermions' propagating direction with a half-translation that moves $y\to y+1$. In this case, the equal-time commutation relation \eqref{eqn:Comm1} needs to be modified to \begin{align}
\left[ \phi^a_y(x), \phi^{a'}_{y'}(x') \right] &= i\pi (-1)^{\text{max}\{y,y'\}} \big[ \d_{yy'} \d^{aa'} \text{sgn}(x'-x) \nonumber \\
& \quad + \d_{yy'} \text{sgn}(a-a') + \text{sgn}(y-y') \big], \label{eqn:Comm2}
\end{align} to accommodate the antiferromagnetic symmetry \begin{align}
\m{T} c^a_y \m{T}^{-1} &= (-1)^y c^{a \dagger}_{y+1}, \quad \m{T} \phi^a_y \m{T}^{-1} = \phi^a_{y+1} + \pi y. \label{eqn:AFTR}
\end{align} 
The discretization of surface state by a coupled wire construction and its effect on symmetries was explained by the symmetry extension pattern discussed in Ref.~\onlinecite{SeibergWitten2016gappedboundary, Witten2016ParityAnomaly} when gapped symmetric boundary states are constructed. 
The AFTR symmetry protects an odd number of surface massless Majorana fermions from single-body backscattering. There can be additional global symmetries, such as mirror, that further protects an arbitrary number of surface Majorana's. In this work, we do not focus on a particular symmetry, but instead concentrate on the many-body gapping potential based on a fractionalization scheme (see Fig.~\ref{fig:surfaceCW}(c)) that can preserve a range of symmetries. In this section, we also require the many-body gapping potential to respect the local $\mathbb{Z}_2$ gauge symmetry \eqref{Z2gaugesymm}.

\subsection{Gapping potentials for surface Majorana fermions}
The simplest gapping terms are single-body backscattering ones, such as
\begin{align}
\m{H}_{\mathrm{dimer}} &= iu\sum_y\sum_{a=1}^r\psi^a_y\psi^{r+a}_{y+1}
\end{align}
that dimerize Majorana channels and introduce mass to all Majorana fermions. Unfortunately, these single-body dimerizations do not respect the local $\mathbb{Z}_2$ symmetry \eqref{Z2gaugesymm}. Nevertheless, they illustrate the idea of decomposition of the degrees of freedom along each wire: $N=2r=r+r$. In each pair, the two sets of Majorana fermions $\psi^1_y,\ldots,\psi^r_y$ and $\psi^{r+1}_y,\ldots,\psi^{2r}_y$ are backscattered independently to adjacent wires in the opposite directions. By introducing this single-body backscattering term, we explicitly break and split the $SO(2r)_1$ symmetry into $SO(r)_1\times SO(r)_1$ along each wire.

With this idea in mind, we can introduce a second type of gapping terms, which preserve the local $\mathbb{Z}_2$ symmetry \eqref{Z2gaugesymm}. From the decomposition of the $SO(2r)$ WZW Kac-Moody algebra (also known as conformal embedding)
\begin{align}
SO(2r)_1 &\supset SO(r)^A_1 \times SO(r)^B_1, \label{eqn:SOdecom1}
\end{align}
we can introduce the two-body Kac-Moody current backscattering interactions
\begin{align}
\m{H}_{\mathrm{dimer}} &= u \sum_y \mathbf{J}^{SO(r)^B}_y \cdot \mathbf{J}^{SO(r)^A}_{y+1}  
\label{eqn:JJint}
\end{align}
for positive $u$ (see figure~\ref{fig:surfaceCW}(c)). The $A$ sector contains $\psi^1_y,\ldots,\psi^r_y$ and the $B$ sector contains $\psi^{r+1}_y,\ldots,\psi^{2r}_y$. We will show that \eqref{eqn:JJint} introduces a non-vanishing excitation energy gap in the next subsection.

The $SO(2r)_1$ WZW theory along the $y$-th wire is generated by the chiral current operator
\begin{align}
J^{(a,b)}_y &= (-1)^y i \psi^a_y \psi^b_y.
\end{align}
Based on \eqref{eqn:SOdecom1}, we can decompose the current operators into two sets: SO$(r)^A_1$ contains $J^{(a,b)}$ for $1 \leq a<b \leq r$ and SO$(r)^B_1$ contains $J^{(a,b)}$ for $r+1 \leq a<b \leq 2r$. We can see that these two sets of operators decouple in the sense that their operator product expansions (OPE) are trivial up to non-singular terms. Moreover, the Sugawara energy-momentum tensor~\cite{bigyellowbook} of the total SO$(2r)_1$ algebra decomposes into two decoupled parts, \begin{align}T_{SO(2r)_1}&=T_{SO(r)^A_1}+T_{SO(r)^B_1},\\T_{SO(r)^A_1} &= \frac{1}{2(r-1)}\sum_{1\leq a<b\leq r}J^{(a,b)}J^{(a,b)}\nonumber\\&=-\frac{1}{2} \sum^r_{a=1} \psi^a \p \psi^a, \\T_{SO(r)^B_1} &= \frac{1}{2(r-1)}\sum_{r+1\leq a<b\leq2r}J^{(a,b)}J^{(a,b)}\nonumber\\&=-\frac{1}{2} \sum^{2r}_{a=r+1} \psi^a \p \psi^a.
\end{align}
The interaction \eqref{eqn:JJint} can be expressed using the Majorana fermions
\begin{align}
\m{H}_{\mathrm{dimer}} &= u \sum_y \sum_{1 \leq a<b \leq r} \psi^{r+a}_y \psi^{r+b}_y \psi^a_{y+1} \psi^b_{y+1}. \label{eqn:JJint2}
\end{align}
We notice in passing the following observations. First, it breaks $O(2r)$ symmetry into $O(r)^A \times O(r)^B$, which transforms
\begin{align}
\psi^a_y &\to (\m{O}^{(-1)^y})^a_b \psi^b_y, \quad \psi^{r+a}_y \to (\m{O}^{(-1)^y})^a_b \psi^{r+b}_y,
\end{align}
where $\m{O}$ is an $r \times r$ orthogonal matrix. 
Second, there are alternative interaction terms, such as $\psi^a_y \psi^b_y \psi^{r+a}_{y+1} \psi^{r+b}_{y+1}$, that can compete with \eqref{eqn:JJint2}. However, as long as mirror symmetry is broken, one of these can be dominant and lead to a finite energy gap. Third, \eqref{eqn:JJint2} is marginally relevant. The renormalization group (RG) equation for $u$ is~ \cite{Cardybook}
\begin{align}
\frac{du}{d\l} &= + 4\pi (r-2) u^2,
\end{align}
showing that the interaction strength is growing at low energy limit when $r>2$, which is the case that we discuss. 

\subsubsection{Excitation energy gap}
We now review that \eqref{eqn:JJint2} introduces a non-vanishing excitation energy gap. A proof can also be found in Ref.~\onlinecite{Sahoo2016coupledwireTSC}. We focus on a single coupled pair of counter-propagating $SO(r)_1$ channels (see figure~\ref{fig:surfaceCW}(c)) at some even $y$.  After relabeling $\psi^{r+a}_y = \psi^a_R$ and $\psi^a_{y+1} = \psi^a_L$ for $a=1,\ldots,r$, the interaction term between the $y$-th and $(y+1)$-th wires becomes the $O(r)$ Gross-Neveu (GN) model~\cite{GrossNeveu}
\begin{align}
\m{H}_{GN} &= -\frac{u}{2} (\boldsymbol\psi_R \cdot \boldsymbol\psi_L)^2. \label{eqn:GN1}
\end{align}
It is known that the GN model has an energy gap when $r>2$. For even $r=2n$, we can use \eqref{eqn:Bosonization1} to pair Majorana fermions into Dirac fermions
\begin{align}
c^a_{R/L} &= \frac{1}{\sqrt{2}} (\psi^{2a-1}_{R/L} + i \psi^{2a}_{R/L}) \sim e^{i \phi^a_{R/L}}, \quad a=1,\dots,n.
\end{align}
Eq.~\eqref{eqn:GN1} bosonizes into \begin{align}
\m{H}_{GN} &\sim u \sum^n_{a=1} \p_x \phi^a_R \p_x \phi^a_L - u \sum_{a_1 \neq a_2} \sum_\pm \cos{\left( 2\T^{a_1} \pm  2\T^{a_2} \right)} \nonumber \\
&= u \sum^n_{a=1} \p_x \phi^a_R \p_x \phi^a_L - u \sum_{\boldsymbol\alpha \in \D} \cos{( \boldsymbol\alpha \cdot \mathbf{2\T})}, \label{eqn:GN2}
\end{align} where $\mathbf{2\T} = (2\T^1, \dots, 2\T^n)$, $2\T^a = \phi^a_R- \phi^a_L$, and $\boldsymbol\alpha=(\alpha_1,\ldots,\alpha_n)^T$ are the $SO(2n)$ roots that lives in \begin{align}\Delta=\left\{\boldsymbol\alpha\in\mathbb{Z}^n:|\boldsymbol\alpha|^2=2\right\}.\end{align} 
As a matter of fact, a subset of the sine-Gordon terms in \eqref{eqn:GN2} will be sufficient in introducing an energy gap. We take
\begin{align}
-u \sum^n_{I=1} \cos{(\boldsymbol\alpha^I \cdot \mathbf{2\T})} &= -u \sum^n_{I=1} \cos{ \left( \sum^n_{J=1} K_{IJ} (\phi'^J_R - \phi'^J_L) \right)} \nonumber \\
&= -u \sum^n_{I=1} \cos{(\mathbf{n}^T_I \mathsf{K} \mathbf{\Phi})}, \label{eqn:GN3}
\end{align}
where $\boldsymbol\alpha^I=(\alpha_1^I,\ldots,\alpha_n^I)$ are the $n$ linearly independent simple roots of $SO(2n)$ (c.f.~\eqref{simplerootsSO6} for $SO(6)$) \begin{align}R_{SO(2n)}=\begin{pmatrix}--&\boldsymbol\alpha^1&--\\\vdots&\vdots&\vdots\\--&\boldsymbol\alpha^n&--\end{pmatrix}=\begin{pmatrix}1&-1&0&\ldots&0&0\\0&1&-1&\ldots&0&0\\\vdots&\vdots&\vdots&\ddots&\vdots&\vdots\\0&0&0&\ldots&1&-1\\0&0&0&\ldots&1&1\end{pmatrix}.\label{simplerootsSO2n}\end{align} Here $K=(K_{IJ})_{n\times n}=R_{SO(2n)}R_{SO(2n)}^T$ is the Cartan matrix of $SO(2n)$, and $\mathsf{K}=K\oplus(-K)=\mathrm{diag}(K,-K)$ includes both the $R$ and $L$ movers. $\mathbf{\Phi} = (\boldsymbol\phi'_R, \boldsymbol\phi'_L)^T$, for $\boldsymbol\phi'_{R/L}=(\phi'^1_{R/L},\ldots,\phi'^n_{R/L})$ and $\phi^a_{R/L}=(R_{SO(2n)}^T)^a_J\phi'^J_{R/L}=(\alpha^J_a)\phi'^J_{R/L}$, are the bosonized variables in the Chevalley basis, and $\mathbf{n}_J = (\mathbf{e}_J, \mathbf{e}_J)^T$. The ``$p\dot{q}$" component of the Lagrangian density expressed in terms of the Chevalley basis is
\begin{align}
\m{L}_0 =\frac{1}{2\pi}\sum_{a=1}^n\p_x\phi^a_R\p_t\phi^a_R-\p_x\phi^a_L\p_t\phi^a_L= \frac{1}{2\pi} \p_x \mathbf{\Phi}^T \mathsf{K} \p_t \mathbf{\Phi}.
\end{align}
The $n$ vectors $\mathbf{n}_I$ are linearly independent and satisfy ``Haldane's nullity gapping condition"~\cite{Haldane1995Haldanecondition}
\begin{align}
\mathbf{n}^T_I \mathsf{K} \mathbf{n}_J &=0, \quad \mbox{for }I,J=1,\dots,n. \label{eqn:HaldaneD}
\end{align}
This shows \eqref{eqn:GN3} introduces a finite excitation energy gap. 

The additional linearly dependent sine-Gordon terms in \eqref{eqn:GN2} are complementary when $u>0$, and they collectively pin the non-competing ground state expectation values $\boldsymbol\alpha\cdot\langle2\boldsymbol\Theta\rangle\in2\pi\mathbb{Z}$. This defines the ``Haldane's dual lattice" (c.f.~(\ref{BCC}) for $SO(6)$) \begin{gather}\begin{split}\mathcal{L}_{\boldsymbol\Theta}&\equiv\left\{2\boldsymbol\Theta:\boldsymbol\alpha\cdot2\boldsymbol\Theta\in2\pi\mathbb{Z}\mbox{ for all }\boldsymbol\alpha\in\Delta\right\}\\&=2\pi\mathrm{BCC}^n=\mathrm{span}_{\mathbb{Z}}\left\{2\pi\boldsymbol\beta_1,\ldots,2\pi\boldsymbol\beta_n\right\},\end{split}\label{BCCn}\end{gather} where the simple dual roots $\boldsymbol\beta_I$ are 
\begin{align}\boldsymbol\beta_I&=\frac{1}{n!\det(R_{SO(2n)})}\varepsilon_{IJ_1\ldots J_{n-1}}\boldsymbol\alpha^{J_1}\wedge\ldots\wedge\boldsymbol\alpha^{J_{n-1}}\nonumber\\R^\vee_{SO(2n)}&=\begin{pmatrix}--&\boldsymbol\beta_1&--\\\vdots&\vdots&\vdots\\--&\boldsymbol\beta_n&--\end{pmatrix}\nonumber\\&=\begin{pmatrix}1&&&&&&\\1&1&&&&&\\1&1&1&&&&\\\vdots&\vdots&\vdots&\ddots&&&\\1&1&1&\ldots&1&&\\1/2&1/2&1/2&\ldots&1/2&1/2&-1/2\\1/2&1/2&1/2&\ldots&1/2&1/2&1/2\end{pmatrix}.\end{align} The dual lattice $\mathcal{L}_{\boldsymbol\Theta}$ (up to a factor of $2\pi$) is the body-centered cubic lattice (BCC) in $n$ dimensions, whose lattice vectors have either all integral or all half-integral entries. The mutual commutativity between the angle variables $2\T^a$ ensures that \eqref{eqn:GN2} introduces a finite excitation energy gap. Details of the Haldane's nullity gapping condition for the K-matrix formalism is reviewed in Appendix~\ref{append:HaldaneCondition}.

When $r=2n+1$ is odd, Eq.~(\ref{eqn:JJint}) can still be applied to introduce a finite energy gap. However, the gapping Hamiltonian here cannot be fully bosonized because of the extra odd Majorana fermion. Since this situation has been discussed in detail in Ref.~\onlinecite{Sahoo2016coupledwireTSC}, it will not be repeated here.

\subsubsection{Quasiparticle excitations}
The quasiparticle excitations of the sine-Gordon gapping potential \eqref{eqn:GN2} take a similar structure to that of the $SO(6)$ case described in section~\ref{sec:Generalcoupledwire}. Here, we only present the main results. A quasiparticle excitation at $(x_0,y_0)$ can be created by a fractional vertex operator $V_{y_0}^{C,\boldsymbol\gamma}(x_0)=\exp\left[i\gamma_a\phi^{C,a}_{y_0}(x_0)\right]$, where $C=A,B$, $a=1,\ldots,n$ and $\phi^{A,a}=\phi^a$, $\phi^{B,a}=\phi^{n+a}$ are the bosonized variables for the Dirac fermions $c^b=(\psi^{2b-1}+i\psi^{2b})/\sqrt{2}\sim e^{i\phi^b}$, for $b=1,\ldots,r=2n$. The vector $\boldsymbol\gamma=(\gamma_1,\ldots,\gamma_n)$ that corresponds to deconfined excitations can take all integral or all half-integral entries, and therefore it lives on the BCC dual lattice $\mathcal{L}_{\boldsymbol\Theta}$ defined in \eqref{BCCn}. There are four primary fields of the $SO(2n)_1$ WZW CFT that generate all deconfined excitations. Each primary field is a super-selection sector of vertex operators that form an irreducible representation of the $SO(2n)_1$ algebra (c.f.~\eqref{SO6EVOPE}). The first is the trivial vacuum excitation 1 that corresponds to the trivial representation of $SO(2n)$. The fermionic primary field $\psi$ consists of the vertex operators $\{e^{i\phi^{C,1}_{y_0}(x_0)},\ldots,e^{i\phi^{C,n}_{y_0}(x_0)}\}$. Each of the vertex operators has conformal scaling dimension $h_\psi=(-1)^{y_0}/2$, and creates a $2\pi$ kink to the ground state expectation value $\langle2\boldsymbol\Theta_{y_0\pm1/2}\rangle$ at $x_0$ (c.f.~\ref{2pikinkSO6}), where the sign depends on $C=A,B$. Each of the two spinor primary fields $s_\pm$ consists of the set of fractional vertex operators $e^{i\boldsymbol\varepsilon\cdot\boldsymbol\phi^C_{y_0}(x_0)}/2$, where the vector $\boldsymbol\varepsilon=(\varepsilon_1,\ldots,\varepsilon_n)$ has unit entries and $\prod_a\varepsilon_a=1$ for $s_+$ and $-1$ for $s_-$. The two super-sectors corresponds to the even and odd spinor representations of $SO(2n)$. Each of the vertex operators has a conformal scaling dimension $h_{s_\pm}=(-1)^{y_0}n/8$, and creates a $\pi$-kink of $\langle2\boldsymbol\Theta_{y_0\pm1/2}\rangle$ at $x_0$ (c.f.~\ref{pikinkSO6}).

The four primary fields $1,\psi,s_+,s_-$ in $SO(2n)_1$ follow a set of pair operator product expansion formulas. Consequently, the four types of quasiparticle excitations in the coupled wire model follow the corresponding fusion rules \begin{align}\begin{split}\psi\times\psi=1,\quad\psi\times s_\pm=s_\mp,\\s_\pm\times s_\pm=\left\{\begin{array}{*{20}l}1,\quad\mbox{for even $n$}\\\psi,\quad\mbox{for odd $n$}\end{array}\right..\end{split}\end{align}

\subsection{Duality transformation of the Hamiltonian and the symmetry}
We now study the duality properties of the gapped surface. The duality transformation here generalizes that in Ref.~\onlinecite{Mross2016DiracQED3}.
\begin{align}
\tilde{\phi}^a_y(x) &= \sum_{y'} \text{sgn}(y-y') (-1)^{y'} \phi^a_{y'}(x), \label{eqn:Dual}
\end{align}
where the flavor index $a$ is a spectator in the transformation. Physically it means that we bring two flux quanta from positive and negative infinities to each flavor of chiral fermions independently. Equivalently, we define the duality according to a particular $U(1)^r$subgroup of $SO(2r)$. We can check that the dual field $\tilde{\phi}^a_j$ preserves the commutation relation of the original boson field, up to a minus sign
\begin{align}
\big[ \tilde{\phi}^a_y(x), \tilde{\phi}^{a'}_{y'}(x') \big] &= - \big[ \phi^a_y(x), \phi^{a'}_{y'}(x') \big].\label{dualityETCR}
\end{align}
Physically, the dual fermion defined by $\tilde{\psi}^a_y(x) = e^{i \tilde{\phi}^a_y(x)}$ still satisfies the correct fermion anticommutation relation, but it has the opposite chirality of the original fermion on each wire. After duality transformation, the kinetic energy term in Eq.(\ref{eqn:Hparallel}) becomes highly nonlocal in terms of the dual bosons. This can be resolved by introducing emergent gauge fields $a^a_j(x)$ for each flavor of bosons (c.f.~the review in Sec.\ref{sec:Dirac-QED3}), and such description will not be repeated here. Instead, we focus on the gapping terms. 
Under Eq.~(\ref{eqn:Dual}), we have
\begin{align}
\tilde{\phi}^a_{y+1} - \tilde{\phi}^a_y &= (-1)^{y+1} (\phi^a_{y+1}-\phi^a_y). \label{eqn:dualityrelation}
\end{align}
Thus the sine-Gordon term in \eqref{eqn:GN2} keeps its original form, namely,
\begin{align}
-u\sum_{\boldsymbol\alpha\in\Delta}\cos{ \big( \boldsymbol\alpha \cdot 2\tilde{\boldsymbol\Theta} \big)} &= -u\sum_{\boldsymbol\alpha\in\Delta} \cos{ \big(\boldsymbol\alpha \cdot 2\boldsymbol\Theta \big)}, \label{eqn:sGselfdual}
\end{align}
where $2 \tilde{\T}^a \equiv \tilde{\phi}^a_j - \tilde{\phi}^a_{j+1}$. The sine-Gordon gapping potential is therefore self-dual.

There is a comment on this duality transformation. This self-dual interaction is a special case of the more general case, where the coefficient $u$ of the current-current interaction is complex valued, as we discussed in Sec.~\ref{sec:Generalcoupledwire}. Without loss of generality, we assume that $|u|=1$. Thus we can write $u=e^{i \t}$. Eq.~(\ref{eqn:sGselfdual}) corresponds to $\t=0$. As we vary $\t$, in addition to cosine terms, there are sine terms from current-current interaction, which flip sign under duality transformation, seen from Eq.~(\ref{eqn:dualityrelation}). Thus the ground state structure would rotate in the Haldane lattice space correspondingly. Then the duality transformation is a reflection with respect to the real axis in the complex $u$ plane. When $\d=\pi$, the interaction becomes self-dual again. But now the system becomes gapless. Therefore, the phase diagram on the $u$ plane is a unit circle centered at the origin, with self-dual points located at $\t=0,\pi$. All the points describe a gapped system except $\t=\pi$.

We notice in passing the duality transformation of the antiferromagnetic time-reversal symmetry. Under the definition \eqref{eqn:AFTR},
\begin{align}
\m{T} \tilde{\phi}^a_y \m{T}^{-1} &= - \tilde{\phi}^a_{y+1} - \frac{\pi}{2} (-1)^{y+1}. \label{eqn:AFTRDual}
\end{align}
The additional minus sign in front of $\tilde{\phi}^a_{y+1}$, when compared with \eqref{eqn:AFTR}, means $\mathcal{T}:\tilde{c}\to\tilde{c}$ now preserves dual Dirac fermion number, whereas $\mathcal{T}:c\to c^\dagger$ flips the original ones. The AFTR symmetry therefore carries an additional particle-hole component when transferred across the duality.

We conclude this section by making the following remarks. First, the duality transformation defined in Eq.(\ref{eqn:Dual}) is not unique. There are alternative duality transformations that converge to the same equal-time commutation relation \eqref{dualityETCR}. Second, the duality transformation \eqref{eqn:Dual} does not work for the gapped phase of $SO(4)$. From Ref.~\onlinecite{Sahoo2016coupledwireTSC}, we see that $SO(4)$ requires special attention because the usual decomposition $SO(4)\sim SO(2)\times SO(2)$ leads to Gross-Neveu interactions that only renormalize the boson velocities without introducing an energy gap. For this purpose, an alternative decomposition is needed -- $SO(4)\sim SU(2)\times SU(2)$, and it leads to a special gapping potential. The $SU(2)$ gapping potential is not self-dual under \eqref{eqn:Dual}, and in fact, the dual theory is highly non-local. We suspect the $SO(4)\sim SU(2)\times SU(2)$ fractionalization is self-dual under some alternative duality transformation that is out of the scope of this work.

\section{A-series: \texorpdfstring{$U(N)_1$}{u(N)} surface theory}
\label{sec:A-series}
\subsection{Surface gapless Dirac Hamiltonian via coupled wire construction and decomposition}
In this section, we discuss the $U(N)_1$ theories constructed from $N$ Dirac fermions. The $U(4)_1$ prototype was discussed in Sec.~\ref{sec:Generalcoupledwire}. Here we describe the general situations. The $N$ surface Dirac fermions, with Hamiltonian \begin{align}\mathcal{H}_0=iv\sum_{a=1}^N\sum_{s,s'=\uparrow,\downarrow}{c^a_s}^\dagger(\sigma_x\partial_x+\sigma_y\partial_y)_{ss'}c^a_{s'},\end{align} can be supported by a topological bulk such as a reflection-symmetric topological crystalline insulator with mirror Chern number $N$~\cite{TeoFuKane08,Fu_first_TCI}. By introducing alternating symmetry breaking Dirac mass on the surface, \begin{align}\delta V=\pm m\sum_{a=1}^N\sum_{s,s'=\uparrow,\downarrow}{c^a_s}^\dagger(\sigma_z)_{ss'}c^a_{s'},\end{align} the gapless electronic degrees of freedom are localized along an array of one-dimensional interfaces (see Fig.~\ref{fig:surfaceCW}(a)). Each interface, that is sandwiched between adjacent stripes with opposite Dirac masses, hosts $N$ chiral Dirac fermions that co-propagate in a single direction~\cite{TeoKane}.

The Hamiltonian that describes the 1D arrays of low-energy Dirac channels is
\begin{align}
\m{H}_{D,0} &= \sum^\infty_{y=-\infty} i v_x (-1)^y \mathbf{c}^\dagger_y \p_x \mathbf{c}_y, \label{eqn:BareHD0}
\end{align}
where $\mathbf{c}_y=(c^1_y,\ldots,c^N_y)$ is an $N$-component chiral Dirac fermion. After bosonizing  these Dirac fermions via $c^a_y = e^{i \phi^a_y}$, we can write Eq.~(\ref{eqn:BareHD0}) in the same form as Eq.~(\ref{eqn:Hparallel}), namely,
\begin{align}
\m{H}_{D,0} &= V_x \sum_y \sum^N_{a=1} \p_x \phi^a_y \p_x \phi^a_y,
\end{align}
where $V_x$ is some non-universal velocity. We can decompose a $U(N)_1$ theory into a $U(1)$ charge sector and an $SU(N)$ spin sector. This decomposition makes the physics richer than that of the D-series, which we will show later. The $U(1)$ charge sector is represented by the diagonal \begin{align}\Phi^\rho_y=N\tilde\phi^\rho_y=\phi^1_y+\ldots+\phi^N_y\label{U1chargeboson}\end{align} and the neutral $SU(N)$ sector is represented by \begin{align}\Phi_{y,I}=\sum_{J=1}^{N-1}K^{SU(N)}_{IJ}\tilde\phi^J_y = \sum^N_{a=1} \alpha^I_a \phi^a_y,\label{SUNneutralboson}\end{align} where $I=1,\dots,N-1$. Here $\boldsymbol\alpha^I=(\alpha^I_1,\ldots,\alpha^I_N)$, for $\alpha^a_I = \d^a_I - \d^a_{I+1}$, are the simple roots of $SU(N)$. The Cartan matrix of $SU(N)$ is the inner product $K^{SU(N)}_{IJ}=\boldsymbol\alpha^I\cdot\boldsymbol\alpha^J$. The roots of $SU(N)$ form the collection of integral vectors \begin{align}\Delta_{SU(N)}=\left\{\boldsymbol\alpha\in\mathbb{Z}^N:|\boldsymbol\alpha|^2=2,\sum_{a=1}^N\alpha_a=0\right\}.\label{SUNroots}\end{align} Details can be found in Appendix \ref{append:Liealgebra}. 

The ``$p\dot{q}$" term of the Lagrangian density decomposes into \begin{align}\mathcal{L}_0&=\frac{1}{2\pi}\sum_y(-1)^y\sum_{a=1}^N\partial_x\phi^a_y\partial_t\phi^a_y\label{SUNL0}\\&=\frac{1}{2\pi}\sum_y(-1)^yN\partial_t\tilde\phi^\rho_y\partial_x\tilde\phi^\rho_y\nonumber\\&\;\;\;+\frac{1}{2\pi}\sum_y(-1)^y\sum_{I,J=1}^{N-1}K^{SU(N)}_{IJ}\partial_t\tilde\phi^I_y\partial_x\tilde\phi^J_y.\nonumber\end{align} In this section, we focus on the partition $N=p+q$ that splits \begin{align}U(N)_1 & \supset U(p)_1\times U(q)_1\nonumber\\ & \supset\underbrace{\left(U(1)_p\times SU(p)_1\right)}_{A-\mathrm{sector}}\times\underbrace{\left(U(1)_q\times SU(q)_1\right)}_{B-\mathrm{sector}}.\label{UNpartition}\end{align} The partition separates the Dirac fermions into two groups. The $A$ sector consists of $c^1,\ldots,c^p$, and the $B$ sector consists of $c^{p+1},\ldots,c^{p+q}$. We label the bosonized variables by $\phi^{A,a}=\phi^a$ for $a=1,\ldots,p$ and $\phi^{B,b}=\phi^{p+b}$ for $b=1,\ldots,q$. The Lagrangian density \eqref{SUNL0} splits into \begin{align}\mathcal{L}_0&=\mathcal{L}^A_0+\mathcal{L}^B_0\\\mathcal{L}^C_0&=\frac{1}{2\pi}\sum_y(-1)^y\sum_{c=1}^r\partial_x\phi^{C,c}_y\partial_t\phi^{C,c}_y\nonumber\\&=\frac{1}{2\pi}\sum_y(-1)^yr\partial_t\tilde\phi^{C,\rho}_y\partial_x\tilde\phi^{C,\rho}_y\nonumber\\&\;\;\;+\frac{1}{2\pi}\sum_y(-1)^y\sum_{I,J=1}^{r-1}K^{SU(r)}_{IJ}\partial_t\tilde\phi^{C,I}_y\partial_x\tilde\phi^{C,J}_y.\nonumber\end{align} where the charged and neutral bosons $\tilde\phi^{C,\rho}$ and $\tilde\phi^{C,I}$ are defined similarly to \eqref{U1chargeboson} and \eqref{SUNneutralboson}, for $C=A,B$ and $r=p,q$ respectively. As there are no cross terms in the Lagrangian density, the $A$ and $B$ sectors are completely decoupled from one another.

If the surface Dirac fermions are supported from a mirror-symmetric topological bulk, the Dirac channels are related by reflection \begin{align}Mc^a_yM^{-1}=(-1)^yic^a_{-y},\quad M\phi^a_yM^{-1}=\phi^a_{-y}+(-1)^y\frac{\pi}{2}.\end{align} The surface array also admits an emergent anti-ferromagnetic time-reversal (AFTR) symmetry (c.f.~\eqref{eqn:AFTR} in Sec.~\ref{sec:Dseries}) \begin{align}\mathcal{T}c^a_y\mathcal{T}^{-1}=(-1)^yc^a_{y+1},\quad\mathcal{T}\phi^a_y\mathcal{T}^{-1}=-\phi^a_y+\frac{1-(-1)^y}{2} \pi.\end{align} The symmetries obey the algebraic relation $M^2=(-1)^F$, $\mathcal{T}^2=(-1)^F\mathrm{translation}_{y\to y+2}$ and $TMT^{-1}M^{-1}=\mathrm{translation}_{y\to y+2}$, where $(-1)^F$ is the fermion parity number operator. Mirror and AFTR symmetry preserving surface many-body gapping coupled wire models can be found in Ref.~\onlinecite{Hong2017TCIsurface,ParkRazaGilbertTeo2018}. Unlike the D-series discussion in Sec.~\ref{sec:Dseries}, in this section, we focus on symmetry breaking many-body gapping potentials that support to fractional quasiparticle excitations. For instance, the wire partition \eqref{UNpartition} respects neither one of the symmetries.

\subsection{Gapping terms for surface Dirac fermions}
We now discuss symmetry breaking gapping interactions to \eqref{eqn:BareHD0}. The array of Dirac channels can acquire a finite excitation energy gap by backscattering dimerizations between adjacent wires. The simplest ones are the single-body dimerizations \begin{align}\m{H}_{\mathrm{1-body}}&=m\sum_{y'}\left[\sum^p_{a=1}{c^a_{2y'-1}}^\dagger c^a_{2y'}+h.c.\right.\nonumber\\&\;\;\;\;\left.+\sum^q_{b=1}{c^{p+b}_{2y'}}^\dagger c^{p+b}_{2y'+1}+h.c.\right].\end{align} It partitions the $N$ Dirac channels in a given wire into $p+q$, and backscatters the two sectors in opposite directions. The backscatterings are therefore non-competing and introduce a single-body mass gap. In this section, we focus on many-body backscattering dimerizations based on the decomposition \eqref{UNpartition}. It partitions the $N$ Dirac fermions in any given wire into the $U(1)_p$ and $U(1)_q$ charged sectors and the $SU(p)_1$ and $SU(q)_1$ neutral sectors. By backscattering these decoupled sectors independently, the potentials \begin{align}\mathcal{H}&=\mathcal{H}^A_\rho+\mathcal{H}^B_\rho+\mathcal{H}^A_{SU(p)_1}+\mathcal{H}^q_{SU(q)_1},\label{HGNUN}\\\mathcal{H}^A_\rho&=-v_A\sum_{y'}\cos\left(\Phi^{A,\rho}_{2y'}-\Phi^{A,\rho}_{2y'+1}\right),\nonumber\\\mathcal{H}^B_\rho&=-v_B\sum_{y'}\cos\left(\Phi^{B,\rho}_{2y'-1}-\Phi^{B,\rho}_{2y'}\right),\nonumber\\\mathcal{H}^A_{SU(p)_1}&=u_A\sum_{y'}{\bf J}^{SU(p)}_{2y'}\cdot{\bf J}^{SU(p)}_{2y'+1},\nonumber\\\mathcal{H}^B_{SU(q)_1}&=u_B\sum_{y'}{\bf J}^{SU(q)}_{2y'-1}\cdot{\bf J}^{SU(q)}_{2y'},\nonumber\end{align} introduce a finite excitation energy gap to the coupled wire model. Here $\Phi^{A,\rho}_y=\phi^1_y+\ldots+\phi^p_y$ and $\Phi^{B,\rho}_y=\phi^{p+1}_y+\ldots+\phi^{p+q}_y$ are the bosonized variables that generate the charged $U(1)$ sectors, where $N=p+q$. The neutral sectors are generated by the $SU(r)_1$ Kac-Moody currents~\cite{bigyellowbook} \begin{align}J^{\boldsymbol\alpha,SU(p)}_y &= \sum_{a,a'=1}^p{c^a_y}^\dagger t^{\boldsymbol\alpha}_{aa'} c^{a'}_y,\nonumber\\J^{\boldsymbol\alpha,SU(q)}_y &= \sum_{b,b'=1}^q{c^{p+b}_y}^\dagger t^{\boldsymbol\alpha}_{bb'} c^{p+b'}_y,\end{align} where the fundamental matrix representations $t^{\boldsymbol\alpha}_{cc'}$ of $SU(r)$, for $r=p,q$, are listed in Appendix~\ref{append:Liealgebra}.

The $SU(r)_1$ backscattering dimerizations can be expressed in terms of bosonized variables. \begin{align}\mathcal{H}^A_{SU(p)_1}&=u_A\sum_{y'}\left[\sum_{a,a'=1}^pV^A_{aa'}\partial_x\phi^a_{2y'}\partial_x\phi^{a'}_{2y'+1}\right.\nonumber\\&\;\;\;\;\left.-\sum_{\boldsymbol\alpha\in\Delta_{SU(p)}}\cos\left(\boldsymbol\alpha\cdot2\boldsymbol\Theta^A_{2y'+1/2}\right)\right],\label{HGNSUp}\\\mathcal{H}^B_{SU(q)_1}&=u_B\sum_{y'}\left[\sum_{b,b'=1}^qV^B_{bb'}\partial_x\phi^{p+b}_{2y'-1}\partial_x\phi^{p+b'}_{2y'}\right.\nonumber\\&\;\;\;\;\left.-\sum_{\boldsymbol\alpha\in\Delta_{SU(q)}}\cos\left(\boldsymbol\alpha\cdot2\boldsymbol\Theta^B_{2y'-1/2}\right)\right],\label{HGNSUq}\end{align} where $2\boldsymbol\Theta^A_{2y'+1/2}=(2\Theta^{A,1}_{2y'+1/2},\ldots,2\Theta^{A,p}_{2y'+1/2})$ has entries $2\Theta^{A,a}_{2y'+1/2}=\phi^a_{2y'}-\phi^a_{2y'+1}$, and $2\boldsymbol\Theta^B_{2y'-1/2}=(2\Theta^{B,1}_{2y'-1/2},\ldots,2\Theta^{B,q}_{2y'-1/2})$ has entries $2\Theta^{B,b}_{2y'-1/2}=\phi^{p+b}_{2y'}-\phi^{p+b}_{2y'}$. Here the velocity terms $V^C_{cc'}$ originate from the backscatterings of the Cartan generators $H^I\sim\boldsymbol\alpha^I\cdot\partial\boldsymbol\phi$ of $SU(r)_1$, where $\boldsymbol\alpha_I$ are the simple roots of $SU(r)$ presented below eq.\eqref{SUNneutralboson}. The sine-Gordon terms are responsible in introducing a finite excitation energy gaps in the neutral sectors, and they originate from the backscatterings of the raising and lowering operator $E^{\boldsymbol\alpha}\sim e^{i\boldsymbol\alpha\cdot\boldsymbol\phi}$, where $\boldsymbol\alpha$ are the root vectors in $\Delta_{SU(r)}$ defined in \eqref{SUNroots}.

Similar to the D-series, the potentials \eqref{HGNSUp} and \eqref{HGNSUq} consists of more sine-Gordon terms than necessary in order to introduce a finite excitation gap. Instead of summing over all root vectors $\boldsymbol\alpha$ in $\Delta_{SU(r)}$, it suffices to include only a set of linearly independent simple roots $\boldsymbol\alpha^1,\ldots,\boldsymbol\alpha^{r-1}$, where we choose \begin{align}R_{SU(r)}&=\begin{pmatrix}--&\boldsymbol\alpha^1&--\\\vdots&\vdots&\vdots\\--&\boldsymbol\alpha^{r-1}&--\end{pmatrix}_{(r-1)\times r}\nonumber\\&=\begin{pmatrix}1&-1&0&\ldots&0&0\\0&1&-1&\ldots&0&0\\\vdots&\vdots&\vdots&\ddots&\vdots&\vdots\\0&0&0&\ldots&1&-1\end{pmatrix}\label{simplerootsSUr}\end{align} so that $K=(K_{IJ})_{n\times n}=R_{SU(r)}R_{SU(r)}^T$ is the Cartan matrix of $SU(r)$. All roots $\boldsymbol\alpha$ are integer combinations of the simple ones, and given $u_A,u_B$ are positive, the redundant sine-Gordon gapping terms are non-competing. Together with the gapping Hamiltonians $\mathcal{H}_\rho$ in \eqref{HGNUN} for the charged sectors, they collectively pin the ground state expectation values of the angle variables to live in the dual lattice \begin{align}\mathcal{L}_{\boldsymbol\Theta}\equiv\left\{2\boldsymbol\Theta\in\mathbb{R}^r:\begin{array}{*{20}c}\boldsymbol\alpha\cdot2\boldsymbol\Theta,\boldsymbol\alpha^0\cdot2\boldsymbol\Theta\in2\pi\mathbb{Z}\\\boldsymbol\alpha\in\Delta_{SU(r)},\boldsymbol\alpha^0=(1,\ldots,1)\end{array}\right\}\label{Urduallatice}\end{align} so that all sine-Gordon terms in \eqref{HGNUN}, \eqref{HGNSUp} and \eqref{HGNSUq} are simultaneously minimized (c.f.~the discussion on $U(4)$ in Sec.~\ref{sec:Generalcoupledwire}).

The dual lattice decomposes into the orthogonal $U(1)$ and $SU(r)$ sectors \begin{align}\mathcal{L}_{\boldsymbol\Theta}&=\mathcal{L}_{\boldsymbol\Theta}^{U(1)}\oplus\mathcal{L}_{\boldsymbol\Theta}^{SU(r)},\\\mathcal{L}_{\boldsymbol\Theta}^{U(1)}&=\mathrm{span}_{\mathbb{Z}}\{2\pi\boldsymbol\beta_0\},\nonumber\\\mathcal{L}_{\boldsymbol\Theta}^{SU(r)}&=\mathrm{span}_{\mathbb{Z}}\{2\pi\boldsymbol\beta_1,\ldots,2\pi\boldsymbol\beta_{r-1}\},\nonumber\end{align} where the primitive reciprocal vectors of $\mathcal{L}_{\boldsymbol\Theta}^{U(1)}$ and $\mathcal{L}_{\boldsymbol\Theta}^{SU(r)}$ are \begin{align}\boldsymbol\beta_I&=\frac{1}{(r-1)!}\varepsilon_{IJ\ldots K}\frac{\boldsymbol\alpha^J\wedge\ldots\wedge\boldsymbol\alpha^K}{\boldsymbol\alpha^0\cdot(\boldsymbol\alpha^1\wedge\ldots\wedge\boldsymbol\alpha^{r-1})},\label{primitivedualrootsSUr}\\\boldsymbol\beta_0&=\frac{1}{r}(1,\ldots,1)\nonumber\end{align} so that $\boldsymbol\alpha^\mu\cdot\boldsymbol\beta_\nu=\delta^\mu_\nu$, for $\mu,\nu=0,1,\ldots,r-1$. Here, the entries of the reciprocal vectors $\boldsymbol\beta_I=(\beta_I^1,\ldots,\beta_I^r)$ of $SU(r)$ 
take the explicit form \begin{align}\beta_I^a=\left\{\begin{array}{*{20}l}(r-I)/r,&\mbox{if $a\leq I$}\\-I/r,&\mbox{if $a>I$}\end{array}\right.\label{primitivedualrootsSUrentries}\end{align} for $I=1,\ldots,r-1$.

\subsubsection{A \texorpdfstring{$\mathbb{Z}_2$}{Z2} twist}
Contrary to the D-series, here there is an alternative choice of gapping potentials, which involves the product of the neutral and charged sectors \begin{align}\mathcal{H}_{\mathbb{Z}_2}&=\mathcal{H}^A_{\mathbb{Z}_2}+\mathcal{H}^B_{\mathbb{Z}_2},\label{eqn:ProdsGGN}\\
\mathcal{H}^A_{\mathbb{Z}_2}&=-u_A\sum_{y'}\cos\left(\Phi^{A,\rho}_{2y'}-\Phi^{A,\rho}_{2y'+1}\right)\nonumber\\&\quad\quad\quad\quad\quad\times\sum_{I=1}^{p-1}\cos\left(\boldsymbol\alpha_{SU(p)}^I\cdot2\boldsymbol\Theta^A_{2y'+1/2}\right),\nonumber\\\mathcal{H}^B_{\mathbb{Z}_2}&=-u_B\sum_{y'}\cos\left(\Phi^{B,\rho}_{2y'-1}-\Phi^{B,\rho}_{2y'}\right)\nonumber\\&\quad\quad\quad\quad\quad\times\sum_{I=1}^{q-1}\cos\left(\boldsymbol\alpha_{SU(q)}^I\cdot2\boldsymbol\Theta^B_{2y'-1/2}\right),\nonumber\end{align} where $2\boldsymbol\Theta^A_{2y'+1/2}=(2\Theta^{A,1}_{2y'+1/2},\ldots,2\Theta^{A,p}_{2y'+1/2})$ has entries $2\Theta^{A,a}_{2y'+1/2}=\phi^a_{2y'}-\phi^a_{2y'+1}$, and $2\boldsymbol\Theta^B_{2y'-1/2}=(2\Theta^{B,1}_{2y'-1/2},\ldots,2\Theta^{B,q}_{2y'-1/2})$ has entries $2\Theta^{B,b}_{2y'-1/2}=\phi^{p+b}_{2y'}-\phi^{p+b}_{2y'}$. Here, unlike the previous sine-Gordon terms in $\mathcal{H}^C_{SU(r)_1}$ in \eqref{HGNSUp} and \eqref{HGNSUq}, the $\mathbb{Z}_2$ terms $\mathcal{H}^C_{\mathbb{Z}_2}$ in \eqref{eqn:ProdsGGN} consist of sums of only the simple roots $\boldsymbol\alpha_{SU(r)}^I$ of $SU(r)$ (see \eqref{simplerootsSUr}).

Eq.\eqref{eqn:ProdsGGN} introduces a finite excitation energy gap. To see this, we notice each product of cosine terms generates two sine-Gordon terms using the combine angle formula \begin{align}&\cos\left(\Phi^{C,\rho}_y-\Phi^{C,\rho}_{y+1}\right)\cos\left(\boldsymbol\alpha_{SU(r)}^I\cdot2\boldsymbol\Theta^C_{y+1/2}\right)\nonumber\\&=\cos\left(\boldsymbol\alpha_{U(1)_r}^0\cdot2\boldsymbol\Theta^C_{y+1/2}\right)\cos\left(\boldsymbol\alpha_{SU(r)}^I\cdot2\boldsymbol\Theta^C_{y+1/2}\right)\nonumber\\&=\frac{1}{2}\cos\left[\left(\boldsymbol\alpha_{U(1)_r}^0+\boldsymbol\alpha_{SU(r)}^I\right)\cdot2\boldsymbol\Theta^C_{y+1/2}\right]\nonumber\\&\;\;\;+\frac{1}{2}\cos\left[\left(\boldsymbol\alpha_{U(1)_r}^0-\boldsymbol\alpha_{SU(r)}^I\right)\cdot2\boldsymbol\Theta^C_{y+1/2}\right],\end{align} where $\boldsymbol\alpha_{U(1)_r}^0=(1,\ldots,1)$ is the $r$-dimensional charge vector and $\boldsymbol\alpha_{SU(r)}^I$, for $I=1,\ldots,r$ are the simple roots of $SU(r)$. The combined angle variables satisfy the ``Haldane nullity" gapping condition~\cite{Haldane1995Haldanecondition} \begin{align}&\left[\left(\boldsymbol\alpha_{U(1)_r}^0+s\boldsymbol\alpha_{SU(r)}^I\right)\cdot2\boldsymbol\Theta^C_{y+1/2},\right.\nonumber\\&\;\;\;\left.\left(\boldsymbol\alpha_{U(1)_{r'}}^0+s'\boldsymbol\alpha_{SU(r')}^{I'}\right)\cdot2\boldsymbol\Theta^{C'}_{y'+1/2}\right]=0\end{align}  where $C,C'=A$ ($B$) and $r,r'=p$ ($q$) for even (odd) $y,y'$ respectively, and $s,s'=\pm$. There are $2r-2$ sine-Gordon terms between adjacent wires at each $y+1/2$. This provides more than enough sine-Gordon terms, when $r\geq2$, to introduce an energy gap for the $r$ pairs of counter-propagating channels. The redundant terms are non-competing and they collectively pin the bosonized angle variables $2\boldsymbol\Theta^C_{y+1/2}$ to the energy-minimizing ground state expectation values in the dual lattice \begin{align}\mathcal{L}^{\mathbb{Z}_2}_{\boldsymbol\Theta}\equiv\left\{2\boldsymbol\Theta\in\mathbb{R}^r:\left(\boldsymbol\alpha_{U(1)_r}^0\pm\boldsymbol\alpha_{SU(r)}^I\right)\cdot2\boldsymbol\Theta\in2\pi\mathbb{Z}\right\}.\label{Z2duallattice}\end{align} 

$\mathcal{L}^{\mathbb{Z}_2}_{\boldsymbol\Theta}$ contains twice as many lattice points as the original dual lattice $\mathcal{L}_{\boldsymbol\Theta}$ in \eqref{Urduallatice} for the previous coupled wire model \eqref{HGNUN}, and consequently, there are twice as many ground states between each adjacent wires. The scalar products $\boldsymbol\alpha^\mu\cdot2\boldsymbol\Theta$ can now either be all even or all odd multiples of $2\pi$, for $\mu=0,1,\ldots,r-1$. Therefore, the dual lattice admits a $\mathbb{Z}_2$ grading \begin{align}\mathcal{L}^{\mathbb{Z}_2}_{\boldsymbol\Theta}=\mathcal{L}^0_{\boldsymbol\Theta}+\mathcal{L}^1_{\boldsymbol\Theta}.\label{Z2grading}\end{align} The even lattice $\mathcal{L}^0_{\boldsymbol\Theta}=\mathcal{L}_{\boldsymbol\Theta}$ is identical to the dual lattice defined in \eqref{Urduallatice}. The odd lattice $\mathcal{L}^1_{\boldsymbol\Theta}$ displaces from the even one by half a lattice spacing \begin{align}\mathcal{L}^1_{\boldsymbol\Theta}=2\pi\boldsymbol\beta_{1/2}+\mathcal{L}^0_{\boldsymbol\Theta}.\label{oddlattice}\end{align} Here $\boldsymbol\beta_{1/2}$ can be chosen to be any vector so that $\boldsymbol\alpha^0\cdot\boldsymbol\beta_{1/2}$ and $\boldsymbol\alpha^I\cdot\boldsymbol\beta_{1/2}$ are all half integers. For example, one can take the entries of $\boldsymbol\beta_{1/2}=(\beta_{1/2}^1,\ldots,\beta_{1/2}^r)$ to be $\beta_{1/2}^a=(2+r-2ar+r^2)/(4r)$ so that $\boldsymbol\alpha^\mu\cdot\boldsymbol\beta_{1/2}=1/2$ for $\mu=0,1,\ldots,r-1$. The Hamiltonian $\mathcal{H}_{\mathbb{Z}_2}$ in \eqref{eqn:ProdsGGN}, the half dual vector $\boldsymbol\beta_{1/2}$ as well as the odd lattice $\mathcal{L}^1_{\boldsymbol\Theta}$ all depend explicitly on the choice of simple roots $\boldsymbol\alpha^I_{SU(r)}$. They therefore explicitly breaks the $SU(r)$ symmetry. Distinct choices of simple roots correspond to inequivalent ground states with distinct odd angle expectation values $\mathcal{L}^1_{\boldsymbol\Theta}$.

At this point, one can also consider gapping potentials that sum over all roots of $SU(r)$. \begin{align}\mathcal{H}_{\mathrm{even}}&=\mathcal{H}^A_{\mathrm{even}}+\mathcal{H}^B_{\mathrm{even}},\label{eqn:ProdsGGN2}\\\mathcal{H}^A_{\mathrm{even}}&=-u_A\sum_{y'}\cos\left(\Phi^{A,\rho}_{2y'}-\Phi^{A,\rho}_{2y'+1}\right)\nonumber\\&\quad\quad\quad\quad\quad\times\sum_{\boldsymbol\alpha\in\Delta_{SU(p)}}\cos\left(\boldsymbol\alpha\cdot2\boldsymbol\Theta^A_{2y'+1/2}\right),\nonumber\\\mathcal{H}^B_{\mathrm{even}}&=-u_B\sum_{y'}\cos\left(\Phi^{B,\rho}_{2y'-1}-\Phi^{B,\rho}_{2y'}\right)\nonumber\\&\quad\quad\quad\quad\quad\times\sum_{\boldsymbol\alpha\in\Delta_{SU(q)}}\cos\left(\boldsymbol\alpha\cdot2\boldsymbol\Theta^B_{2y'-1/2}\right).\nonumber\end{align} In this case, the Hamiltonian still introduces a finite excitation energy gap. However, the additional non-simple root terms put extra restrictions to the ground state expectation values of $2\boldsymbol\Theta^C_{y+1/2}$. The angle values minimize energy only when $\left(\boldsymbol\alpha_{U(1)_r}^0\pm\boldsymbol\alpha_{SU(r)}\right)\cdot2\boldsymbol\Theta$ are all integer multiples of $2\pi$, for {\em all} roots $\boldsymbol\alpha_{SU(r)}$. This rules out the odd solutions in $\mathcal{L}^1_{\boldsymbol\Theta}$ for $r\geq3$. For instance, $\boldsymbol\alpha^1+\boldsymbol\alpha^2$ is also a root vector, and the above restriction implies $\boldsymbol\alpha_{U(1)_r}^0\cdot2\boldsymbol\Theta$ as well as $\boldsymbol\alpha_{SU(r)}\cdot2\boldsymbol\Theta$ to be full integer multiples of $2\pi$. The energy-minimizing angle variables to Hamiltonian \eqref{eqn:ProdsGGN2} therefore must be even and live exclusively in $\mathcal{L}^0_{\boldsymbol\Theta}$. This is not unexpected since the exactly solvable Hamiltonian \eqref{eqn:ProdsGGN2} preserves the $SU(r)$ symmetry and so must its ground state. For instance, the angle values that belong to the $SU(r)$-breaking odd lattice $\mathcal{L}^1_{\boldsymbol\Theta}$ in \eqref{oddlattice} correspond to confined excitations that cost linearly diverging energy. 

On the other hand, one can also consider another set of gapping potentials \begin{align}\mathcal{H}_{\mathrm{odd}}&=\mathcal{H}^A_{\mathrm{odd}}+\mathcal{H}^B_{\mathrm{odd}},\label{eqn:ProdsGGN3}\\\mathcal{H}^A_{\mathrm{odd}}&=u_A\sum_{y'}\cos\left(\Phi^{A,\rho}_{2y'}-\Phi^{A,\rho}_{2y'+1}\right)\nonumber\\&\quad\quad\quad\quad\times\sum_{\boldsymbol\alpha\in\Delta_{SU(p)}}(-1)^{\mathrm{Tr}(\boldsymbol\alpha)}\cos\left(\boldsymbol\alpha\cdot2\boldsymbol\Theta^A_{2y'+1/2}\right),\nonumber\\\mathcal{H}^B_{\mathrm{odd}}&=u_B\sum_{y'}\cos\left(\Phi^{B,\rho}_{2y'-1}-\Phi^{B,\rho}_{2y'}\right)\nonumber\\&\quad\quad\quad\quad\times\sum_{\boldsymbol\alpha\in\Delta_{SU(q)}}(-1)^{\mathrm{Tr}(\boldsymbol\alpha)}\cos\left(\boldsymbol\alpha\cdot2\boldsymbol\Theta^B_{2y'-1/2}\right),\nonumber\end{align} where $(-1)^{\mathrm{Tr}(\boldsymbol\alpha)}$ is even (odd) if $\boldsymbol\alpha=a_1\boldsymbol\alpha^1+\ldots+a_{r-1}\boldsymbol\alpha^{r-1}$ is an even (resp.~odd) combination of the simple roots, for $\mathrm{Tr}(\boldsymbol\alpha)=a_1+\ldots+a_{r-1}$. Contrary to the even Hamiltonian \eqref{eqn:ProdsGGN2}, the odd Hamiltonian \eqref{eqn:ProdsGGN3} here has minimum energy when the angle variables live inside the odd lattice $\mathcal{L}^1_{\boldsymbol\Theta}$ in \eqref{oddlattice} that breaks $SU(r)$.

These gapping potentials can be continuously deformed into one another, for example, via linear interpolation \begin{align}\mathcal{H}_t=(1-t)\mathcal{H}_{\mathrm{even}}+t\mathcal{H}_{\mathrm{odd}}.\label{Htransition}\end{align} The ground states between an adjacent pair of wires are specified by the even (odd) dual lattice $\mathcal{L}^0_{\boldsymbol\Theta}$ ($\mathcal{L}^1_{\boldsymbol\Theta}$) when $t<1/2$ ($t>1/2$) respectively. At the transition at $t=1/2$, the Hamiltonian only carries sine-Gordon terms from roots that are odd combinations of the simple ones. Consequently, the ground states are identical to that of $\mathcal{H}_{\mathbb{Z}_2}$ and corresponds to the same $\mathbb{Z}_2$ graded angle expectation value structure $\mathcal{L}^{\mathbb{Z}_2}_{\boldsymbol\Theta}$ in \eqref{Z2duallattice} and \eqref{Z2grading}. This transition is analogous to Zeeman transition across the ordered phase of the Ising model
\begin{align}
H_{\mathrm{Ising}} &= -J \sum_i \s^z_i \s^z_{i+1} - h \sum_i \s^x_i - B \sum_i \s^z_i,
\end{align}
where $B$ is the magnetic field for the Zeeman coupling. When $J>h$ and $B=0$, the ordered phase has two degenerate ground states specified by $\langle\sigma^z_i\rangle=\pm1$. The Zeeman coupling $B$ introduces a preference of up spins versus down ones, and breaks the degeneracy. Here, the parameter $t-1/2$ in \eqref{Htransition} takes a similar role as the Zeeman field $B$.

In general, there is an intricate phase diagram when the strengths and signs of the sine-Gordon terms $\cos\left(\boldsymbol\alpha\cdot2\boldsymbol\Theta\right)$ can vary from one to another. There are multiple distinct $\mathbb{Z}_2$ critical phases, where the ground states between an adjacent pair of wires take a $\mathbb{Z}_2$ graded structure. In the thermodynamic limit with an infinite number of wires, this introduces a diverging ground state degeneracy. This signifies a gap-closing critical transition between distinct 2D gapped phases. On the other hand, the diverging degeneracy could also be lifted if the theory is coupled with a $\mathbb{Z}_2$ gauge theory (similar to the one studied for the D-series in section~\ref{sec:Dseries}). These discussions are out of the scope of this article and we refer them to future works.

\subsubsection{Quasiparticle excitations}
The deconfined quasiparticle excitations of the coupled wire model \eqref{HGNUN} are kinks of the angle variables $\langle2\boldsymbol\Theta^C_{y+1/2}\rangle$. Similar discussions were provided for $U(4)$ in Sec.~\ref{sec:Generalcoupledwire}. Here, we summarized the results for the general $U(N)$. The ground state expectation values $\langle2\boldsymbol\Theta^C_{y+1/2}\rangle$ belong in the dual lattice $\mathcal{L}_{\boldsymbol\Theta}$ defined in \eqref{Urduallatice}. A quasiparticle excitation at $(x_0,y_0)$ is a kink where the angle variable $\langle2\boldsymbol\Theta^C_{y_0+1/2}(x)\rangle$ jumps discontinuously from one value to another in $\mathcal{L}_{\boldsymbol\Theta}$ when $x$ passes across $x_0$. A quasiparticle excitation can be created by acting a vertex operator $V_{y_0}^{C,\boldsymbol\gamma}(x_0)=\exp\left[i\gamma_a\phi^{C,a}_{y_0}(x_0)\right]$ on a ground state, where $C=A$ ($B$) and $a=1,\ldots,r$ for $r=p$ ($q$). These vertex operators are classified according to the primary fields of the $U(1)_r\times SU(r)_1$ Kac-Moody algebra. Each primary field is a super-selection sector of vertex operators that form an irreducible representation of the $U(1)\times SU(r)$ (c.f.~\eqref{SO6EVOPE} for $SO(6)_1$). For example, the smallest primary field $[1]_\rho$ for the $U(1)_r$ charge sector is the single vertex operator $e^{i\tilde\phi^{C,\rho}_{y_0}(x_0)}$, where $\tilde\phi^{C,\rho}=\Phi^{C,\rho}/r=(\phi^{C,1}+\ldots+\phi^{C,r})/r$. It creates a fractional excitation with spin (equivalently, conformal scaling dimension) $1/2r$. General primary field excitations $[m]_\rho$ in the $U(1)_r$ sector are generated by higher order copies $e^{im\tilde\phi^{C,\rho}_{y_0}(x_0)}$. They carry spin $m^2/2r$ and follow the fusion rule $[m]_\rho\times[m']_\rho=[m+m']_\rho$.

There are $r$ primary fields in the $SU(r)_1$ sector. Examples were presented for the $SU(4)_1$ case in Sec.~\ref{sec:Generalcoupledwire}. Here, we demonstrate the general case. We begin with the smallest non-trivial primary field, denoted by $E^1$, that corresponds to the fundamental representation of $SU(r)$. The super-selection sector $E^1$ consists of the collection of vertex operators \begin{align}E^{C,1}_{y_0}(x_0)\sim\mathrm{span}_{\mathbb{C}}\left\{e^{i\boldsymbol\gamma\cdot\boldsymbol\phi^C_{y_0}(x_0)}:\boldsymbol\gamma=\sigma(\boldsymbol\beta_1),\sigma\in S_r\right\}\end{align} where $\boldsymbol\beta_1$ is the primitive dual root $(r-1,-1,\ldots,-1)/r$ (see \eqref{primitivedualrootsSUr} and \eqref{primitivedualrootsSUrentries}), and $\sigma$ permutes the entries of the $r$-dimensional vector. The super-selection sector irreducibly represents $SU(r)_1$ in the sense that it is closed under operator products with the $SU(r)_1$ currents (c.f.~\eqref{SO6EVOPE}). Since all entries of $\boldsymbol\beta_1$ is identical except one, there are exactly $r$ permutations $\sigma(\boldsymbol\beta)$. Therefore $E^1$ are generated by $r$ vertex operators, which form the fundamental representation of $SU(r)$.

In general, the primary field $E^c$, for $c=1,\ldots,r-1$, is the super-selection sector \begin{align}E^{C,c}_{y_0}(x_0)\sim\mathrm{span}_{\mathbb{C}}\left\{e^{i\boldsymbol\gamma\cdot\boldsymbol\phi^C_{y_0}(x_0)}:\boldsymbol\gamma=\sigma(\boldsymbol\beta_c),\sigma\in S_r\right\},\end{align} where the simple dual root $\boldsymbol\beta_c$ was defined in \eqref{primitivedualrootsSUr} and \eqref{primitivedualrootsSUrentries}. There are exactly $C^r_c=r!/[c!(r-c)!]$ entry permutations and therefore $E^c$ forms a $C^r_c$ dimensional irreducible representation of $SU(r)_1$. Since $\sigma(\boldsymbol\beta_c)$ has $c$ entries being $(r-c)/r$ and $r-c$ entries being $-c/r$, the primary field has spin (equivalently, conformal scaling dimension) \begin{align}h_{E^c}=\frac{c(r-c)^2+(r-c)c^2}{2r^2}=\frac{(r-c)c}{2r}.\end{align} Lastly, the trivial primary field is $E^0=1$. The primary fields obey the fusion rules \begin{align}E^c\times E^{c'}=E^{[c+c']_{\mathrm{mod}\mbox{ }r}}.\end{align}

\subsection{Duality transformation}
We generalize the duality properties of the coupled wire model from that of the free Dirac fermion reviewed in Sec.~\ref{sec:Dirac-QED3}.
Under the duality transformation 
\begin{align}
\tilde{\phi}^a_y &= \sum_{y'} \mathrm{sgn}(y-y')(-1)^{y'} \phi^a_{y'}, \label{eqn:Dual1}
\end{align}
the angle variables in the sine-Gordon terms in \eqref{HGNUN}, \eqref{HGNSUp} and \eqref{HGNSUq} are self-dual up to a sign
\begin{align}
2\tilde{\T}^{A,a}_{2y'+1/2} &\equiv \tilde{\phi}^a_{2y'} - \tilde{\phi}^a_{2y'+1} = -2 \T^{A,a}_{2y'+1/2}.
\end{align}
Therefore, the sine-Gordon terms in \eqref{HGNUN}, \eqref{HGNSUp} and \eqref{HGNSUq} are also self-dual
\begin{align}
\tilde{\mathcal{H}}^A_\rho&=-v_A\sum_{y'}\cos\left(\boldsymbol\alpha_{U(1)_p}^0\cdot2\tilde{\boldsymbol\Theta}_{2y'+1}\right)\nonumber\\&=-v_A\sum_{y'}\cos\left(\boldsymbol\alpha_{U(1)_p}^0\cdot2\boldsymbol\Theta_{2y'+1}\right)=\mathcal{H}^A_\rho\\
\tilde{\mathcal{H}}^A_{SU(p)_1}&=-u_A\sum_{y'}\sum_{\boldsymbol\alpha\in\Delta_{SU(p)}}\cos\left(\boldsymbol\alpha\cdot2\tilde{\boldsymbol\Theta}^A_{2y'+1/2}\right)\nonumber\\&=-u_A\sum_{y'}\sum_{\boldsymbol\alpha\in\Delta_{SU(p)}}\cos\left(\boldsymbol\alpha\cdot2\boldsymbol\Theta^A_{2y'+1/2}\right)\nonumber\\&=\mathcal{H}^A_{SU(p)_1}
\end{align}
Similarly, the sine-Gordon terms for the $B$ sector are also self-dual.

Lastly, we consider vertex operators that correspond to primary fields and create quasiparticle excitations. The duality transformation \eqref{eqn:Dual1} can be re-expressed in terms of the angle variables $2\T^a_y(x)$ as
\begin{align}\begin{split}
\tilde{\phi}^a_{2y}(x) =  \phi^a_{2y+1}(x) + \sum_{y'}\mathrm{sgn}(y-y') 2\T^{A,a}_{2y'+1/2}(x)\nonumber\\\tilde{\phi}^a_{2y+1}(x) =  \phi^a_{2y}(x) + \sum_{y'}\mathrm{sgn}(y-y') 2\T^{A,a}_{2y'+1/2}(x)\end{split},
\end{align}
and similarly for the $B$ sector. We see that the dual vertex operators are dressed with non-local strings, similar to \eqref{eqn:DiracDual2} in section~\ref{sec:Dirac-QED3}. When acting on a ground state, the angle variables $2\T^{A,a}_{2y'+1/2}$ are pinned and can be replaced by their ground state expectation values. The non-local string therefore condenses into the ground state leaving only complex phases behind.

Like $D$ series, if we extend the coupling constants of the sine-Gordon terms $u_{A/B}$ to be complex valued, then the ground state manifold changes continuously as we vary the phases $\t_{A/B}$ of $u_{A/B}$. The self-dual points are $\t_{A/B}= 0,\pi$ and duality transformation (\ref{eqn:Dual1}) on the complex $u$ plane is a reflection with respect to the real axis. The ground state manifold can be visualized for SU(3) or SU(4) cases and it should be true for the general SU($N$) theories.

\section{E-series: \texorpdfstring{$(E_{8,7,6})_1$}{E876} surface theory}
\label{sec:E-series}

The exceptional Lie algebra $E_6$, $E_7$ and $E_8$ are the remaining simply-laced Lie algebra in the {\it ADE} classification. We first discuss the $E_8$ algebra. In addition to the conventional topological insulators that host protected Dirac surface states, topological paramagnets~\cite{VishwanathSenthil12,WangPotterSenthil13} are alternative time reversal and charge $U(1)$ symmetry preserving topological states enabled by interactions. These are short-ranged entangled SPT states in three dimensions that do not exhibit bulk quasiparticle fractionalization or topological order. However, they do carry anomalous surface states that cannot be supported in a pure two dimensional system. We are interested in the {\it efmf} topological paramagnetic state. Like a conventional topological insulator, its surface state is unstable against time reversal breaking perturbations. A finite excitation energy gap can be introduced on the surface by a magnetic order without requiring surface topological order or fractionalization. The {\it efmf} topological paramagnet is distinct from a conventional topological insulator in that a magnetic surface domain wall -- a line interface that separates two time reversal breaking gapped surface domains with opposite magnetic orientations -- hosts quasi-one-dimensional low-energy electronic degrees of freedom that are chiral only in energy but has no electric charge transport. Electronic quasiparticles are chiral in the sense that they propagate in a single forward direction along the line interface. They collectively account for a chiral heat current $I_{\mathrm{eng}}=I_{\mathrm{eng}}^R-I_{\mathrm{eng}}^L$ that obey the differential thermal conductance $\kappa=dI_{\mathrm{eng}}/dT=c(\pi^2k_B^2/3h)T$ in low temperature $T$, where the central charge is $c=8$. However, electric charge transport is non-chiral in that the chiral electric current $I=I^R-I^L$ does not response to change of electric potential, $\sigma=dI/dV=0$. These low-energy degrees of freedom can be effectively described by a $1+1$D $E_8$ Kac-Moody CFT at level 1. They can be described by the bosonized Lagrangian density \begin{align}\mathcal{L}_0&=\frac{1}{2\pi}\sum_{a=1}^8\partial_t\phi^a\partial_x\phi^a-\sum_{a,b=1}^8V_{ab}\partial_x\phi^a\partial_x\phi^b\nonumber\\&=\frac{1}{2\pi}\sum_{I,J=1}^8(K_{E_8})_{IJ}\partial_t\phi'^I\partial_x\phi'^J-\sum_{I,J=1}^8V'_{IJ}\partial_x\phi'^I\partial_x\phi'^J,
\end{align} where the ``Cartan-Weyl" and ``Chevalley" bosonized variables $\phi$ and $\phi'$ are related by the basis transformation \begin{align}\phi'^I&=(K_{E_8}^{-1})^{IJ}\Phi_J=\sum_{a=1}^8(R_{E_8}^{-1})^I_a\phi^a,\nonumber\\R_{E_8} &= \begin{pmatrix}--&\boldsymbol\alpha^1&--\\\vdots&\vdots&\vdots\\--&\boldsymbol\alpha^8&--\end{pmatrix} \nonumber \\
&= \begin{pmatrix}
1 & -1 & 0 & 0 & 0 & 0 & 0 & 0 \\
0 & 1 & -1 & 0 & 0 & 0 & 0 & 0 \\
0 & 0 & 1 & -1 & 0 & 0 & 0 & 0 \\
0 & 0 & 0 & 1 & -1 & 0 & 0 & 0 \\
0 & 0 & 0 & 0 & 1 & -1 & 0 & 0 \\
0 & 0 & 0 & 0 & 0 & 1 & -1 & 0 \\
0 & 0 & 0 & 0 & 0 & 0 & 1 & -1 \\
-\frac{1}{2} & -\frac{1}{2} & -\frac{1}{2} & -\frac{1}{2} & -\frac{1}{2} & \frac{1}{2} & \frac{1}{2} & \frac{1}{2}
\end{pmatrix},\label{simplerootsE8}\end{align} and the Cartan matrix of $E_8$ \begin{align}K_{E_8}=R_{E_8}R_{E_8}^T\end{align} (see Eq.~(\ref{eqn:CartanE876}) in Appendix~\ref{append:Liealgebra} for an explicit expression) has determinant 1 and is invertible.

Here, it is important to realize that the neutral fermionic vertex operators $e^{i\phi^a}$ are non-local and fractional. They are not the primary field excitations of the $E_8$ CFT, which only supports local integral excitations. Instead, the low-energy physical excitations are generated by the local bosonic vertex operators $e^{i\Phi_I}=e^{iK_{IJ}\phi'^J}=e^{iR^I_a\phi^a}$. Since $K_{E_8}$ has integral inverse, $e^{i\phi'^I}=e^{i(K^{-1})^{IJ}\Phi_J}$ are also local and bosonic. These are even integral combinations of electrons/holes, each of which is assumed to carry net zero electric charge. All odd combinations of electrons/holes correspond to gapped fermionic excitations. They do not contribute to the low-temperature chiral energy transport and are not described by the low-energy effective $E_8$ CFT.

The $E_8$ Kac-Moody currents consist of the 8 Cartan generators $\partial\Phi_I$ and the 240 roots $E^{\boldsymbol\alpha}=e^{i\boldsymbol\alpha\cdot\boldsymbol\phi}$. The 240 roots can be decomposed into the 112 $SO(16)$ roots and 128 spinor representations of $SO(16)$. \begin{align}\Delta_{E_8}=\Delta_{SO(16)}\cup\Delta_{s_-}\end{align} The 112 root vectors in $\Delta_{SO(16)}$ were defined in \eqref{simplerootsSO2n} in Sec.~\ref{sec:Dseries}, and they take the form $\boldsymbol\alpha=\pm{\bf e}_a\pm{\bf e}_b$, where $a\neq b$. In this paper, we adopt the convention where the $E_8$ roots extends from that of $SO(16)$ by its {\em odd} spinors $s_-$. The 128 odd spinor vectors in $\Delta_{s_-}$ take the form $\boldsymbol\alpha=(\varepsilon_1,\ldots,\varepsilon_8)/2$ where $\varepsilon_a=\pm1$ and $\varepsilon_1\ldots\varepsilon_8=-1$. All 240 roots of $E_8$ are integral combinations of the simple ones defined by the row vectors of $R_{E_8}$ in \eqref{simplerootsE8}. Since $e^{iR^I_a\phi^a}$ are bosonic integral combinations of local electrons, so are all the $E_8$ current operators.

We consider time reversal breaking stripes with alternating magnetic orientation on the surface of the {\it efmf} topological paramagnet (c.f.~figure~\ref{fig:surfaceCW}). This reduces the low-energy electronic degrees of freedom to an array of $E_8$ wires with alternating propagating directions. Similar to the D-series coupled wire model discussed in section~\ref{sec:Dseries}, the $E_8$ array exhibits an emergent antiferromagnetic time reversal symmetry, which composes of a time reversal and a half-translation $y\to y+1$. AFTR preserving fractionalization $E_8\sim SO(8)\times SO(8)$ and gapping interactions were studied in ref.~\onlinecite{Sahoo2016coupledwireTSC}. Instead, in this section, we focus on AFTR symmetry breaking gapping interactions based on asymmetric partitions of the $E_8$ current algebra. In particular, we concentrate on the conformal embeddings \begin{align}E_8\supseteq E_7\times SU(2),\quad E_8\supseteq E_6\times SU(3)\label{E8decomposition}\end{align} that involve the other two exceptional simply-laced Lie algebras. The coupled wire model is constructed by backscattering the two decoupled components $E_7$ and $SU(2)$ (or $E_6$ and $SU(3)$) on each wire to adjacent wires in opposite directions.

Before discussing these surface models, we first consider a set of simple gapping potentials that fully dimerizes the $E_8$ wires. \begin{align}\mathcal{H}_{\mathrm{dimer}}&=u\sum_{y'}{\bf J}^{E_8}_{2y'-1}\cdot{\bf J}^{E_8}_{2y'}\nonumber\\&=u\sum_{y'}\sum_{a=1}^8\partial_x\phi^a_{2y'-1}\partial_x\phi^a_{2y'}\nonumber\\&\;\;\;-u\sum_{y'}\sum_{\boldsymbol\alpha\in\Delta_{E_8}}\cos\left(\boldsymbol\alpha\cdot2\boldsymbol\Theta_{2y'-1/2}\right),\label{HE8}\end{align} where the sine-Gordon angle parameter is $\Theta^a_{2y'-1/2}=\phi^a_{2y'-1}-\phi^a_{2y'}$. Similar to the coupled wire models in the previous sections, to simultaneously minimize the sine-Gordon terms in \eqref{HE8}, the angle parameters take ground state expectation values inside the dual lattice (c.f.~\eqref{BCCn}) \begin{align}
\m{L}^{E_8}_{\b{\T}} &\equiv \{ 2\b{\T}: \b{\a} \cdot 2\b{\T} \in 2\pi \mathbb{Z},\quad\boldsymbol\alpha\in\Delta_{E_8} \}, \nonumber \\
&= \text{span}_{\mathbb{Z}} \{ 2\pi \b{\beta_1}, \dots, 2\pi \b{\beta_8} \}, \\
\boldsymbol\beta_I&=\frac{1}{8!}\varepsilon_{IJ_1\ldots J_7}\frac{\boldsymbol\alpha^{J_1}\wedge\ldots\wedge\boldsymbol\alpha^{J_7}}{\boldsymbol\alpha^1\cdot(\boldsymbol\alpha^2\wedge\ldots\wedge\boldsymbol\alpha^8)},\nonumber\end{align}
where $\boldsymbol\alpha^1,\ldots,\boldsymbol\alpha^8$ are the simple roots in \eqref{simplerootsE8}. The primitive dual root vectors satisfy $\boldsymbol\beta_I\cdot\boldsymbol\alpha^J=\delta_I^J$, i.e.~$R^\vee_{E_8}R_{E_8}^T=\openone_{8\times8}$, and they take the explicit form \begin{align}R^\vee_{E_8} &= \begin{pmatrix}--&\boldsymbol\beta_1&--\\\vdots&\vdots&\vdots\\--&\boldsymbol\beta_8&--\end{pmatrix} \nonumber \\
&= -\frac{1}{2} \begin{pmatrix}
-1 & 1 & 1 & 1 & 1 & 1 & 1 & 1 \\
0 & 0 & 2 & 2 & 2 & 2 & 2 & 2 \\
1 & 1 & 1 & 3 & 3 & 3 & 3 & 3 \\
2 & 2 & 2 & 2 & 4 & 4 & 4 & 4 \\
3 & 3 & 3 & 3 & 3 & 5 & 5 & 5 \\
2 & 2 & 2 & 2 & 2 & 2 & 4 & 4 \\
1 & 1 & 1 & 1 & 1 & 1 & 1 & 3 \\
2 & 2 & 2 & 2 & 2 & 2 & 2 & 2
\end{pmatrix}.
\end{align}
The dual lattice is self-dual up to a $2\pi$ multiplicative factor in the sense that $\text{span}_{\mathbb{Z}} \{ \b{\a}^1,\dots,\b{\a}^8 \} = \text{span}_{\mathbb{Z}} \{ \b{\beta}_1,\dots, \b{\beta}_8 \}$ because \begin{align}R^\vee_{E_8}={R_{E_8}^T}^{-1}=K_{E_8}^{-1}R_{E_8}\end{align} and $K_{E_8}$ has integral inverse. This is consistent with the fact that the root lattice of $E_8$ is unimodular. Consequently, all deconfined excitations of the coupled wire model \eqref{HE8} that correspond to kinks of $\langle2\boldsymbol\Theta_{2y'-1/2}\rangle\in\mathcal{L}^{E_8}_{\boldsymbol\Theta}$ are local and can be created by integral combination of electron/hole operators.

\subsection{\texorpdfstring{$E_7\times SU(2)$}{E7 x SU(2)}}
We now construct the coupled wire model that utilizes the partition $E_8 \supset E_7 \times SU(2)$ and describes a gapped symmetry breaking surface of a topological paramagnet. Each $E_8$ wire on the 2D surface array (c.f.~figure~\ref{fig:surfaceCW}(c)) is decomposed into a $E_7$ and a $SU(2)$ Kac-Moody CFT at level 1. These two sectors decouple from each other and act on orthogonal Hilbert spaces. This motivates the gapping Hamiltonian \begin{align}\mathcal{H}&=u\sum_{y'}{\bf J}^{E_7}_{2y'-1}\cdot{\bf J}^{E_7}_{2y'}+{\bf J}^{SU(2)}_{2y'}\cdot{\bf J}^{SU(2)}_{2y'+1}\label{CWE70}\end{align} that backscatters the two decoupled currents from a wire into adjacent wires in opposite directions. In the following, we define the current embeddings of ${\bf J}^{E_7}$ and ${\bf J}^{SU(2)}$ into $E_8$.

We begin with the new set of simple root vectors of $E_7 \times SU(2)$
\begin{align}
\b{\a}^I &= \mathbf{e}_{I+1} - \mathbf{e}_{I+2}, \quad I=1,\dots,6, \nonumber \\
\b{\a}^7 &= \frac{1}{2} (-1,-1,-1,-1,-1,1,1,1), \nonumber \\
\b{\a}^8 &= \frac{1}{2} (-1,1,1,1,1,1,1,1), \label{eqn:E7SimpleRoots}
\end{align}
where $\b{\a}^1,\ldots,\b{\a}^7$ are the simple root vectors of $E_7$ and $\b{\a}^8$ generates $SU(2)$. It is easy to see that the Cartan $K$-matrix splits \begin{align}K_{E_7\times SU(2)}=\left(\b{\a}^I\cdot\b{\a}^J\right)_{8\times8}=\begin{pmatrix}K_{E_7}&0\\0&K_{SU(2)}\end{pmatrix},\end{align} where the explicit form of $K_{E_7}$ can be found in  Eq.~(\ref{eqn:CartanE876}) in Appendix~\ref{append:Liealgebra} and $K_{SU(2)}=2$. The $E_7$ root system can be embedded in $E_8$ by taking the subset \begin{align}\Delta_{E_7}=\left\{\boldsymbol\alpha\in\Delta_{E_8}:\boldsymbol\alpha\cdot\boldsymbol\alpha^8=0\right\}\subseteq\Delta_{E_8}.\end{align} The 126 roots in $\Delta_{E_7}$ is an extension of the 42 roots of $SU(7)$ -- a subgroup of $E_7$ -- by the weight vectors of the irreducible representations ${\bf 7}$, $\overline{\bf 7}$, ${\bf 35}$, and $\overline{\bf 35}$. \begin{align}\Delta_{E_7}=\iota\Delta_{SU(7)}+{\bf 7}+\overline{\bf 7}+{\bf 35}+\overline{\bf 35}.\end{align} To illustrate this, we embed the root system of $SU(7)$ (see \eqref{SUNroots}) in that of $SO(16)\subseteq E_8$ by putting the 7 dimensional root vectors $\boldsymbol\alpha\in\Delta_{SU(7)}$ in the 8 dimensional space, \begin{align}\begin{split}\iota:\boldsymbol\alpha=(\alpha_1,\ldots,\alpha_7)\mapsto\iota\boldsymbol\alpha=(0,\alpha_1,\ldots,\alpha_7),\\\iota\Delta_{SU(7)}=\left\{\iota\boldsymbol\alpha:\boldsymbol\alpha\in\Delta_{SU(7)}\right\}.\end{split}\label{iotaE7}\end{align} Next, we observe that certain sub-collections of the $E_8$ roots form the weight vectors of the representations ${\bf 7}$, $\overline{\bf 7}$, ${\bf 35}$, and $\overline{\bf 35}$. They are given by \begin{align}{\bf 7}&=\left\{{\bf e}_1+{\bf e}_j:j=2,\ldots,8\right\},\nonumber\\\overline{\bf 7}&=\left\{-{\bf e}_1-{\bf e}_j:j=2,\ldots,8\right\},\nonumber\\{\bf 35}&=\left\{\frac{1}{2}(1,s_2,\ldots,s_8):s_2,\ldots,s_8=\pm1,\sum_{j=2}^8s_j=1\right\},\nonumber\\\overline{\bf 35}&=\left\{\frac{1}{2}(-1,s_2,\ldots,s_8):s_2,\ldots,s_8=\pm1,\sum_{j=2}^8s_j=-1\right\}.\nonumber\end{align} Each of these weight vectors is orthogonal to $\boldsymbol\alpha^8$ and therefore decouples from the $SU(2)$. While ${\bf 7}$ and $\overline{\bf 7}$ can be embedded in the root system of $SO(16)$, ${\bf 35}$ and $\overline{\bf 35}$ can only be embedded in $E_8$ as they consists of half-integral vectors. Each of these collections of weight vectors $\boldsymbol\gamma^a$ corresponds to a super-selection sector of vertex operators $\mathrm{span}\{e^{i\boldsymbol\gamma^a\cdot\boldsymbol\phi}\}$ that transforms closely and irreducibly under the $SU(7)_1$ Kac-Moody algebra (c.f.~\eqref{SO6EVOPE}). Each sector splits into $\eta\otimes E^c$, where $E^c$ is a primary field of $SU(7)_1$ and $\eta$ is a primary field of the coset $(E_7)_1/SU(7)_1$, so that the combined spin (conformal scaling dimension) is 1. 

The coupled wire model \eqref{CWE70} can be expressed as a sum of sine-Gordon gapping interactions \begin{align}\mathcal{H}&=-u\sum_{y'}\sum_{\boldsymbol\alpha\in\Delta_{E_7}}\cos\left(\boldsymbol\alpha\cdot2\boldsymbol\Theta_{2y'-1/2}\right)\nonumber\\&\;\;\;-u\sum_{y'}\cos\left(\boldsymbol\alpha^8\cdot2\boldsymbol\Theta_{2y'+1/2}\right),\label{CWE71}\end{align} where $\boldsymbol\alpha^8$ is the root vector of $SU(2)$ when embedded in $E_8$ and we have suppressed the non-gapping Cartan generator terms that renormalize velocities. Here, $2\boldsymbol\Theta=(2\T^1,\ldots,2\T^8)$ and $2\T^a_{y+1/2}=\phi^a_y-\phi^a_{y+1}$. The sine-Gordon terms in the first line in \eqref{CWE71} dimerize the $E_7$ currents between wire $2y'-1$ and $2y'$ while terms in the second line dimerize the remaining $SU(2)$ currents between wire $2y'$ and $2y'+1$. Together, they introduce a finite excitation energy gap.

Quasiparticle excitation can be created by primary fields of the $E_7$ or the $SU(2)$ sector. The semionic primary field of $SU(2)$ at wire $y$ is the super-selection sector of vertex operators $s\sim\mathrm{span}\{e^{i\boldsymbol\beta_8\cdot\boldsymbol\phi_y},e^{-i\boldsymbol\beta_8\cdot\boldsymbol\phi_y}\}$. Here, the weight vector $\boldsymbol\beta_8=(-1,1,\ldots,1)/4=\boldsymbol\alpha^8/2$ is orthogonal to all $E_7$ roots and has length square $|\boldsymbol\beta_8|^2=1/2$. Consequently, the primary field decouples from the $E_7$ sector and carries conformal scaling dimension $h_s=|\boldsymbol\beta_8|^2/2=1/4$. Each of the vertex operators creates a bulk quasiparticle excitation in the form of a kink of the sine-Gordon angle parameter between wire $y$ and $y+1$ ($y-1$) if $y$ is even (resp.~odd). The anti-semionic primary field of $E_7$ at wire $y$ is the super-selection sector of vertex operators \begin{align}\bar{s}\sim\mathrm{span}\{e^{i\boldsymbol\beta\cdot\boldsymbol\phi_y}:\boldsymbol\beta\in\mathcal{S}_{E_7}\},\end{align} where $\mathcal{S}_{E_7}$ is the collection of dual vectors \begin{align}\mathcal{S}_{E_7}&=\{\boldsymbol\beta_8-{\bf e}_I-{\bf e}_J:2\leq I<J\leq8\}\nonumber\\&\quad\quad\cup\{-\boldsymbol\beta_8+{\bf e}_I+{\bf e}_J:2\leq I<J\leq8\}\nonumber\\&\quad\quad\cup\{\boldsymbol\beta_8+{\bf e}_1-{\bf e}_I:2\leq I\leq8\}\nonumber\\&\quad\quad\cup\{-\boldsymbol\beta_8-{\bf e}_1+{\bf e}_I:2\leq I\leq8\}.\end{align} This collection of 56 vertex operators form the 56 dimensional irreducible representation of $E_7$ and corresponds to the only non-trivial primary field of $E_7$ at level 1. All weight vectors in $\mathcal{S}_{E_7}$ are orthogonal to $\boldsymbol\alpha^8$ and they all have length square $|\boldsymbol\beta|^2=3/2$. Therefore the primary field $\bar{s}$ decouples from $SU(2)$ and carries conformal scaling dimension $h_{\bar{s}}=3/4$. It creates a kink of the sine-Gordon angle parameter between wire $y$ and $y-1$ ($y+1$) if $y$ is even (resp.~odd).

\subsection{\texorpdfstring{$E_6\times SU(3)$}{E6 x SU(3)}}
The discussion of $E_6\times SU(3)$ resembles that of $E_7\times SU(2)$. The gapping Hamiltonian takes the current backscattering form \begin{align}\mathcal{H}&=u\sum_{y'}{\bf J}^{E_6}_{2y'-1}\cdot{\bf J}^{E_6}_{2y'}+{\bf J}^{SU(3)}_{2y'}\cdot{\bf J}^{SU(3)}_{2y'+1}.\label{CWE60}\end{align} $E_6$ and $SU(3)$ are embedded in the $E_8$ by setting the simple roots 
\begin{align}
\b{\a}^I &= \mathbf{e}_{I+2} - \mathbf{e}_{I+3}, \quad I=1,\dots,5,\nonumber \\
\b{\a}^6 &= \frac{1}{2} (-1,-1,-1,-1,-1,1,1,1), \nonumber \\
\b{\a}^7 &= (1,-1,0,0,0,0,0,0), \nonumber \\
\b{\a}^8 &= \frac{1}{2} (-1,1,1,1,1,1,1,1). \label{eqn:E6SimpleRoots}
\end{align}
The Cartan $K$-matrix \begin{align}K_{E_6\times SU(3)}=\left(\b{\a}^I\cdot\b{\a}^J\right)_{8\times8}=\begin{pmatrix}K_{E_6}&0\\0&K_{SU(3)}\end{pmatrix}\end{align} splits, and therefore the $E_7$ and $SU(2)$ sectors decouple. The explicit form of the Cartan matrices of $E_6$ and $A_2=SU(3)$ can be found in Eq.~(\ref{eqn:CartanE876}) Appendix~\ref{append:Liealgebra}. The $SU(3)$ root system, as embedded in $E_8$, consists of vectors in \begin{align}\Delta_{SU(3)}=\{\pm\b{\a}^7,\pm\b{\a}^8,\pm(\b{\a}^7+\b{\a}^8)\}\subseteq\Delta_{E_8}.\end{align} Like the roots of $E_7$, the roots of $E_6$ are the orthogonal complement of $SU(3)$ in $E_8$, \begin{align}\Delta_{E_6}=\{\boldsymbol\alpha\in\Delta_{E_8}:\boldsymbol\alpha\cdot\boldsymbol\alpha^7=\boldsymbol\alpha\cdot\boldsymbol\alpha^8=0\}\subseteq\Delta_{E_8}.\end{align} The 72 roots of $E_6$ extend the 30 roots of $SU(6)$ by including weight vectors of the irreducible representations ${\bf 1}$, $\overline{\bf 1}$, ${\bf 20}$, and $\overline{\bf 20}$. \begin{align}\Delta_{E_6}=\iota\Delta_{SU(6)}+{\bf 1}+\overline{\bf 1}+{\bf 20}+\overline{\bf 20}.\end{align} Here, $\iota$ embeds the $SU(6)$ roots into $E_8$ (c.f.~\eqref{iotaE7} for the $E_7$ case) so that the embedded simple $SU(6)$ roots are $\boldsymbol\alpha^1,\ldots,\boldsymbol\alpha^5$. The four irreducible representations of $SU(6)$ involved in the extension have weight vectors \begin{align}{\bf 1}&=\{{\bf e}_1+{\bf e}_2\},\quad\overline{\bf 1}=\{-{\bf e}_1-{\bf e}_2\},\nonumber\\{\bf 20}&=\left\{\frac{1}{2}(1,1,s_3,\ldots,s_8):s_3,\ldots,s_8=\pm1,\sum_{j=3}^8s_j=0\right\},\nonumber\\\overline{\bf 20}&=\left\{\frac{-1}{2}(1,1,s_3,\ldots,s_8):s_3,\ldots,s_8=\pm1,\sum_{j=3}^8s_j=0\right\}.\nonumber\end{align}

Up to non-gapping boson velocity terms, the coupled wire model \eqref{CWE60} can be expressed as a sum of sine-Gordon potentials \begin{align}\mathcal{H}&=-u\sum_{y'}\sum_{\boldsymbol\alpha\in\Delta_{E_6}}\cos\left(\boldsymbol\alpha\cdot2\boldsymbol\Theta_{2y'-1/2}\right)\nonumber\\&\;\;\;-u\sum_{y'}\sum_{\boldsymbol\alpha\in\Delta_{SU(3)}}\cos\left(\boldsymbol\alpha\cdot2\boldsymbol\Theta_{2y'+1/2}\right),\label{CWE61}\end{align} where $2\boldsymbol\Theta=(2\T^1,\ldots,2\T^8)$ and $2\T^a_{y+1/2}=\phi^a_y-\phi^a_{y+1}$. The sine-Gordon terms in the first line in \eqref{CWE61} dimerize the $E_6$ currents between wire $2y'-1$ and $2y'$ while terms in the second line dimerize the remaining $SU(3)$ currents between wire $2y'$ and $2y'+1$. They can be shown to introduce a finite excitation energy gap. The proof is similar to the previous cases for the A and D-series and will be omitted.

Quasiparticle excitations, in the form of kinks of angle parameters in \eqref{CWE61}, can be created by primary fields $e^{i\boldsymbol\beta\cdot\boldsymbol\phi}$ in the $E_6$ and $SU(3)$ Kac-Moody CFT at level 1. We begin with the $SU(3)$ sector. The fundamental representation corresponds to the primary field $e^+\sim\mathrm{span}\{e^{i\boldsymbol\beta\cdot\boldsymbol\phi}:\boldsymbol\beta\in\mathcal{S}_{SU(3)}\}$, where the weight vectors are \begin{align}\mathcal{S}_{SU(3)}&=\{-\boldsymbol\beta_7,\boldsymbol\beta_8,\boldsymbol\beta_7-\boldsymbol\beta_8\},\\\boldsymbol\beta_8&=\frac{1}{3}(0,0,1,\ldots,1),\quad\boldsymbol\beta_7=\boldsymbol\alpha_7+\boldsymbol\alpha_8-\boldsymbol\beta_8.\nonumber\end{align} The anti-fundamental representation corresponds to the Hermitian conjugation $e^-=(e^+)^\dagger$. Both primary fields carry conformal scaling dimension $h_{e^\pm}=1/3$.

The fundamental 27-dimensional representation of $E_6$ corresponds to the primary field $\overline{e^+}\sim\mathrm{span}\{e^{i\boldsymbol\beta\cdot\boldsymbol\phi}:\boldsymbol\beta\in\mathcal{S}_{E_6}\}$, where the weight vectors are \begin{align}\mathcal{S}_{E_6}&=\{-\boldsymbol\beta_8+{\bf e}_I+{\bf e}_J:3\leq I<J\leq8\}\nonumber\\&\quad\quad\cup\{\boldsymbol\beta_6-{\bf e}_I+{\bf e}_8:I=3,\ldots,8\}\nonumber\\&\quad\quad\cup\{-\boldsymbol\beta_1-{\bf e}_I+{\bf e}_8:I=3,\ldots,8\},\\\boldsymbol\beta_6&=\frac{1}{6}(-3,-3,1,1,1,1,1,-5),\nonumber\\\boldsymbol\beta_1&=\frac{1}{6}(-3,-3,5,-1,-1,-1,-1,-1).\nonumber\end{align} The anti-fundamental representation corresponds to the primary field $\overline{e^-}\sim\mathrm{span}\{e^{-i\boldsymbol\beta\cdot\boldsymbol\phi}:\boldsymbol\beta\in\mathcal{S}_{E_6}\}=(\overline{e^+})^\dagger$. The two primary fields both share the same conformal scaling dimension $h_{\overline{e^\pm}}=2/3$.

\section*{Duality properties for E-series}
The ground state structure of E-series has similar behaviors like A- and D-series, namely, if we extend the coupling constant of the sine-Gordon terms to complex regime, duality transformation acts as a reflection with respect to the real axis of the complex plane of the coupling constant. Although we cannot visualize it due to the high dimensionality of the root systems, it is reasonable to conclude that all the points on the complex plane describe a gapped surface except those points on the negative real axis.

\section{Conclusions and discussions}
\label{sec:DisCon}	
We systematically studied Abelian surface topological orders that fall under the {\it ADE} classification of simply-laced Lie algebras, as well as their symmetries and dualities properties via coupled wire models. A summary was given in Sec.~\ref{sec:introsummary} in the introduction. Here, we further elaborate on particular results that were not covered in Sec.~\ref{sec:introsummary}. The SPT/SET surface degrees of freedom were first projected onto an array of wires with alternating propagating directions by a generic symmetry-breaking surface stripe order. These chiral wires were then decomposed and backscattered to neighboring wires, thereby obtaining a finite excitation energy gap. We derived the exactly solvable ground state structures as well as the properties quasiparticle excitations by studying the inter-wire sine-Gordon Hamiltonians of the bosonized variables. Specifically, for the $D$ series, the antiferromagnetic time-reversal symmetry defined in Ref.~\onlinecite{Sahoo2016coupledwireTSC} was dualized to a particle-hole-like symmetry. For the $A$ series, the mixing between the $U(1)$ charge and the neutral $SU(N)$ sectors allowed us to construct a theory that supports $\pi$-fluxes that mimics a $\mathbb{Z}_2$ orbifold/gauge theory. Throughout the $ADE$ discussions, we noticed that all the current backscattering interactions were self-dual in the sense that their dualized gapping terms had the same form as their original ones, except for the special $D$ series case of $SO(4)$ which required alternative treatment and was out of the scope of this paper.

This paper provides several future directions. (1) Based on the {\it ADE} classifications that are explored here as parent states, it is interesting to study the descendant topological states, for instance, $E_8$ quantum Hall state.~\cite{LopesQuitoHanTeo2019E8G2F4} (2) Our analysis can be systematically generalized to non-simply-laced affine Lie algebras. There has already been some specific progress in this direction~\cite{Sahoo2016coupledwireTSC,Cheng2018STOTCSC}. (3) The general ground state degeneracy (GSD) and modular properties when the model is compactified on a closed surface need to be carefully addressed in future works. This is especially the case for the non-Abelian theories. GSD of orbifold structures that support $\pi$-fluxes, similar to those appeared in the $A$ series, should also be explicitly analyzed. (4) The duality analysis suggests the coupled wire models are particular exact solvable points that belong in a moduli space of surface states that bridges between different dual phases through phase transitions. It would be interesting to explore these moduli spaces of surface states in a controlled but perhaps non-exactly solvable coupled wire manner. Moreover, it would be interesting to utilize the coupled-wire constructions to establish the dualities between non-Abelian gauge theories proposed recently~\cite{Hsin2016SUduality,Aharony2017SOSPduality}. (5) Topological phases and dualities in (3+1)D systems can also be studied using the coupled-wire construction. There have already been several attempts~\cite{Raza2017DSMcoupledwire,IadecolaNeupertChamonMudry16,IadecolaNeupertChamonMudry17} in particular situations, and it would be interesting to perform a systematic exploration that encompasses and classifies phases with similar properties.

\acknowledgments
We are thankful to Mayukh Khan and Taylor Hughes for insightful discussions that inspired the application of the $ADE$ classification to surface topological orders. JCYT is supported by the National Science Foundation under Grant No.~DMR-1653535.

\appendix
\section{Gapping conditions for K-matrix formalism}
\label{append:HaldaneCondition}

\subsection{Gapping terms for the general K-matrix theory}
We briefly review the gapping condition and the gapping term for the general K-matrix theory. Assume we have two effective Lagrangians on the boundary of a (2+1)D system:
\begin{align}
\m{L}_L &= -\frac{1}{4\pi} K^L_{IJ} \p_t \phi^L_I \p_x \phi^L_J + V^L_{IJ} \p_x \phi^L_I \p_x \phi^L_J, \nonumber \\
\m{L}_R &= \frac{1}{4\pi} K^R_{IJ} \p_t \phi^R_I \p_x \phi^R_J + V^R_{IJ} \p_x \phi^R_I \p_x \phi^R_J,
\end{align}
where $K^R$ and $K^L$ have the same dimension $N$ and signature, and $V^R$ and $V^L$ are some symmetric non-universal potentials. Define $\mathsf{K} \equiv K^R \oplus (-K^L)$. The completely gapping condition or Haldane's nullity condition~\cite{Haldane1995Haldanecondition} is that there exists $N$ 2N-component linearly independent integer vectors $\b{\ell}_i=(\b{\ell}^R_i,\b{\ell}^L_i)^T$, called null vectors, satisfying 
\begin{align}
\b{\ell}_i^T \mathsf{K} \b{\ell}_j &= 0,\quad i,j=1,\dots,N. \label{eqn:HaldaneCond2}
\end{align}
Then the whole gapping term is written as
\begin{align}
\m{H}_{gapping} &= \sum_{i=1}^N C_i \cos{\left( \b{\ell}^T_i \mathsf{K} \mathbf{\Phi} + \alpha_i \right)},
\end{align}
where $\mathbf{\Phi}=(\b{\phi}^R,\b{\phi}^L)^T$ and $\alpha_i$ are some undetermined variables, which can be fixed by the specific theory. 

Actually if we only impose that we can pin the gapping terms simultaneously to their minima, we only need $\b{n}^T_i \mathsf{K}^{-1} \b{n}_j =0$ for $i,j=1,\dots,N$, where $\b{n}_i$ are integer vectors. However, if we further require that the gapping terms are composed of local operators, we need $\b{n}_i = \mathsf{K} \b{\ell}_i$, which gives Eq.(\ref{eqn:HaldaneCond2}). 

One corollary is that when $K^R=K^L$, then we can always choose $\mathbf{l}^R_i = \mathbf{l}^L_i$ to gap out the whole system, as long as there are enough linearly independent $N$-component integer vectors $\mathbf{l}^R_i$. 

\subsection{Gapping conditions in different basis}
For a general $K$-matrix theory with simply-laced algebra, we can write the kinetic term in two equivalent ways
\begin{align}
\m{L}_0 &= \frac{1}{4\pi} \int dxdt \ K_{IJ} \p_x \phi'^I \p_t \phi'^J,
\end{align}
with the canonical quantization
\begin{align}
[\phi'^I(x), \p_{x'} \phi'^J(x')] &= 2\pi i K^{-1}_{IJ} \d(x-x'). \label{eqn:KComm1}
\end{align}
We can choose simple roots for the current algebra $\alpha_I$ such that $\alpha_I \cdot \alpha_J = K_{IJ}$. We denote
\begin{align}
R &= \begin{pmatrix}
---- & \alpha_1 & ---- \\
 & \vdots & \\
---- & \alpha_r & ----
\end{pmatrix}
\end{align}
as the matrix formed by these simple roots, where $r$ is the rank of the Lie algebra. Then we have $R R^T = K$. Now we make a basis transformation
\begin{align}
\phi^I &= \sum_J R_{JI} \phi'^J. \label{eqn:BasisTransA1}
\end{align}
Then we can check that Eq.~(\ref{eqn:KComm1}) becomes
\begin{align}
[\phi^I(x), \p_{x'} \phi^J(x')] &= 2\pi i \d^{IJ} \d(x-x'),
\end{align}
where we have used $R^T K^{-1} R =1$, which is obvious. If $\b{\ell}^i= (\b{\ell}^i_R,\b{\ell}^i_L)$ is a set of $2r$-component Haldane null vectors, they should satisfy the nullity condition
\begin{align}
(\b{\ell}^i)^T \mathsf{K} \b{\ell}^j &= 0, \quad \forall i,j=1,\dots, r, 
\end{align}
in the $\phi'^I$ basis, where $\mathsf{K}= K \oplus (-K)$, or 
\begin{align}
\b{\ell}^i_R \cdot \b{\ell}^j_R - \b{\ell}^i_L \cdot \b{\ell}^j_L &= 0, \quad \forall i,j=1,\dots,r,
\end{align}
in the $\phi^I$ basis.

\section{Simply-laced Lie algebras and their representations}
\label{append:Liealgebra}
We review the simply-laced Lie algebras, namely, {\it ADE} classifications, and their representations here. \cite{bigyellowbook} ``Simply-laced'' means that all roots $\bm{\alpha}$ of the corresponding algebras have identical length, which are usually normalized to be $|\bm{\alpha}|=\sqrt{2}$. Let $r$ be the rank of an algebra $G$, namely, the maximal number of mutually commuting generators of $G$. Then in {\it Cartan-Weyl} basis, we have
\begin{align}
\left[H^i, E^{\bm{\alpha}}\right] &= \alpha^i E^{\bm{\alpha}}, \nonumber \\
\left[E^{\bm{\alpha}}, E^{-\bm{\alpha}}\right] &= \frac{2}{|\bm{\alpha}|^2} \sum^r_{i=1} \alpha^i H^i = \sum^r_{i=1} \alpha^i H^i, \nonumber \\
\left[E^{\bm{\alpha}}, E^{\bm{\beta}}\right] &\propto \begin{cases}
E^{\bm{\alpha+\beta}} \quad &\text{if}\  \bm{\alpha+\beta} \in \D, \\
0 \quad &\text{otherwise}
\end{cases} \quad \text{for} \ \bm{\alpha} \neq \bm{\beta}.
\end{align}
All roots of $G$ can be obtained from $r$ simple roots $\bm{\alpha}_1, \dots, \bm{\alpha}_r$ by linear combinations.  The choice of simple roots is not unique. For $SU(r+1)$ algebras, it can be chosen as 
\begin{align}
\bm{\alpha}_I &= \mathbf{e}_I - \mathbf{e}_{I+1}, \quad I = 1,\dots,r,
\end{align}
where $\bm{e}_I$ are unit basis vectors of $\mathbb{R}^{r+1}$. For $SO(2r)$ algebras, it can be chosen as
\begin{align}
\bm{\alpha} &= \begin{cases}
\mathbf{e}_I- \mathbf{e}_{I+1} \quad &\text{for} \ I=1,\dots,r-1, \\
\mathbf{e}_{r-1} + \mathbf{e}_r \quad &\text{for} \ I=r,
\end{cases}
\end{align}
where $\bm{e}_I$ are unit basis vectors of $\mathbb{R}^{r}$. For E-series, simple roots are usually taken at one's convenience. We have shown some particular choices in the main text. 

The fundamental representation $t^a$ of $SU(r+1)$ algebra have properties
\begin{align}
\text{Tr} (t^a t^b) &= \d^{ab}, \nonumber \\
\sum_a t^a_{ij} t^a_{kl} &= \d_{il} \d_{jk} - \frac{1}{r+1} \d_{ij} \d_{kl}, \\
\sum_{a,b} f_{abc} f_{abd} &= 2(r+1) \d_{cd}, \nonumber
\end{align}
where $f_{abc}$ are the structure constants of the $SU(r+1)$ algebra. The vector representation of $SO(2r)$ Lie algebra has an explicit matrix representation 
\begin{align}
t^a_{ij} \equiv t^{rs}_{ij} &= i (\d^r_i \d^s_j - \d^r_j \d^s_i), \quad 1 \leq r<s \leq 2r, \nonumber \\
\text{Tr}(t^a t^b) &= 2\d_{ab}, \\
\sum_a t^a_{ij} t^a_{kl} &= 2(-\d_{ik} \d_{jl} + \d_{il} \d_{jk}),
\end{align}
and the structure constant can be written as
\begin{align}
f_{abc} \equiv f_{(rs)(pq)(mn)} &= (\d_{rm} \d_{nq} \d_{sp} - \d_{ms} \d_{nq} \d_{rp}) \nonumber \\
& \quad+ (\d_{mp} \d_{sq} \d_{nr} - \d_{np} \d_{sq} \d_{rm})\nonumber \\
& \quad + (\d_{pr} \d_{ns} \d_{mq} - \d_{rq} \d_{ns} \d_{mp}).
\end{align}

The Cartan matrix $K$ of the algebra $G$ is an $r \times r$ matrix defined by 
\begin{align}
K_{IJ} &= \frac{2 \bm{\alpha}^T_I \bm{\alpha}_J}{|\bm{\alpha}_J|^2} = \sum^r_{i=1} \frac{2\alpha^i_{\ I} \alpha^i_{\ J}}{|\bm{\alpha}_J|^2} = \sum^r_{i=1} \alpha^i_{\ I} \alpha^i_{\ J}.
\end{align}
It is easy to see that the Cartan matrix for simply-laced algebras are symmetric. Cartan matrices for simply-laced algebras are listed below. 
\newpage
\begin{align}
K_{SU(r+1)} &= \begin{pmatrix}
2 & -1 & 0 & \cdots & 0 & 0 \\
-1 & 2 & -1 & \cdots & 0 & 0 \\
0 & -1 & 2 & \cdots & 0 & 0 \\
\vdots & \vdots & \vdots & \ddots & \vdots & \vdots \\
0 & 0 & 0 & \cdots & 2 & -1 \\
0 & 0 & 0 & \cdots & -1 & 2
\end{pmatrix},\nonumber \\
K_{SO(2r)} &= \begin{pmatrix}
2 & -1 & 0 & \cdots 0 & 0 & 0 \\
-1 & 2 & -1 & \cdots 0 & 0 & 0 \\
0 & -1 & 2 & \cdots 0 & 0 & 0 \\
\vdots & \vdots & \vdots & \ddots & \vdots & \vdots \\
0 & 0 & 0 & \cdots 2 & -1 & -1 \\
0 & 0 & 0 & \cdots -1 & 2 & 0 \\
0 & 0 & 0 & \cdots -1 & 0 & 2
\end{pmatrix}, \quad (r \geq 4), \nonumber \\
K_{E_8} &= \begin{pmatrix}
2 & -1 & 0 & 0 & 0 & 0 & 0 & 0 \\
-1 & 2 & -1 & 0 & 0 & 0 & 0 & 0 \\
0 & -1 & 2 & -1 & 0 & 0 & 0 & 0 \\
0 & 0 & -1 & 2 & -1 & 0 & 0 & 0 \\
0 & 0 & 0 & -1 & 2 & -1 & 0 & -1 \\
0 & 0 & 0 & 0 & -1 & 2 & -1 & 0 \\
0 & 0 & 0 & 0 & 0 & -1 & 2 & 0 \\
0 & 0 & 0 & 0 & -1 & 0 & 0 & 2
\end{pmatrix}, \nonumber \\
K_{E_7} &= \begin{pmatrix}
2 & -1 & 0 & 0 & 0 & 0 & 0 \\
-1 & 2 & -1 & 0 & 0 & 0 & 0 \\
0 & -1 & 2 & -1 & 0 & 0 & 0 \\
0 & 0 & -1 & 2 & -1 & 0 & -1 \\
0 & 0 & 0 & -1 & 2 & -1 & 0 \\
0 & 0 & 0 & 0 & -1 & 2 & 0 \\
0 & 0 & 0 & -1 & 0 & 0 & 2
\end{pmatrix}, \nonumber \\
K_{E_6} &= \begin{pmatrix}
2 & -1 & 0 & 0 & 0 & 0 \\
-1 & 2 & -1 & 0 & 0 & 0 \\
0 & -1 & 2 & -1 & 0 & -1 \\
0 & 0 & -1 & 2 & -1 & 0 \\
0 & 0 & 0 & -1 & 2 & 0 \\
0 & 0 & -1 & 0 & 0 & 2
\end{pmatrix}. \label{eqn:CartanE876}
\end{align}
Sometimes it is convenient to use {\it Chevalley} basis as it is directly related to the Cartan matrix:
\begin{align}
h^I &= \frac{2}{|\bm{\alpha}_I|^2} \sum^r_{i=1} \alpha^i_{\ I} H^i = \sum^r_{i=1} \alpha^i_{\ I} H^i,
\end{align}
with the commutation relations
\begin{align}
\left[ h^I, E^{\pm \bm{\alpha}_J} \right] &= \pm K_{IJ} E^{\pm \bm{\alpha}_J} , \quad \left[ E^{\bm{\alpha}_J}, E^{-\bm{\alpha}_J} \right] = \d^{IJ} h^J.
\end{align}

In this paper, we are focused on the level-1 algebras of $ADE$ classifications, in which there exist free field representations. To be specific, $SO(2r)_1$ algebras (D-series), can be constructed by $2r$ independent Majorana fermions $\psi^i$ with operator product expansions (OPEs)
\begin{align}
\psi_i (z) \psi_j(w) &\sim \frac{\d_{ij}}{z-w}, \quad i,j=1,\dots,2r.
\end{align}
The current operators can be constructed with these free Majorana fermions as
\begin{align}
J^a(z) &= \frac{1}{2} \sum_{i,j} (\psi_i t^a_{ij} \psi_j)(z),
\end{align}
where normal ordering is assumed. One can check that these currents satisfy the current algebra
\begin{align}
J^a(z) J^b(w) &\sim \frac{k\d_{ab}}{(z-w)^w} + \sum_c \frac{if_{abc} J^c(w)}{(z-w)},
\end{align}
where $f_{abc}$ are called structure constants. 

For $SU(r+1)_1$ algebras (A-series), we can use $r$ independent free bosons $\phi^i$ with OPEs
\begin{align}
\phi^i(z) \phi^j(w) &\sim -\d_{ij} \ln{(z-w)}, \quad i,j=1,\dots,r.
\end{align}
The currents in Cartan-Weyl basis can be constructed as
\begin{align}
H^j(z) &= i \p \phi^j(z), \nonumber \\
E^{\b{\alpha}}(z) &= c_{\b{\alpha}} e^{i \b{\alpha} \cdot \b{\phi}(z)},
\end{align}
where $c_{\b{\alpha}}$ is a correction factor ensuring the correct OPEs. This bosonic construction also works for D-series if we pair up Majorana fermions and then bosonize them. 
For $(E_8)_1$ algebras (E-series), we can follow the same construction as in A-series with 8 independent free bosons to construct the currents, with the vector and spinor representations of $SO(16)$ algebra introduced in the main text. $E_7$ and $E_6$ algebras can be constructed from the corresponding conformal embeddings, respectively.

%

\end{document}